\documentclass{aa}  
\usepackage{amssymb}
\usepackage{xcolor}
\usepackage{graphicx}
\usepackage{hyperref}
\newcommand*{\bba}{$^{\scriptstyle 3\mathrm{D}}$B{\sc {arolo}}}
\newcommand*{\galfit}{G{\sc {alfit}}}
\newcommand*{\cannubi}{C{\sc {annubi}}}
\newcommand*{\stardust}{S{\sc {tardust}}}

\newcommand*{\casa}{C{\sc {asa}}}
\newcommand*{\grizli}{G{\sc {rizli}}}
\def\CIi{[CI]($^3P_1 - ^3P_0$)}
\def\CIii{[CI]($^3P_2  - ^3P_1$)}

\usepackage{txfonts}
\begin{document}

   \title{The ALMA-ALPAKA survey I}
   \subtitle{High-resolution CO and [CI] kinematics of star-forming galaxies at $z= 0.5$ - 3.5}

\author{F. Rizzo
        \inst{1, 2}
        \and
        F. Roman-Oliveira
        \inst{3}
        \and
        F. Fraternali
        \inst{3}
        \and
        D. Frickmann
        \inst{1, 2}
        \and
        F. M. Valentino
        \inst{4, 1}
        \and
        G. Brammer
        \inst{1, 2}
        \and
        A. Zanella
        \inst{5}
        \and
        V. Kokorev
        \inst{3}
        \and
        G. Popping
        \inst{4}
        \and
        K. E. Whitaker
        \inst{6, 1}
        \and
        M. Kohandel
        \inst{7}
        \and
        G. E. Magdis
        \inst{1, 8, 2} 
        \and
        L. Di Mascolo
        \inst{9, 10, 11}
        \and
        R. Ikeda
        \inst{12, 13}
        \and
        S. Jin
        \inst{1, 8}
        \and
        S. Toft
        \inst{1, 2}
        }  

   \institute{Cosmic Dawn Center (DAWN)
   \and Niels Bohr Institute, University of Copenhagen, Jagtvej 128, 2200 Copenhagen N, Denmark\\
              \email{francesca.rizzo@nbi.ku.dk}
              \and
             Kapteyn Astronomical Institute, University of Groningen, Landleven 12, 9747 AD, Groningen, The Netherlands
             \and
             European Southern Observatory, Karl-Schwarzschild-Str. 2, D-85748 Garching bei München, Germany
             \and
             Istituto Nazionale di Astrofisica, Vicolo dell'Osservatorio 5, 35122, Padova, Italy
             \and
             Department of Astronomy, University of Massachusetts, Amherst, MA 01003, USA
             \and
             Scuola Normale Superiore, Piazza dei Cavalieri 7, I-56126 Pisa, Italy
             \and
             DTU-Space, Technical University of Denmark, Elektrovej 327, DK-2800, Kgs. Lyngby, Denmark
             \and
             Astronomy Unit, Department of Physics, University of Trieste, via Tiepolo 11, Trieste 34131, Italy
            \and 
            INAF -- Osservatorio Astronomico di Trieste, via Tiepolo 11, Trieste 34131, Italy
            \and 
            IFPU -- Institute for Fundamental Physics of the Universe, Via Beirut 2, 34014 Trieste, Italy
            \and
            Department of Astronomy, School of Science, SOKENDAI (The Graduate University for Advanced Studies), 2-21-1 Osawa, Mitaka, Tokyo 181-8588, Japan
            \and
            National Astronomical Observatory of Japan, 2-21-1 Osawa, Mitaka, Tokyo 181-8588, Japan 
             }

  \abstract
   {Spatially-resolved studies of the kinematics of galaxies provide crucial insights into their assembly and evolution, enabling to infer the properties of the dark matter halos, derive the impact of feedback on the interstellar medium (ISM), measure and characterize the outflow motions. To date, most of the kinematic studies at $z = 0.5 - 3.5$ were obtained using emission lines tracing the warm, ionized gas (e.g., H$\alpha$, [OII], [OIII]). But, whether these provide an exhaustive or only a partial view of the dynamics of galaxies and of the properties of the ISM is still debated. Complementary insights on the cold gas kinematics are therefore needed.}
   {We present the ``Archival Large Program to Advance Kinematic Analysis'' (ALPAKA), a project aimed at gathering high-resolution observations of CO and [CI] emission lines of star-forming galaxies at $z = 0.5 - 3.5$ from the Atacama Large Millimeter Array (ALMA) public archive. With $\approx$ 147 hours of total integration time, ALPAKA assembles $\sim$ 0.25$\arcsec$ observations for 28 star-forming galaxies, the largest sample with spatially-resolved cold gas kinematics as traced by either CO or [CI] at $z \gtrsim 0.5$, spanning 7 Gyr of cosmic history. A large fraction of ALPAKA galaxies (19/28) lie in overdense regions (clusters, groups, and protoclusters).}
   {
   By combining multi-wavelength ground- and space-based ancillary data, we derive the stellar masses ($M_{\star}$) and star-formation rates (SFR) for the ALPAKA targets. We exploit the ALMA data to infer the dynamical state of the ALPAKA galaxies and derive their rotation curves and velocity dispersion profiles using \bba.}
   {ALPAKA probes the massive ($M_{\star}\gtrsim 10^{10}$ M$_{\odot}$), actively star-forming (SFR $\approx 10 - 3000$ M$_{\odot}$\,yr$^{-1}$) part of the population of galaxies at $z \sim 0.5 - 3.5$. 
   Based on our kinematic classification, we find that 19/28 ALPAKA galaxies are rotating disks, 2 are interacting systems, while for the remaining 7 sources the classification is uncertain. The disks have velocity dispersion values that are 
   typically larger in the innermost regions than in the outskirts, with a median value for the entire disk sample of
   35$^{+11}_{-9}$ km\,s$^{-1}$. Despite the bias of our sample towards galaxies hosting very energetic mechanisms, the ALPAKA disks have high ratios of ordered-to-random motion ($V/\sigma$) with a median value of 9$^{+7}_{-2}$.  
   }
   {}

   \keywords{Galaxies: evolution -- 
   Galaxies: high-redshift --
   Galaxies: ISM --
    Galaxies: kinematics and dynamics --
    Galaxies: photometry --
    Galaxies: structure}

   \maketitle
%
\section{Introduction} \label{sec:intro}
According to the current paradigm of galaxy formation and evolution, the assembly of galaxies is regulated by a variety of physical processes: the interplay between dark and baryonic matter, gas accretion, galaxy mergers, star formation, and stellar and active galactic nucleus (AGN) feedback \citep{Mo_2010, Cimatti_2019, Vogelsberger_2020}. From the theoretical perspective, state-of-the-art cosmological simulations reproduce most of the global properties of galaxies at different cosmic epochs \citep[e.g.,][]{Schaye_2015, Nelson_2018, Pillepich_2018, Roca_2021, Kannan_2022, Pallottini_2022}. However, this success does not necessarily imply that we understand how galaxies form and evolve in detail, as the modeling of processes acting on scales below the resolution of simulations (e.g., stellar and AGN feedback, star formation) relies on assumptions and calibrations for the so-called sub-grid models \citep[e.g.,][]{Naab_2017, Vogelsberger_2020}.
As a consequence, the kpc and sub-kpc spatial distributions of the baryonic and dark matter within galaxies strongly vary with the adopted models \citep{Kim_2016, Roca_2021}. 

From the observational perspective, the role played by various processes in driving galaxy evolution is still unclear, even in redshift ranges that have been widely studied in the last decades (e.g., $z = 0.5 - 3$). Understanding the assembly of galaxies requires simultaneous high-resolution, multi-wavelength studies of their morphologies and kinematics. The morphological analysis of the distribution of stars \citep[e.g.,][]{Lang_2014, vanderwel_2014, Mowla_2019, Ferreira_2022, Kartaltepe_2022} and gas/dust \citep[e.g.,][]{Calistro_2018, Gullberg_2019, Hodge_2019, Rujopakarn_2019, Kaasinen_2020, Tadaki_2020, Puglisi_2021, Ikeda_2022} within galaxies allows one to trace the build up of structures.
Galaxy kinematics, on the other hand, provide constraints on the impact of gas accretion, mergers, outflows, and environmental mechanisms on the growth of galaxies \citep[e.g.,][]{Ubler_2018, Loiacono_2019, Concas_2022, Bacchini_2023, Roman_2023}. For instance, the prevalence of rotating disks among the star-forming population at $z \sim 1 - 3$ has been established as a key evidence that their assembly is mainly driven by smooth gas accretion as opposed to mergers \citep[e.g.,][]{Wisnioski_2015, Stott_2016, Forster_2020}. In addition, for galaxy disks, measurements of gas rotation allow one to study the dynamics and infer the content and distribution of dark matter within galaxies \citep[e.g.,][]{Straatman_2017, Posti_2019, Price_2021, Sharma_2022, Mancera_2022, Lelli_2023}. In contrast, measurements of gas velocity dispersion provide key insights into the mechanisms injecting turbulence into the interstellar medium \citep[ISM; e.g., stellar feedback, release of gravitational energy through accretion from the cosmic web, radial flows within galaxies;][]{Krumholz_2018, Ubler_2019, Kohandel_2020, Rizzo_2021, Jimenez_2022, Rathjen_2022}. 

To date, most of the information on galaxy kinematics at $z \sim 0.5 - 3.5$ have been derived from integral field unit (IFU) observations of optical emission lines tracing warm ionized gas \citep[e.g., H$\alpha$, {[OIII]}, {[OII]};][]{Forster_2006, Stott_2016, Turner_2017, Forster_2018, Wisnioski_2019}. A common finding of these studies is that the turbulence within both dusty starburst and main-sequence star-forming galaxies systematically increases from $z = 0$ to $z \sim 3$ \citep{Forster_2006, Stott_2016, Turner_2017, Forster_2018, Wisnioski_2019, Birkin_2023} and the degree of rotation support ($V/\sigma$) decreases from typical values of 10 in local spiral galaxies to 
 2 at $z = 3.5$ \citep{Gnerucci_2011, Turner_2017}.

At $z \gtrsim 4$, high-resolution galaxy kinematics have been studied for tens of galaxies, using Atacama Large Millimeter Array (ALMA) observations targeting the [CII]-158 $\mu$m emission line, a cold gas tracer \citep[e.g.,][]{Neeleman_2020, Rizzo_2020, Fraternali_2021, Jones_2021, Tsukui_2021, Herrera_2022, Posses_2023, Roman_2023}. Most of these studies, mainly obtained on targets pre-selected as dusty starbursts, found that a large fraction of star-forming galaxies at $z = 4 - 5$ are dynamically cold disks with $V/\sigma$ ratios as high as 10 \citep[e.g.,][]{Rizzo_2020, Fraternali_2021, Lelli_2021, Rizzo_2021, Roman_2023}.

Three reasons could be responsible for the discrepancies in the two redshift ranges (i.e., $z = 0.5 - 3.5$ and $z = 4 - 5$): spectral resolution, beam-smearing corrections and the use of different tracers of the ISM phases. On the one hand, IFU observations are characterized by spectral resolutions that are typically a factor of 3 worse than those achievable with ALMA. Furthermore, discrepancies between different studies may arise due to the techniques employed for deriving beam-smearing corrected kinematic parameters. The beam-smearing effect causes, in fact, a degeneracy between the rotation velocity and velocity dispersion, mainly resulting in inflated values of $\sigma$ \citep[e.g.,][]{Begeman_1989}. In the past decade, different techniques have been developed to correct for beam-smearing effects. They rely on either applying forward-modeling approaches to fit the data cubes \citep[e.g.,][]{Bouche_2015, DiTeodoro_2015} or correcting the velocity dispersion maps a-posteriori \citep[e.g.,][]{Swinbank_2012, Stott_2016, Johnson_2018}. Using ALMA mock data at different angular resolutions, \citet{Rizzo_2022} showed that the a-posteriori correction is suboptimal, resulting in overestimation of $\sigma$ up to 200\% when the galaxy is spatially resolved with only a few resolution elements. Instead, the typical errors of the velocity dispersion using forward-modeling techniques are of the order of 25\% \citep{DiTeodoro_2015, Rizzo_2022}.
On the other hand, the distinct gas phases probed by different emission lines may also play a critical role. Kinematic studies at $z \approx 0$ show that the velocity dispersion is typically higher when measured with warm ionized gas as opposed to molecular or atomic gas \citep{Levy_2018, Girard_2021}. At high-$z$, simultaneous studies of the cold and warm kinematics on the same galaxies are still in their infancy and limited in statistics \citep[e.g.,][]{Ubler_2018, Molina_2019, Lelli_2023}. Furthermore, the warm gas might be more affected by the presence of gas in outflows than the cold gas\citep{Lelli_2018, Ejdetjarn_2022, Kretschmer_2022}. Should this be the case, the rotation velocity and velocity dispersion from warm gas might be uncertain as these two quantities are typically derived assuming that the gas moves in circular orbits, while radial or vertical motions could bias their measurements. 

Systematic, high-resolution kinematic measurements obtained using cold gas tracers are needed not only to gain a comprehensive insight into the ISM properties of $z \sim 0.5 - 3.5$ galaxies but also to make an accurate comparison with the dynamical properties of galaxies at $z \gtrsim 4$. Unfortunately, because of the sensitivity and frequency coverage of state-of-the-art facilities, high-resolution [CII] observations at $z \lesssim 3.5$ are challenging or not feasible, due to the low atmospheric transmission \citep{Carilli_2013}. To study the $z \sim 0.5 - 3.5$ kinematics using cold gas tracers, one can target carbon monoxide, CO, transitions or fine structure lines from atomic carbon, [CI]. However, emission lines from CO and [CI] are typically fainter than [CII] \citep{Carilli_2013, Bernal_2022}. To date, the only instrument capable of achieving the combined sensitivity and angular resolution requirements of $\lesssim 0.3 \arcsec$ needed for studying the CO or [CI] kinematics at $z \gtrsim 0.5$ is ALMA. However, 
even with ALMA, CO and [CI] observations are extremely time consuming. For this reason, studies of the kinematics using cold gas tracers at cosmic noon have been presented only for a handful of galaxies \citep[e.g.,][]{Molina_2019, Noble_2019, Kaasinen_2020, Xiao_2022, Lelli_2023}. Finally, since the published kinematic measurements are derived using different algorithms and assumptions, a systematic compilation and comparison of the results is not straightforward. 

Here we present the project ``ALMA Archival Large Program to Advance Kinematic Analysis" (ALMA-ALPAKA), aimed at filling all the above gaps by collecting high data quality emission line observations of $z = 0.5 - 3.5$ galaxies from the ALMA public archive. We note that the use of archival data does not have an impact on the originality of the present work as, to date, the ALMA observations for $\approx 67\%$ of the ALPAKA galaxies have not been published in any studies. With an increase in sample size by a factor of 3 compared to the total number of targets found in all previous works \citep[e.g.,][]{Molina_2019, Noble_2019, Kaasinen_2020, Xiao_2022, Lelli_2018, Lelli_2023}, ALPAKA will be used to obtain the first high-resolution characterization of the morpho-kinematic properties of galaxies at $z \sim 0.5 - 3.5$ using millimeter to optical observations.

In this paper, the first of a series, we present the ALPAKA sample, discuss the sample selection (Sect.\,\ref{sec:sample}) and describe the ALMA and \textit{Hubble} Space Telescope (HST) data analyzed in this work (Sect.\,\ref{sec:observations}). In Sect.\,\ref{sec:gloabl}, we derive the global properties of the ALPAKA targets using ancillary data. The kinematic modelling and assumptions, the identification of the disks and the description of their kinematic parameters is presented in Sect.\,\ref{sec:kinematic}. In Sect.~\ref{sec:details}, we describe the kinematics of each ALPAKA target.
In Sect.\,\ref{sec:discussion}, we discuss the potential bias due to selection effects and the dynamical properties of the subsample of ALPAKA disk galaxies. Finally, in Sect.\,\ref{sec:conclusion}, we summarize the main findings and describe the main objectives of the ALPAKA project.

Throughout this paper, we assume a $\Lambda$CDM cosmology with parameters from \citet{Planck} and a Chabrier Initial Mass Function \citep{Chabrier}.

\section{Sample description} \label{sec:sample}
\subsection{Selection criteria}
\begin{figure}[htbp!]
    \begin{center}  \includegraphics[width=0.95\columnwidth]{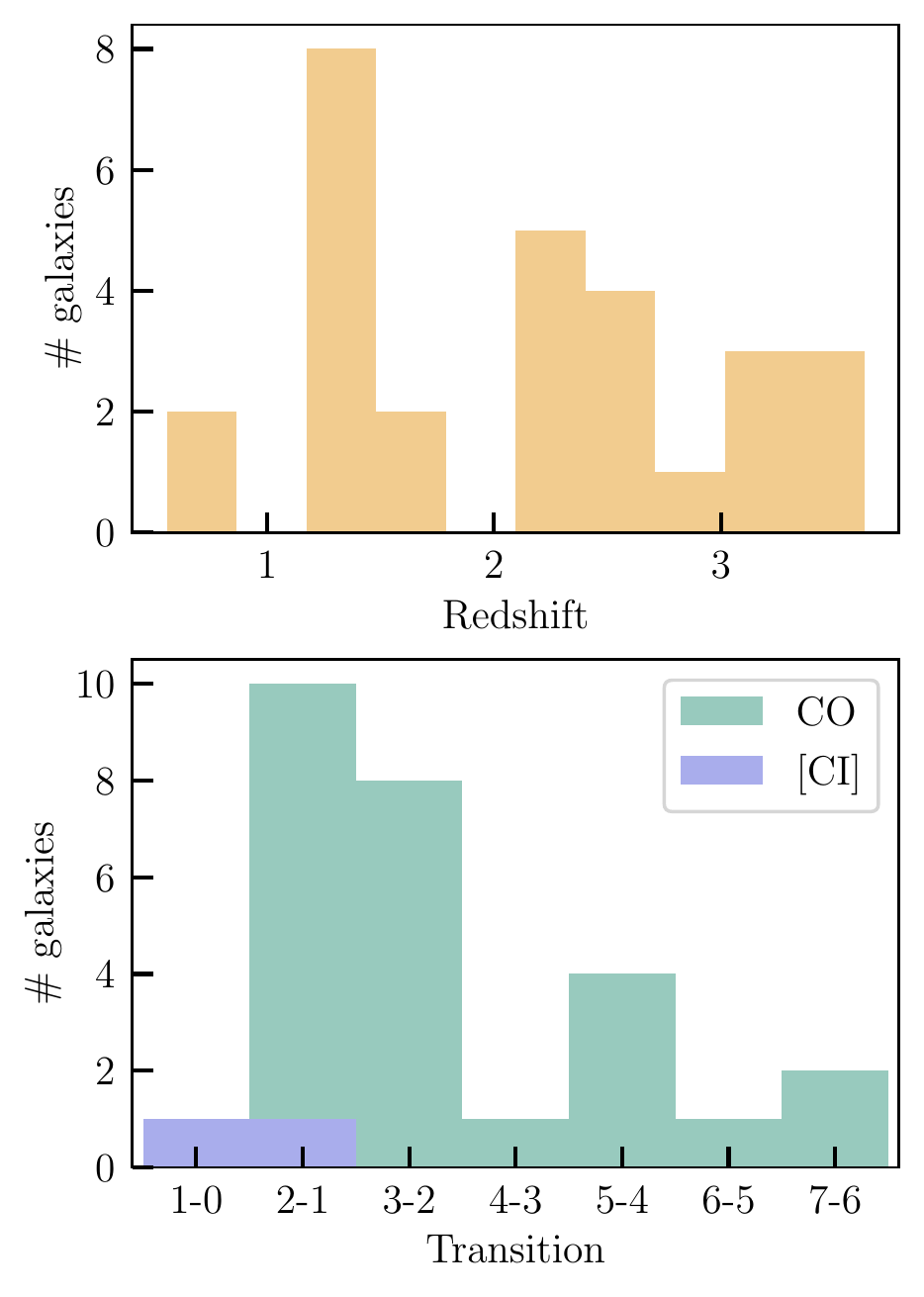}
        \caption{Distribution of the redshift (upper panel) and the rotational transitions of CO and atomic fine structure line transitions of [CI] (bottom panel) for the ALPAKA data set.}  
        \label{fig:zdistr}
    \end{center}
\end{figure}

The ALPAKA project is designed to study the kinematic and dynamical properties of galaxies at $z = 0.5 - 3.5$. To this end, we collected the sample by selecting data publicly available from the ALMA archive and with spectral setup covering CO and/or [CI] emission lines. We queried the database at the end of August 2022 to select galaxies with spectroscopic redshift in the range $0.5 - 4$, with angular resolution $\lesssim 0.5\arcsec$, spectral resolution $\lesssim 50$ km s$^{-1}$ and medium to high signal-to-noise ratio (SNR), i.e., data cubes with a SNR $\gtrsim$ 3.5 per channel in at least 5 spectral channels. These requirements guarantee data quality sufficient to infer robust kinematic parameters \citep{Rizzo_2022}. The resulting sample consists of 28 galaxies, whose ID, name, coordinates, and redshifts are reported in Table \ref{tab:tab1} (see Sect. \ref{sec:details} for details on each target). In Fig.~\ref{fig:zdistr} (upper panel), we show the histogram with the redshift distribution of the ALPAKA sample. We note that about half of the sample (12/28) covers the lowest redshift range ($z \lesssim 1.6$) with two targets at $z \approx 0.6$, while the remaining 16 galaxies are at $z \gtrsim 2$. The peak of 8 galaxies in the redshift bin centred at $z = 1.5$ is due to targets belonging to the same cluster (see Sect.~\ref{sec:target_ch}). 

\subsection{Target characterization} \label{sec:target_ch}
High-resolution ALMA observations of high-$z$ galaxies require long integration times. Therefore, to optimize the observing strategy, such observations are usually performed on sources for which the fluxes of the target emission lines are known through previous low angular resolution and shallow campaigns. Despite the fact that our targets were selected using an agnostic approach regarding the multi-wavelength coverage, we find that all ALPAKA galaxies 
lie in broadly studied survey areas. In Table \ref{tab:tab1}, we list the main features of the targets as reported in the literature: 
 the fields in which they lie, and information regarding the environment and the presence of AGN (see Sect. \ref{sec:details} for details). The latter are based on previous diagnostic features (e.g., rest-frame UV or mid-infrared spectra, X-ray luminosities) depending on spectroscopic coverage of the galaxies in different fields. 
Given that the selection criteria we adopt are only based on the data quality of archive data, the ALPAKA sample is heterogeneous: 68\% (19 out of 28) of the targets lie in overdense regions (clusters, groups and protocluster); 7\% (2 out of 28) were previously considered as merging systems and 1 of the two is in a group; 25\% (7 out of 28) host an AGN. Among the AGN hosts, 3 sources are protocluster members. A discussion on the potential impact of AGN feedback on the ISM kinematics properties of the host galaxies is provided in Sect.\ref{sec:discussion}.

\begin{table*}[htbp!] 
\begin{center}
\caption{The ALPAKA sample. }
\label{tab:tab1}
\begin{tabular}{ccccccc}
\\
\hline\hline \noalign{\smallskip}
ID  &  Name & RA (deg) & Dec (deg) & redshift & Field/Survey & Notes \\
\noalign{\smallskip}
\hline
\noalign{\medskip}
1 & GOODS-S 15503 & 53.08205 & -27.83995 & 0.561 & GOODS-S & - \\ 
2 & COSMOS 2989680 & 150.43186 & 2.80261 & 0.623 & COSMOS & Minor merger\\
3 & COSMOS 1648673 & 149.98144 & 2.25321 & 1.445 & COSMOS & AGN(?) \\
4 & ALMA.03  & 333.99393 & -17.62988 & 1.453 & XCS & Cluster member \\
5 & ALMA.010 & 333.98853 & -17.63149 & 1.453 & XCS & Cluster member\\ 
6 & ALMA.08 & 333.99266 & -17.63950 & 1.456 & XCS & Cluster member\\ 
7 & ALMA.01 & 333.99233 & -17.63737 & 1.466 & XCS & Cluster member\\%
8 & ALMA.06 & 333.99880 & -17.63306 & 1.467 & XCS & Cluster member \\
9 & ALMA.013 & 333.99909 & -17.63797 & 1.471 & XCS & Cluster member\\
10 & SHiZELS-19 & 149.79817 & 2.39008 & 1.484 & COSMOS & - \\
11 & SpARCS J0225-371 & 36.44191 & -3.92436 & 1.599 & SpARCS & Cluster member\\%
12 & SpARCS J0224-159 & 36.11320 & -3.40037 & 1.634 & SpARCS & Cluster member\\%
13 & COSMOS 3182 & 150.07594 & 2.21182 & 2.103 & COSMOS & Protocluster member\\
14 & Q2343-BX610 & 356.53934 & 12.82202 & 2.211 & SINS/zC-SINF & AGN(?)\\
15 & GS30274 & 53.13114 & -27.77319 & 2.225 & GOODS-S & AGN \\
16 & HXMM01-a & 35.06938 & -6.02830 & 2.311 & HerMES & Group member\\
17 & HXMM01-b+c & 35.06907 & -6.02904 & 2.308 & HerMES &  Group member, merger\\
18 & HATLAS J084933-W & 132.38994 & 2.24573 & 2.407 & H-ATLAS & AGN, protocluster member\\
19 & CLJ1001-131077 & 150.23728 & 2.33813 & 2.494 & COSMOS & Cluster member\\ 
20 & CLJ1001-130949 & 150.23691 & 2.33579 & 2.504 & COSMOS & Cluster member\\ 
21 & CLJ1001-130891 & 150.23986 & 2.33646 & 2.513 & COSMOS & Cluster member\\ 
22 & Gal3 & 150.33139 & 2.16239 & 2.935 & COSMOS & -\\
23 & ADF22.1 & 334.38507 & 0.29551 & 3.089 & SSA2 & AGN, protocluster member\\
24 & ADF22.5 & 334.38416 & 0.29323 & 3.094 & SSA2 & Protocluster member \\
25 & ADF22.7 & 334.38120 & 0.29946 & 3.088 & SSA2 & AGN, protocluster member \\
26 & Gal5 & 149.87724 & 2.28388 & 3.339 & COSMOS & -\\
27 & Gal4  & 150.27827 & 2.25887 & 3.431 & COSMOS & -\\
28 & W0410-0913 & 62.54425 & -9.21812 & 3.630 & WISE & AGN, protocluster member\\
\noalign{\smallskip}
\hline
\end{tabular}
\tablefoot{Columns 1 and 2: ALPAKA ID and mostly used name for the galaxy. Columns 3 and 4: coordinates of the center used for the kinematic fitting in Sect.\ref{sec:kinematic}. Column 5: redshift from the systemic velocity obtained from the kinematic fitting. Column 6: field or survey where the galaxies lie (GOOD-S: \citet{Giavalisco_2004}, COSMOS: \citet{Scoville_2007}, XCS: \citet{Romer_2001}, SpARCS: \citet{Muzzin_2009}, SINS/zC-SINF: \citet{Forster_2018}, HerMES: \citet{Oliver_2012}, H-ATLAS: \citet{Eales_2010}, SSA2: \citet{Steidel_1998}, WISE: \citet{Wright_2010}). Column 7: indication about the environment and the presence of an AGN taken from the literature (see details in Sect. \ref{sec:details}). The presence of a "?" indicates that for the corresponding galaxy there are hints of the presence of an AGN ( Sect.~\ref{sec:details}).}
\end{center}
\end{table*}

\section{Observations}\label{sec:observations}
 \subsection{ALMA data}\label{sec:alma}
 In Table \ref{tab:tab2}, we list the main properties of the ALMA datasets: ALMA project ID, frequency coverage and emission line for each target. When multiple observations at similar angular resolutions are available (e.g., ID10 and 14), we combine them to increase the SNR. Due to the selection criteria, and to the ALMA sensitivity and frequency coverage, the ALPAKA galaxies have observations of various emission lines: 18 targets have low-J CO transition observations -- CO(2-1) or CO(3-2) --, while for 10 targets only high-J CO transitions or [CI] emission lines are available (see bottom panel in Fig.~\ref{fig:zdistr} and Table \ref{tab:tab2}). The on-source integration times for each target are listed in Table \ref{tab:tab2} and range from 43 minutes to 14 hours, for a total of $\approx$ 90 hours. The total integration time, including overheads, is of 147 hours, corresponding to the duration of an ALMA Large Program.\\
In this paper, we make use of the calibrated measurement sets provided by the European ALMA Regional Centre \citep{Hatziminaoglou_2015}, that calibrated the raw visibility data using the standard pipeline script delivered with the raw observation sets. All of the post-processing steps were handled using the Common Astronomy Software Applications (\casa) suite \citep{McMullin_2007}, version 6. The calibrated data were first inspected to confirm the quality of the pipeline calibration and that no further flagging was required. For data sets containing one single target, we then subtracted the continuum from the line spectral windows using \texttt{UVCONTSUB}. Most of the data are averaged into groups of between two to six channels in order to obtain an average SNR of at least 3.5 per velocity channel in at least 5 spectral channels of the data cube\footnote{We define the SNR per channel, the ratio between the average of the masked line intensity per pixel and the noise.}. 
This procedure results in channels with typical velocity widths ranging from 16 to $\approx$ 39 km s$^{-1}$. The continuum and spectral lines were imaged using the \texttt{TCLEAN} routine in the \casa\ package, assuming a natural weighting of the visibilities to maximise the SNR. Targets with high SNR are imaged using a Briggs weighting of the visibilities \citep[robust parameter set equal to 0.5;][]{Briggs_1995}, in order to enhance the angular resolution of the output images, without significantly degrading their effective sensitivity and meet the requirements on the SNR per channel defined above. The \texttt{CLEAN} algorithm is run down to a flux threshold of 2 $\times$ RMS, where RMS is the root mean square of the data measured within the dirty data cubes. For datasets containing multiple targets, to better account for the source-to-source variation of the continuum signal, we perform the continuum subtraction using the \texttt{IMCONTSUB} task. In Table \ref{tab:tab2}, we present the main properties of the ALPAKA data cubes, their beam size, channel widths, and RMS per spectral channel. The beam of the observations of the ALPAKA targets ranges from 0.1$\arcsec$ to 0.5$\arcsec$ (median value of 0.25$\arcsec$) and the corresponding resolution in physical units varies between 1 kpc and 4 kpc (median value of 2 kpc).

\subsection{HST data}
Spatially resolved observations of the stellar continuum of galaxies allow one to derive not only the structural parameters defining the light distribution (e.g., effective radius, Sérsic index) but also to constrain the geometrical parameters (e.g., inclination angle) of the sources. As discussed in detail in Sect.~\ref{sec:geometry}, accurate measurements of the inclination and position angles provide robust velocity estimates and kinematic characterizations. The comparison between the stellar and gas morphology and the gas kinematics of a galaxy helps to identify any merger or outflow features (see Sect. \ref{sec:kinematic}). Further, combined measurements of the stellar and gas distributions and the rotation curves provide a unique means to infer the dark matter content within galaxies. \\
For 23 ALPAKA galaxies, HST observations are publicly available and are taken from the Complete Hubble Archive for Galaxy Evolution (CHArGE, Brammer et al., in prep.). The latter performs uniform processing of all archival HST imaging and slitless spectroscopy observations of high-$z$ galaxies. In CHArGE, the data were processed with the \grizli\ pipeline, which creates mosaics for all filter exposures that cover a given area of the sky \citep{Brammer_2021}.
All exposures are aligned to each other using different techniques \citep[see][for details]{Kokorev_2022} resulting in a typical astrometric precision $\lesssim$ 100 mas.

\begin{figure*}[htbp!] 
    \begin{center}
        \includegraphics[width=0.98\textwidth]{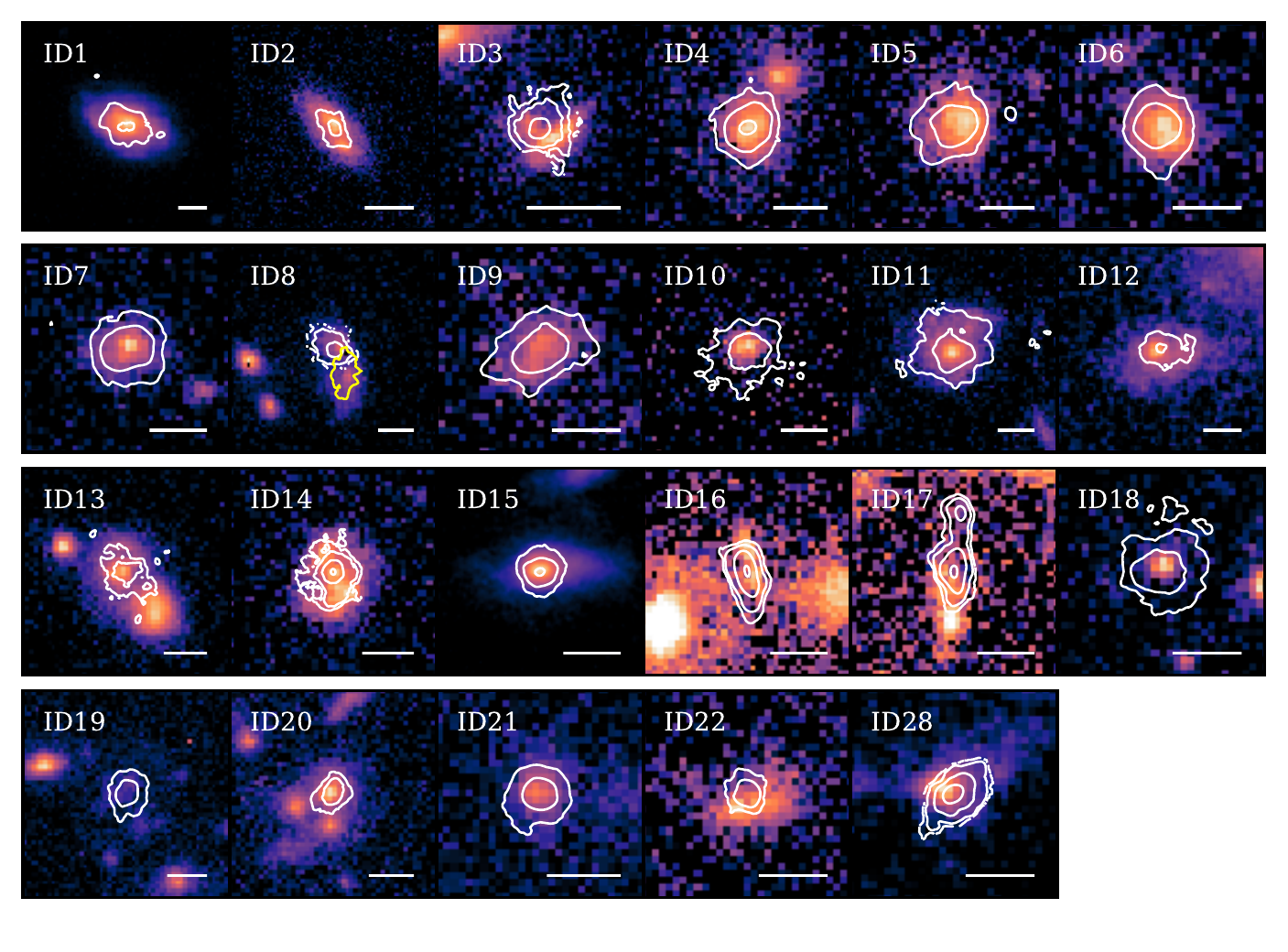}
        \caption{HST images with contours of the CO or [CI] total-flux maps from ALMA for the 23 ALPAKA galaxies with HST observations. The lowest white contour is a "pseudo-contour" 
 -- see Appendix B in \citet{Roman_2023} for details -- at 4 RMS. The additional yellow contour in the ID8 panel shows a second source detected in CO(2-1) (see Sect.~\ref{sec:id4_8} for details).
        The CO transitions and HST filters shown here are listed in Tables \ref{tab:tab2} and \ref{tab:mstar}, respectively. The white bar in the bottom left of each panel shows a scale of 1$\arcsec$.}    
        \label{fig:hstalma1}
    \end{center}
\end{figure*}

For each source we select the reddest HST filter (see Table \ref{tab:mstar}) in which the source is detected in order to minimize biases in the determination of the structural parameters due either to the irregular morphology of the galaxies that host UV-bright star-forming regions \citep{Guo_2018, Zanella_2019} or the presence of dust attenuation and its patchy distribution \citep{Cibinel_2017}. In Table \ref{tab:mstar} we report the central rest-frame wavelength covered by the selected HST filter. For 17 galaxies the HST data cover the rest-frame optical emission ($\gtrsim 4000 \AA$), while 6 galaxies are covered only in the near-UV range. Fig.~\ref{fig:hstalma1} shows the HST images for the 23 ALPAKA targets, while the white contours show the integrated CO or [CI] emission lines from ALMA data. By visually inspecting the overlap of the HST and ALMA data, it is evident that the bulk of the stellar and gas emissions largely overlap. However, for some galaxies (e.g., ID1, 2, 12, 15) the gas emission seems more compact than the stellar counterpart (see Sects.~\ref{sec:details} and \ref{sec:disks}). A detailed comparison between the sizes and morphologies of the stellar and gas emission is contained in the MSc thesis by D. Frickmann\footnote{https://nbi.ku.dk/english/theses/masters-theses/ditlev-frickmann/} and it will be extended and published as part of the ALPAKA series (Sect.~\ref{sec:conclusion}).

\begin{table*}[htbp!]
\begin{center}
\caption{Description of the ALMA observations and datasets of the ALPAKA sample. }
\label{tab:tab2}
\begin{tabular}{ccccccccc}
\hline\hline  \noalign{\smallskip}
ID   & Project ID & Line & Frequency Range & Beam & Channel Width & RMS & Integration time  \\
& & &  GHz & (" $\times$ ") & (km s$^{-1}$) & (mJy/beam) & (hours)\\
\noalign{\smallskip}
\hline
\noalign{\medskip}
1 & 2017.1.01659.S & CO(2-1) & 146.72 - 148.59 & 0.49 $\times$ 0.39 & 39 & 0.056 & 10.48\\ 
2 & 2016.1.00624.S & CO(3-2) & 212.07 - 231.00 & 0.20 $\times$ 0.16 & 32 & 0.23 & 0.72\\
3 & 2016.1.01426.S & CO(5-4) & 217.00 - 236.54 & 0.14 $\times$ 0.10 & 30 & 0.12 & 1.5 \\
4 & 2017.1.00471.S & CO(2-1) & 92.06 - 93.93 & 0.50 $\times$ 0.34 & 25 & 0.096 & 12.8\\ 
5 & 2017.1.00471.S & CO(2-1) & 92.06 - 93.93 & 0.50 $\times$ 0.34 & 25 & 0.094 & 12.8\\ 
6 & 2017.1.00471.S & CO(2-1) & 92.06 - 93.93 & 0.50 $\times$ 0.34 & 25 & 0.1 & 12.8\\  
7 & 2017.1.00471.S & CO(2-1) & 92.06 - 93.93 & 0.50 $\times$ 0.35 & 25 & 0.1 & 12.8\\ 
8 & 2017.1.00471.S & CO(2-1) & 92.06 - 93.93 & 0.50 $\times$ 0.34 & 25 & 0.1 & 12.8\\
9 & 2017.1.00471.S & CO(2-1) & 92.06 - 93.93 & 0.51 $\times$ 0.35 & 25 & 0.1 & 12.8\\
10 & 2017.1.01674.S &  CO(2-1) & 91.42 - 106.72 & 0.30 $\times$ 0.28 & 25 & 0.069 & 3.65\\
& 2015.1.00862.S &  CO(2-1) & 91.77 - 107.49 & comb. & comb. & comb. & 2.49\\
11 & 2017.1.01228.S & CO(2-1) & 85.90 - 101.81 & 0.53 $\times$ 0.42 & 26 & 0.1 & 2.69\\
12 & 2018.1.00974.S & CO(2-1) & 86.72 - 102.53 & 0.39 $\times$ 0.30 & 39 & 0.1 & 2.91\\
13 & 2017.1.00413.S & \CIii & 243.03 - 261.99 & 0.19 $\times$ 0.16 & 18 & 0.15 & 1.09\\
14 & 2019.1.01362.S  & CO(4-3) & 140.58 - 156.27 & 0.18 $\times$ 0.17 & 16 & 0.043 &  13.82\\
& 2017.1.01045.S & CO(4-3) & 140.80 - 156.38 & comb. & comb. & comb. & 3.89\\
& 2013.1.00059.S & CO(4-3) & 140.48 - 156.10 & comb. & comb. & comb. & 1.56 \\
15 & 2018.1.00543.S &  CO(3-2) & 92.34 - 108.18 & 0.28 $\times$ 0.23 & 32 & 0.058 & 5.95\\
16 & 2015.1.00723.S &  CO(7-6) & 226.21 - 245.50 & 0.22 $\times$ 0.20 & 19 & 0.11 & 2.55\\
17 & 2015.1.00723.S &  CO(7-6) & 226.21 - 245.50 & 0.22 $\times$ 0.20 & 19 & 0.11 & 2.55\\
18 & 2018.1.01146.S &  \CIi & 86.51 - 102.39 & 0.25 $\times$ 0.19 & 32 & 0.082 & 1.33\\
19 & 2016.1.01155.S & CO(3-2) & 85.75 - 100.61 & 0.35 $\times$ 0.29 & 35 & 0.12 & 3.32\\ 
20 & 2016.1.01155.S & CO(3-2) & 85.75 - 100.61 & 0.35 $\times$ 0.29 & 35 & 0.12 & 3.32\\ 
21 & 2016.1.01155.S & CO(3-2) & 85.75 - 100.61 & 0.36 $\times$ 0.30 & 35 & 0.13 & 3.32\\ 
22 & 2017.1.01677.S & CO(5-4) & 133.53 - 149.39 & 0.19 $\times$ 0.16 & 32 & 0.13 & 2.31\\
23 & 2018.1.01306.S & CO(3-2) & 84.09 - 99.78 & 0.24 $\times$ 0.20 & 28 & 0.069 & 5.71\\
24 & 2018.1.01306.S & CO(3-2) & 84.09 - 99.78 & 0.24 $\times$ 0.20 & 28 & 0.075 & 5.71\\
25 & 2018.1.01306.S & CO(3-2) & 84.09 - 99.78 & 0.24 $\times$ 0.20 & 28 & 0.074 & 5.71\\
26 & 2017.1.01677.S & CO(5-4) & 129.50 - 144.99 &  0.28 $\times$ 0.21 & 35 & 0.11 & 3.40 \\
27 & 2017.1.01677.S & CO(5-4) & 129.50 - 144.99 & 0.29 $\times$ 0.21 & 36 & 0.12 & 3.40\\
28 & 2017.1.00908.S & CO(6-5) & 135.76 - 151.74 & 0.18 $\times$ 0.15 & 31 & 0.06 & 3.61\\
\noalign{\smallskip}
\hline
\end{tabular}
\tablefoot{For targets ID4 - 9, the observation consists of three mosaic pointings; the corresponding on-source integration time refers to the total value summed over all pointings. For targets with multiple observations (i.e., ID10 and 14), we image the CO after combining the corresponding measurement sets and we list the resulting properties of the cubes (beam, channel width, RMS).\\
Here, we provide the list of principal investigators (PIs) for the ALMA projects employed in this work: 2017.1.01659.S, Chemin, L.; 2016.1.00624.S, Freundlich, J.; 2017.1.00471.S, Hayashi, M.; 2017.1.01674.S, Molina, J.; 2015.1.00862.S, Ibar, E.; 2017.1.01228.S, Noble, A.; 2018.1.00974.S, Noble, A.; 2017.1.00413.S, Barro, G.; 2019.1.01362.S, Herrera-Camus, R.; 2017.1.01045.S,  Brisbin, D.; 2013.1.00059.S, Aravena, M.; 2018.1.00543.S, Herrera-Camus, R.; 2015.1.00723.S, Oteo, I.; 2018.1.01146.S, Nagar, N.; 2016.1.01155.S, Wang, T.; 2017.1.01677.S, Cassata, P.; 2018.1.01306.S, Umehata, H.; 2017.1.00908.S, Assef, R.}
\end{center}
\end{table*}

\section{Global properties} \label{sec:gloabl}
Being in well characterized survey areas, extensive studies of the global, unresolved properties (e.g., stellar mass, $M_{\star}$, and star-formation rate, SFR) of the ALPAKA galaxies are already available. However, these parameters are derived using different algorithms and assumptions. 
For this reason, we collect UV-to-radio photometric data using public multi-wavelength catalogues \citep{Fu_2013, Ivison_2013, Magnelli_2013, Jin_2018, Hayashi_2018, Liu_2019, Weaver_2022} with the aim of fitting the spectral-energy distribution (SED) in a consistent way. The only exceptions are ID11 and ID12, for which photometric data are not publicly available. Throughout the rest of the paper, for these two galaxies, we will refer to the stellar masses and SFR reported in \citet{Noble_2017} and obtained after making assumptions consistent with those used for the rest of our sample \citep[see][for a detailed discussion on the $M_{\star}$ and SFR obtained with different tools]{Kokorev_2021}. 

To derive the stellar masses and SFR for the ALPAKA galaxies, we perform the SED fitting using \stardust\ \citep{Kokorev_2021}. \stardust\ performs a multi-component fit that linearly combines three classes of templates: stellar libraries from an updated version of the Stellar Population Synthesis models described in \citet{Brammer_2008}; AGN torus templates from \citet{Mullaney_2011}; infrared models of dust emission arising from star formation from \citet{Draine_2007, Draine_2014}. These three components  are fitted simultaneously but independently from each other, i.e., without assuming an energy balance between the absorbed UV/optical radiation and the infrared emission. This approach allows one to account for spatial offsets between the stellar and dust distributions within a galaxy \citep[see discussion in][]{Kokorev_2021}. For fitting the ALPAKA targets, we include the AGN templates only for galaxies that are previously identified as AGN hosts (Table \ref{tab:mstar}). 

\begin{figure*}[htbp!]
    \begin{center}    \includegraphics[width=1\textwidth]{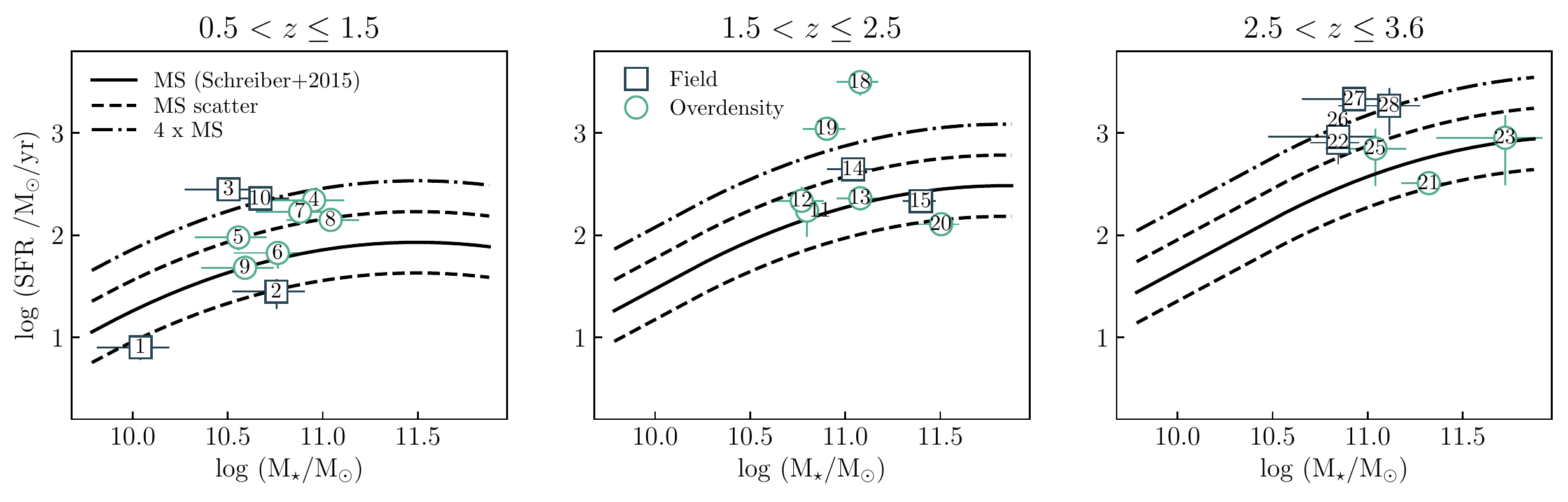}
        \caption{Distribution of the ALPAKA galaxies in the stellar mass (M$_{\star}$) - SFR plane, divided in three redshift bins. The dark green squares show ALPAKA galaxies in the field, while the green circles show galaxies in overdense regions (e.g., clusters, protoclusters, groups). The solid line in the three panels show the empirical main-sequence relations from \citet{Schreiber_2015} at $z = 1.3, 2.2, 3$, which are the average redshifts of ALPAKA targets in the three bins. The dashed and dot-dashed lines show the $\pm 1\sigma$ scatter and the line dividing the main-sequence and starburst galaxies, respectively \citep{Rodighiero_2011}. We note that only the 25 ALPAKA targets with good estimates of both M$_{\star}$ and SFR are shown.}  
        \label{fig:ms}
    \end{center}
\end{figure*}

In Table~\ref{tab:mstar}, we show the best-fit stellar masses, SFR, and the star-formation infrared luminosity ($\mathrm{L_{IR}}$) from \stardust. 
We note that, for AGN-host galaxies, in addition to star-formation, the dusty tori contribute to the total infrared luminosity. However, the values of $\mathrm{L_{IR}}$ listed in Table~\ref{tab:mstar} are obtained after integrating the best-fit star-formation dust model employed by \stardust\ in the range 8 - 1000 $\mu$m. For all ALPAKA galaxies, we consider only the dust-obscured SFR, traced by $\mathrm{L_{IR}}$, and we neglect the contribution from UV emission. For three galaxies --- ID16, 17, 24 --- due to the lack of good coverage in the optical/near-infrared, the uncertainties on the stellar masses are much larger than the parameter values itself. Therefore, the respective estimates of the stellar mass are not stastically meaningful. In addition, due to the lack of a good far-infrared coverage, measuring the SFR for ID24 is challenging. In Fig. \ref{fig:ms}, we show the distribution of the ALPAKA sample in the SFR-$M_{\star}$ plane for the 25 galaxies with a reliable estimate (i.e., uncertainties smaller than the value) of $M_{\star}$. We divide our targets in three redshift bins and compare them with the main-sequence relation at the corresponding redshift. For the latter, we use the parametrization of \citet[][solid lines]{Schreiber_2015} using a Chabrier IMF and show the 0.3 dex scatter at the average redshifts of the galaxies in the three bins, $z = 1.3, 2.2, 3$ (dashed lines). The dot-dashed lines in Fig. \ref{fig:ms} show the empirical lines, located 4 times above the SFR of main sequence that is usually used to divide main-sequence from starburst galaxies \citep{Rodighiero_2011}. According to this definition, a large fraction of ALPAKA galaxies are starbursts (12/25 or 48\%; see column 5 in Table~\ref{tab:mstar}). In the low redshift bin ($z =$0.5 - 1.5), 70\% of ALPAKA galaxies lie within the $\pm1\sigma$ scatter of the main sequence relation, while this fraction falls to 37\% at $z \sim 3$.

\begin{figure*}[htbp!]
    \begin{center}        \includegraphics[width=1\textwidth]{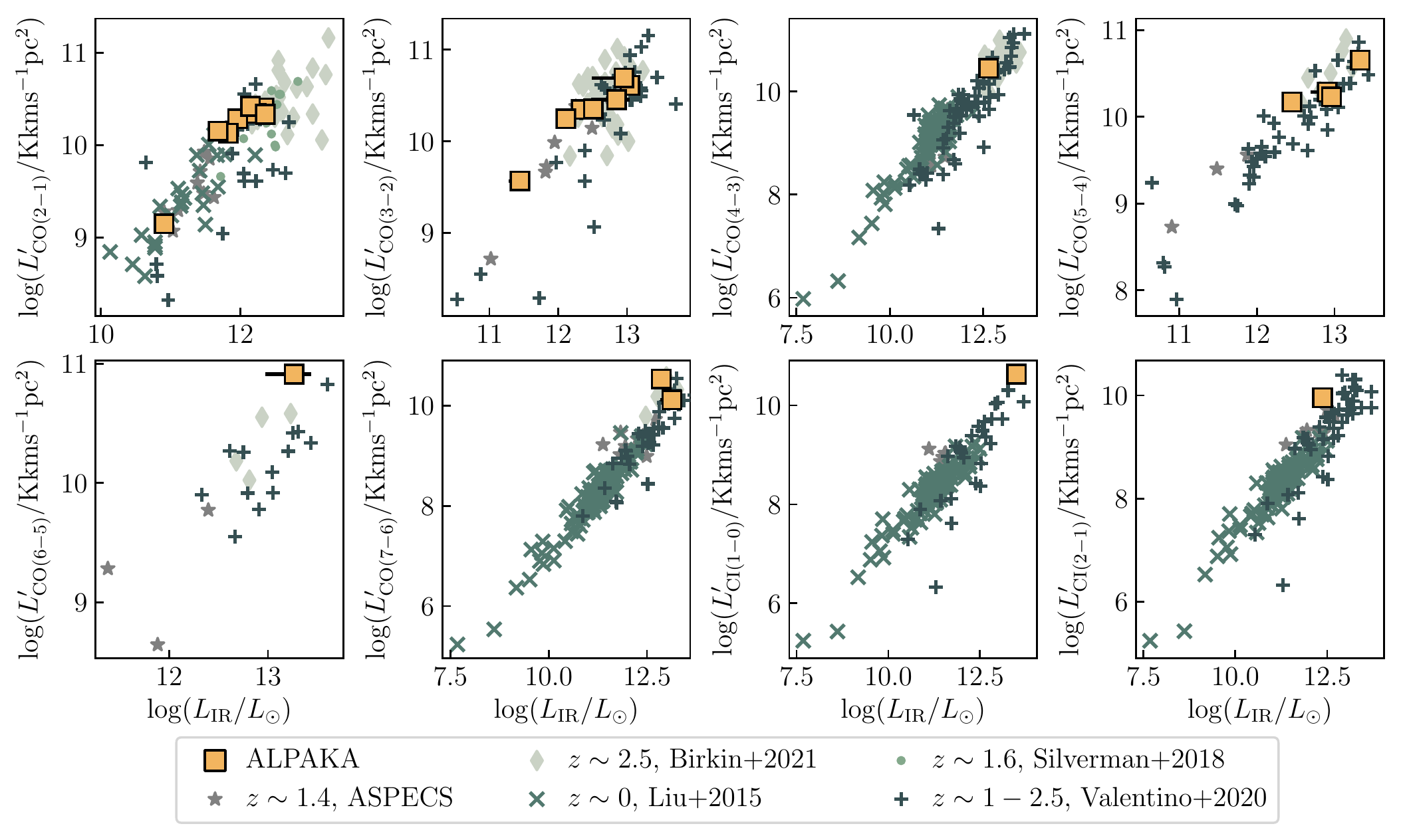}
        \caption{Distribution of the ALPAKA galaxies and samples of local and high-$z$ galaxies from the literature \citep[e.g.,][]{Liu_2015, Silverman_2018, Valentino_2020, Boogaard_2020, Birkin_2021} in the emission line-infrared luminosity planes. The ALPAKA galaxies typically cover the bright part of the distributions, especially for high-J transitions, as spatially resolved ALMA observations are feasible only for very luminous galaxies. } 
        \label{fig:lir}
    \end{center}
\end{figure*}

The ALPAKA sample covers high stellar mass galaxies, $\gtrsim 10^{10} M_{\odot}$ with SFR ranging from 8 to $\sim$3000 M$_{\odot}$\,yr$^{-1}$. This can be ascribed to a selection effect: being discovered as bright sources in the infrared or sub-mm wavelength, ALPAKA galaxies have high SFRs and gas fractions. To visualize this, in Fig.~\ref{fig:lir}, we show the distribution of the ALPAKA sample in the emission line-infrared luminosity planes, an observational proxy of the Kennicutt-Schmidt relation \citep{Schmidt_1959, Kennicutt_1998}, with respect to compilations of local and high-$z$ galaxies from the literature \citep[e.g.,][]{Liu_2015, Silverman_2018, Valentino_2020, Boogaard_2020, Birkin_2021, Valentino_2021}. The CO fluxes and respective luminosities for the ALPAKA sample, as listed in Table~\ref{tab:mstar}, are derived by summing the flux above 3 $\times$ RMS in the high-resolution flux-integrated maps presented in Sect.~\ref{sec:alma}. We check that these values are within $\pm20\%$ from the ones obtained by fitting the total-flux maps with the \texttt{IMFIT} tool within \casa\ that adopts a 2D Gaussian.  
In all cases, these fluxes are consistent with previous estimates, mostly obtained through unresolved observations, showing that we are not missing any significant emission on large scales. The samples of galaxies from the literature comprise: the compilation in \citet{Valentino_2020} consisting of 30 main-sequence galaxies at $z \sim 1$ and 65 submillimeter galaxies (SMGs) and quasars at $z \sim 2.5$ \citep{Walter_2011, Alaghband-Zadeh_2013, Canameras_2018, Yang_2017, Andreani_2018} and 146 local starbursts from \citet{Liu_2015}; 12 starbursts at $z \sim 1.6$ \citep{Silverman_2018}; 22 main-sequence and starburst galaxies at $z \sim 1.4$ from the ALMA Spectroscopic Survey in the Hubble Ultra Deep Field survey \citep{Boogaard_2020}; 47 SMGs at a median $z$ of 2.5 from \citet{Birkin_2021}. Considering the substantial effort in obtaining high SNR spatially resolved ALMA observations at $z \sim 0.5 - 3.5$, the ALPAKA sample typically covers the brightest part of the distributions in each panel of Fig.\ref{fig:lir}, with a median luminosity of $2 \times 10^{10}$ K\,km\,s$^{-1}$ pc$^{2}$. ALPAKA galaxies are, therefore, somewhat biased towards the most actively star-forming galaxy populations (see discussion in Sect. \ref{sec:discussion}). 

\section{Kinematic analysis}\label{sec:kinematic}
In this section, we describe how we analyze the kinematics of the ALPAKA galaxies by fitting the data using rotating disk models (Sec.\,\ref{sec:barolo}). This allows us to provide a first-order description of the gas motions within the ALPAKA sample. In Sect. \ref{sec:kinclassification}, we show how we identify any deviations from pure circular orbits, likely due to radial and vertical motions driven by outflows and interactions. The kinematic properties of each ALPAKA target are described in Sect.~\ref{sec:details}. In Sect.~\ref{sec:disks}, we present the kinematic parameters of the disk subsample.

\begin{table*}[htbp!] 
\begin{center}
\caption{Properties of the ALPAKA sample and description of the HST dataset. 
}
\label{tab:mstar}
\begin{tabular}{cccccccccc}
\\
\hline\hline  \noalign{\smallskip}
ID  &  M$_{\star}$ & SFR & $\Delta_{\mathrm{MS}}$ & Type & L$_{\mathrm{IR}}$ & I$_{\mathrm{line}}$ &  L'$_{\mathrm{CO/[CI]}}$ &  HST Filter & $\lambda_{\mathrm{rest, eff}}$\\
 & $10^{10}$M$_{\odot}$ & M$_{\odot}$ yr$^{-1}$ & & & $10^{12}$ L$_{\odot}$ & Jy km s$^{-1}$ & $10^{10}$ K km s$^{-1}$ pc$^{2}$ & & \AA\\
\noalign{\smallskip}
\hline
\noalign{\medskip}
1 & 1.1 $\pm$ 0.3 & 8 $\pm$ 2 & $0.11 \substack{+ 0.14 \\- 0.15 }$ & MS & 0.8 $\pm$ 0.2 & 0.32 $\pm$ 0.07 & 0.14 $\pm$ 0.03 & F160W & 9794\\ 
2 & 5.7 $\pm$ 0.5 & 28 $\pm$ 9 & $0.22 \substack{+ 0.12 \\- 0.16 }$ &  MS & 2.8 $\pm$ 0.9 & 1.54 $\pm$ 0.05 & 0.37 $\pm$ 0.01 & F814W & 4921\\
3 & 3.2 $\pm$ 1.1 & 281 $\pm$ 18 & $0.89 \substack{+ 0.15 \\- 0.09 }$ & SB & 28 $\pm$ 2 & 3.2 $\pm$ 0.7 & 1.5 $\pm$ 0.3 & F814W & 3254\\
4 & 9 $\pm$ 4 & 220 $\pm$ 50 &  $0.51 \substack{+ 0.15 \\- 0.14 }$ & SB & 22 $\pm$ 5 & 0.88 $\pm$ 0.07 & 2.5 $\pm$ 0.2 & F160W & 6236\\
5 & 3.6 $\pm$ 0.5 & 92 $\pm$ 24 & $0.38 \substack{+ 0.11 \\- 0.13 }$ & MS & 9 $\pm$ 2 & 0.67 $\pm$ 0.05 & 1.9 $\pm$ 0.1 & F160W & 6236\\ 
6 & 5.8 $\pm$ 1.0 & 67 $\pm$ 20 & $0.10 \substack{+ 0.11 \\- 0.17 }$ & MS & 6.7 $\pm$ 2.0  & 0.46 $\pm$ 0.07 & 1.3 $\pm$ 0.2 & F160W & 6210\\ 
7 & 7.6 $\pm$ 0.9 & 170 $\pm$ 30 & $0.42 \substack{+ 0.07 \\- 0.08 }$ & MS & 17 $\pm$ 3 & 0.75 $\pm$ 0.02 & 2.22 $\pm$ 0.06 & F160W & 6185\\%
8 & 11 $\pm$ 3 & 141 $\pm$ 33 & $0.27 \substack{+ 0.11 \\- 0.12 }$ & MS & 14 $\pm$ 3 & 0.9 $\pm$ 0.2 & 2.6 $\pm$ 0.7 & F160W & 6185\\
9 & 3.9 $\pm$ 0.9 & 48 $\pm$ 11 & $0.04 \substack{+ 0.13 \\- 0.14 }$ & MS & 4.8 $\pm$ 1.2 & 0.48 $\pm$ 0.05 & 1.4 $\pm$ 0.1 & F160W & 6185\\
10 & 4.7 $\pm$ 0.4 & 229 $\pm$ 15 & $0.67 \substack{+ 0.04 \\- 0.04 }$ & SB & 23 $\pm$ 2 & 0.72 $\pm$ 0.09 & 2.15 $\pm$ 0.27 & F160W & 6160\\
11$^{*}$ & 6.3 $\pm$ 0.8 & 174 $\pm$ 78 & $0.43 \substack{+ 0.16 \\- 0.25 }$ & SB & - & 1.19 $\pm$ 0.13 & 4.07 $\pm$ 0.44 & F160W & 5899\\%
12$^{*}$ & 5.9 $\pm$ 1.8 & 217 $\pm$ 82 & $ 0.54 \substack{+ 0.16 \\- 0.22 }$ & SB & - & 0.76 $\pm$ 0.14 & 2.75 $\pm$ 0.51 & F160W & 4759\\
13 & 12 $\pm$ 3 & 230 $\pm$ 56  & $ 0.18 \substack{+ 0.13 \\- 0.14 }$ & MS & 24 $\pm$ 10 & 1.96 $\pm$ 0.72 & 0.90 $\pm$ 0.33 & F160W & 5092\\
14 & 11 $\pm$ 3 & 441 $\pm$ 77 & $ 0.45 \substack{+ 0.11 \\- 0.10 }$ & MS & 44 $\pm$ 17 & 1.83 $\pm$ 0.06 & 2.8 $\pm$ 0.1 & F140W & 4279\\
15 & 25 $\pm$ 5 & 215 $\pm$ 40 & $ -0.05 \substack{+ 0.09 \\- 0.09 }$ & MS & 21.5 $\pm$ 5.2 & 0.8 $\pm$ 0.02 & 2.23 $\pm$ 0.06 & F160W & 4737\\
16 & - & 730 $\pm$ 300 & - & - & 73 $\pm$ 13 & 6.3 $\pm$ 0.8 & 3.4 $\pm$ 0.4 & F110W & 3384\\
17 & - & 1360 $\pm$ 680 & - & - & 136 $\pm$ 25 &  2.4 $\pm$ 0.5 & 1.3 $\pm$ 0.3 & F110W & 3384\\
18 & 12 $\pm$ 3 & 3150$\pm$830  & $ 1.22 \substack{+ 0.13 \\- 0.14 }$ & SB & 315 $\pm$ 83 & 2.7 $\pm$ 0.2 &  4.3 $\pm$ 0.4 & F110W & 3285\\
19 & 8 $\pm$ 2 & 1100$\pm$150 & $ 0.86 \substack{+ 0.12 \\- 0.10 }$ & SB & 110 $\pm$ 15 & 1.2 $\pm$ 0.1 & 4.1 $\pm$ 0.3 & F160W & 4377\\ 
20 & 32$\pm$8 & 130$\pm$170 &  $ -0.27 \substack{+ 0.25 \\- 0.38 }$ & MS & 13 $\pm$ 17 & 0.5 $\pm$ 0.1 & 1.8 $\pm$ 0.3 & F160W  & 4365\\ 
21 & 21$\pm$6 & 323$\pm$70  & $ 0.06 \substack{+ 0.11 \\- 0.12 }$ & MS & 32 $\pm$ 7 & 0.6 $\pm$ 0.1 & 2.2 $\pm$ 0.5 & F160W & 4352\\ 
22 & 7$\pm$2 & 804$\pm$313  & $ 0.67 \substack{+ 0.18 \\- 0.22 }$ & SB & 80 $\pm$ 31 & 1.2 $\pm$ 0.4 & 1.9 $\pm$ 0.7 & F160W & 3887\\
23 & 53 $\pm$ 30 & 898$\pm$590  & $ 0.17 \substack{+ 0.24 \\- 0.35 }$ & MS & 90 $\pm$ 59 & 1.0 $\pm$ 0.5 & 4.9 $\pm$ 3.6 & - & - \\
24 & - & - & - & - & - & 0.63 $\pm$ 0.09 & 3.0 $\pm$ 0.4 & - & -\\
25 & 11 $\pm$ 5 & 704 $\pm$ 400 & $ 0.42 \substack{+ 0.27 \\- 0.31 }$ & SB & 70 $\pm$ 40 & 0.6 $\pm$ 0.2 & 2.8 $\pm$ 1.0 & - & - \\
26 & 7 $\pm$ 4 & 916 $\pm$ 70 & $ 0.63 \substack{+ 0.29 \\- 0.16 }$ & SB & 92 $\pm$ 7 & 0.86 $\pm$ 0.06 & 1.7 $\pm$ 0.1 & - & -\\
27 & 8.5 $\pm$ 4.0 & 2150 $\pm$ 210 & $ 0.93 \substack{+ 0.26 \\- 0.15 }$ & SB & 215 $\pm$ 21 & 2.2 $\pm$ 0.3 & 4.5 $\pm$ 0.7 & - & -\\
28 & 13 $\pm$ 6 & 1848 $\pm$ 900 & $ 0.67 \substack{+ 0.27 \\- 0.28 }$ & SB & 184 $\pm$ 90 & 5.20 $\pm$ 0.05 & 8.16 $\pm$ 0.08 & F160W & 3299
\\
\noalign{\smallskip}
\hline
\end{tabular}
\tablefoot{Column 4 shows the distance of the ALPAKA galaxy with respect to the main-sequence relation, $\Delta_{\mathrm{MS}} = \log(\mathrm{SFR/SFR_{MS}})$, where $\mathrm{SFR_{MS}}$ is the SFR defined by \citet{Schreiber_2015} at fixed redshift and stellar mass. Column 5 indicates whether the galaxy is a main-sequence (MS) or a starburst (SB). To be conservative, galaxies with ID4, 11, 12, 25 are classified as starburst as their $\Delta_{\mathrm{MS}}$ values are $\gtrsim \log(4) = 0.6$ within the 1-$\sigma$ uncertainties. Column 7 shows the flux for the CO or [CI] transitions listed in Table \ref{tab:tab1}. The last column shows the rest-frame effective wavelength probed by the HST data.
$^{*}$ For targets 11 and 12, the M$_{\star}$ and SFR values are taken from \citet{Noble_2017}.}
\end{center}
\end{table*}

\subsection{Disk modeling}\label{sec:barolo}
We fitted the kinematics of the ALPAKA galaxies using the software \bba\ \citep{DiTeodoro_2015}. \bba\ produces three dimensional (two spatial, one spectral axis) realizations of a so-called tilted-ring model \citep{Rogstad_1974}. The latter consists of a disk divided into a series of concentric rings, each with its kinematic (i.e., systemic velocity $V_{\mathrm{sys}}$, rotation velocity $V_{\mathrm{rot}}$ and velocity dispersion $\sigma$) and geometric properties (i.e., center, inclination angle $i$ and position angle $PA$\footnote{Within \bba\, the position angle is measured from the major axis on the receding half of the rotating disk, taken anticlockwise from the North direction on the sky.}).
For a thin disk model, the line-of-sight velocity $V_{\mathrm{los}}$ at a radius $R$ is given by
\begin{equation}
V_{\mathrm{los}}(R) = V_{\mathrm{sys}} + V_{\mathrm{rot}}(R) \cos \phi \sin i,
\label{eq:vlos}
\end{equation}
where $\phi$ is the azimuthal angle in the disk plane. \\
The best-fit model is obtained by means of a least-square minimization. At each step of the model optimization and before calculating the residuals between the data and the model, \bba\ convolves the model disk with a Gaussian kernel with sizes and position angle equal to the beam of the corresponding observation. In the case of the ALMA observations, this is set to be equal to the synthesized beam of the cleaned image. This approach allows for a robust recovery of the rotation velocity and velocity dispersion profiles, since it largely mitigates the effects of beam smearing also in the case of data with relatively low angular resolution \citep{DiTeodoro_2015, DiTeodoro_2016, Rizzo_2022}. \bba\ is a suitable tool to model the gas kinematics at the typical spatial resolution and SNR of the galaxies in ALPAKA. This tool has been extensively tested using mock and real data over a wide range of data quality \citep[e.g.,][]{DiTeodoro_2015, Rizzo_2022}. In fact, \bba\ has been shown to recover values of the rotation velocity and dispersion with an accuracy of $\sim$25\% when barely resolved observations (i.e., 3 independent beam along the major axis) at SNR$\gtrsim$3 are employed \citep{DiTeodoro_2015, Rizzo_2022}.

\subsubsection{Geometrical parameters from morphological fitting} \label{sec:geometry}
\bba\ can estimate the geometrical parameters, namely the center, inclination, and position angle of each tilted ring component. However, due to the relatively small number of resolution elements covering the CO/[CI] line emission, we prefer to reduce the number of free parameters by fixing the center and inclination. In particular, estimating the inclination of the galaxies is crucial. Correcting for it can in fact account for a large fraction of the rotation velocity if the galaxies are seen at low inclinations, due to the $\sin i$ dependence of Eq.~(\ref{eq:vlos}). When dealing with low-resolution observations of low-$z$ and high-$z$ galaxies, the inclination is usually fixed to the one estimated from the optical images \citep{deBlok_1996, Lelli_2016, Wisnioski_2019, Kaasinen_2020}. However, since only a fraction of ALPAKA galaxies have HST data covering the rest-frame optical emission, we use two methods for estimating their inclinations:
\begin{itemize}
    \item \galfit\ on HST data. For the 23 galaxies with HST data, we used \galfit\ \citep{Peng_2002} to model their 2D surface brightness using one Sérsic component \citep[][see details in Appendix \ref{sec:galfit}]{Sersic_1968}. \galfit\ fits the center, the three parameters describing the Sersic profile (total magnitude, Sérsic index, effective radius), the position angle $PA_{\mathrm{HST}}$ and the axis ratio $b/a$ between the projected major and minor axis. The latter allows for computing the inclination angles, $i_{\mathrm{HST}} = \arccos(b/a)$. In Table \ref{tab:PA}, we report the best-fit geometrical parameters, $PA_{\mathrm{HST}}$, $i_{\mathrm{HST}}$. In Figs. \ref{fig:galfit1} and \ref{fig:galfit2}, we show the HST data and the corresponding \galfit\ models and residuals.
    \item \cannubi\ on ALMA data. \cannubi\ is a Markov Chain Monte Carlo algorithm that models the geometry of galaxies without assuming parametric descriptions of the surface brightness distribution. \cannubi\ uses \bba\ to fit either the total-flux map or the entire cube using resolution-matched 3D tilted-ring models of rotating disks \citep[see details in][]{Roman_2023, Mancera_2020}. The free parameters of the fit are the center of the disk, its radial extent, the position and inclination angles ($PA_{\mathrm{ALMA}}$, $i_{\mathrm{ALMA}}$). 
\end{itemize}
To perform a brief check of the independence of our results on the specific method used to model each data set, we repeat the analysis of the ALMA data using \galfit. Since \galfit\ is optimized for dealing with optical/near-infrared images (e.g., units in counts, magnitude zero-points, Poissonian error in each pixel), we applied some arbitrary conversion factors to fit the total-flux maps obtained from ALMA data. The resulting best-fit inclinations are consistent within 5\% with the ones obtained with \cannubi. 

\begin{figure*}[htbp!]
    \begin{center}      
    \includegraphics[width=\textwidth]{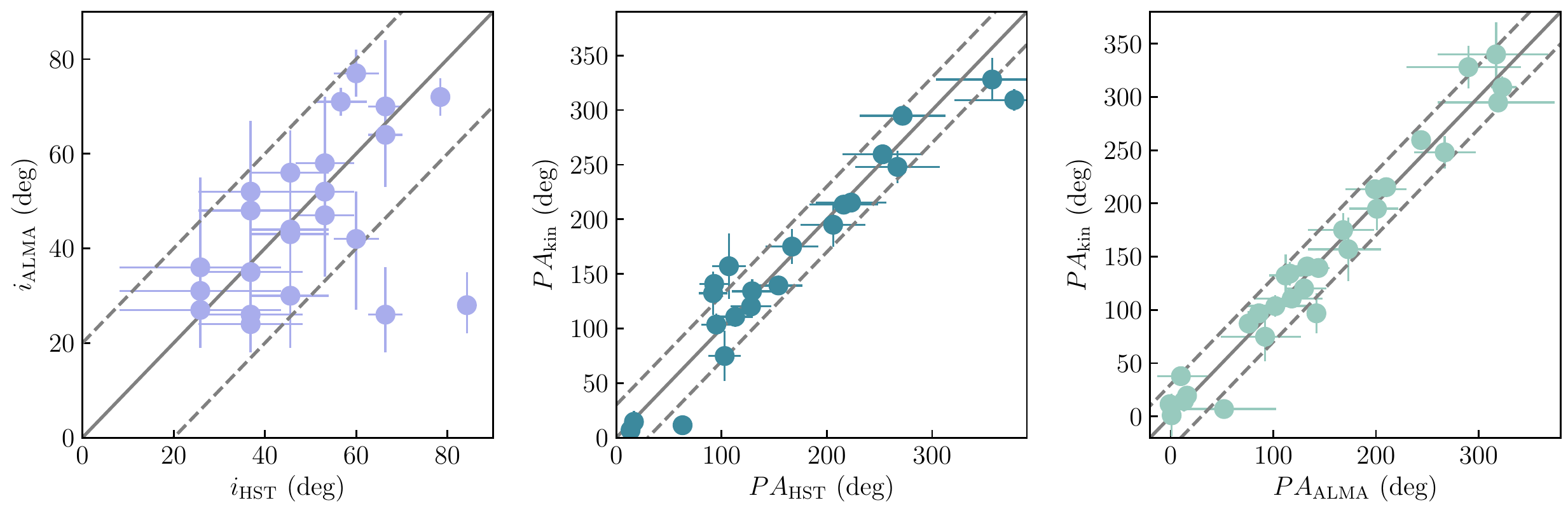}
    \caption{\textit{Left panel:} comparison between the inclination angles derived from the morphological fitting of HST and ALMA data using \galfit\ and \cannubi\, respectively. \textit{Central panel:} comparison between the position angles derived from the morphological fitting of HST data with \galfit\ and kinematic fitting of ALMA data with \bba. \textit{Right panel:} comparison between the position angles derived from the morphological fitting of ALMA data with \cannubi\ and the corresponding kinematic fitting. The gray line shows the 1:1 relation and the gray dotted lines show deviations at $\pm 20 \deg$ (left panel) and $\pm 30 \deg$ (central and right panels).}
        \label{fig:ipa}
    \end{center}
\end{figure*}

Fig.~\ref{fig:ipa}, left panel, shows the distribution of the difference between the inclination angles found with the two methods (upper left panel). For 21 out of the 23 ALPAKA galaxies with HST data, $i_{\mathrm{HST}}$ and $i_{\mathrm{ALMA}}$ values are consistent within the $\pm1.5\sigma$ and the difference between these two angles are within 20 $\deg$. Since the rotation velocity has a dependence on the inclination angle that goes as $1/\sin i$, Eq.(~\ref{eq:vlos}),
an uncertainty of 20 $\deg$ results in relative uncertainties on the rotation velocity of 7\% and 18\% for a nearly edge-on (e.g., 75 $\deg$) and face-on (e.g., 26 $\deg$) galaxy, respectively\footnote{To compute the relative uncertainties on $V_{\mathrm{rot}}$, we used the equation based on the propagation of errors: $\Delta V_{\mathrm{rot}}/V_{\mathrm{rot}} = \Delta i/ \tan i$.}. 
The only galaxies with a difference between the two inclination angles $\gtrsim 20$ $\deg$ are ID3 and ID22, with $i_{\mathrm{HST}}-i_{\mathrm{ALMA}}$ of 54 and 42 $\deg$, respectively. However, the HST data for both ID3 and ID22 cover only their rest-frame near UV/blue optical emission and they are likely not representative of the bulk of the stellar population. In both cases, there is a difference between the CO and HST morphologies. For instance, ID3 has, two UV bright clumps, clearly visible in Fig.\ref{fig:galfit1}, while the CO emission has a smooth distribution and its center is located between them (Fig.\ref{fig:hstalma1}). The CO(5-4) emission from ID22 cover only the innermost regions of the corresponding HST data. 
To summarize, for most of the galaxies with HST data, the value of $i_{\mathrm{HST}}$ are consistent with $i_{\mathrm{ALMA}}$, despite that these two values are derived by fitting two different components of the galaxy (i.e., stellar continuum and gas emission line) and using different tools and assumptions. 

For the rest of this subsection, we attribute to each ALPAKA galaxy the value of $i_{\mathrm{HST}}$ if HST images are available (with the exception of ID3 and 22, see Sect. \ref{sec:details}), and $i_{\mathrm{ALMA}}$ otherwise. In Appendix \ref{app:inclination}, we discuss potential bias towards low inclinations for the ALPAKA sample. 

\begin{table}[t!]
\begin{center}
\caption{Geometric parameters of the ALPAKA galaxies.}
\label{tab:PA}
\begin{tabular}{lllllll}
\hline\hline \noalign{\smallskip}
ID   & $PA_{\mathrm{HST}}$ & $i_{\mathrm{HST}}$ & $PA_{\mathrm{ALMA}}$ & $i_{\mathrm{ALMA}}$ & $PA_{\mathrm{kin}}$ & KC\\
& $\deg$ & $\deg$ & $\deg$ & $\deg$ &  $\deg$ \\
\noalign{\smallskip}
\hline
\noalign{\medskip}
1 & 253 $\pm$ 38 & 53 $\pm$ 6 & 244 $^{+9}_{-9}$ & 47 $^{+6}_{-8}$ & 259 $\pm$7 & D\\ 
2 &223 $\pm$ 33 & 60 $\pm$ 5 & 210 $^{+5}_{-6}$ & 77 $^{+5}_{-5}$ & 215 $\pm$8 & D\\
3$^{*}$ & 129 $\pm$ 19 & 84 $\pm$ 1 & 116 $^{+11}_{-13}$ & 28 $^{+7}_{-6}$ & 134 $\pm$11 & D\\
4 & 167 $\pm$ 25 & 37 $\pm$ 11 & 168 $^{+30}_{-34}$ & 52 $^{+15}_{-24}$ & 175 $\pm$16 & U\\
5 & 357 $\pm$ 54 & 37 $\pm$ 11 &  290 $^{+51}_{-60}$ & 26 $^{+11}_{-8}$ & 328 $\pm$20 & U\\
6 & 216 $\pm$ 32 & 46 $\pm$ 8 & 199 $^{+31}_{-29}$  & 43 $^{+21}_{-19}$ & 213 $\pm$8 & D\\
7 & 272 $\pm$ 41 & 37 $\pm$ 11 &  319 $^{+55}_{-59}$ & 35 $^{+18}_{-13}$ & 294 $\pm$3 & D\\
8 & 206 $\pm$ 31 & 37 $\pm$ 11 & 201 $^{+21}_{-27}$ & 48 $^{+13}_{-21}$ & 195 $\pm$20 & D\\
9 & 128 $\pm$ 19 & 46 $\pm$ 8 & 130 $^{+22}_{-17}$ & 44 $^{+13}_{-17}$ & 120 $\pm$3 & D\\
10 & 103 $\pm$ 15 & 46 $\pm$ 8 & 92 $^{+35}_{-43}$ & 30 $^{+12}_{-11}$ & 74 $\pm$23 & U\\
11 & 113 $\pm$ 17 & 26 $\pm$ 18 & 118 $^{+30}_{-37}$ & 31 $^{+12}_{-11}$ & 111 $\pm$9 & D\\
12 & 95 $\pm$ 14 & 66 $\pm$ 4 & 102 $^{+38}_{-27}$ & 70 $^{+14}_{-15}$ & 103 $\pm 10$ & D\\
13 & -16 $\pm$ 2 & 37 $\pm$ 11 & 10 $^{+26}_{-23}$ & 24 $^{+8}_{-6}$ & 37 $\pm$4 & D\\
14 & 13 $\pm$ 2 & 53 $\pm$ 6 &  323 $^{+14}_{-11}$  & 58 $^{+14}_{-8}$ & 309 $\pm$10 & U\\
15 & 92 $\pm$ 14 & 60 $\pm$ 5 & 112 $^{+21}_{-16}$ & 42 $^{+10}_{-15}$ & 132 $\pm$20  & D\\
16$^{*}$ & 17 $\pm$ 3 & 78 $\pm$ 2 & 13 $^{+3}_{-3}$  & 73 $^{+4}_{-4}$ & 15 $\pm$10  & U\\
17$^{*}$ & 354 $\pm$ 53 & 66 $\pm$ 4 & 359 $^{+7}_{-6}$  & 65 $^{+8}_{-11}$ & 1 $\pm$20  & M\\
18$^{*}$ & 267 $\pm$ 40 & 26 $\pm$ 18 & 267 $^{+30}_{-30}$ & 27 $^{+9}_{-8}$ & 248 $\pm$15 & D\\
19 & 309 $\pm$ 46 & 56 $\pm$ 5 & 359 $^{+5}_{-5}$ &  71 $^{+3}_{-3}$ & 11 $\pm$9 & D\\
20 & 154 $\pm$ 23 & 46 $\pm$ 8 & 144 $^{+11}_{-11}$ & 56 $^{+9}_{-12}$ & 141 $\pm$8 & D\\
21 & 14 $\pm$ 2 & 26 $\pm$ 18 & 52 $^{+51}_{-52}$  & 36 $^{+19}_{-14}$ & 7 $\pm$5 & U\\
22$^{*}$ & 107 $\pm$ 16 & 66 $\pm$ 4 & 173 $^{+32}_{-39}$ & 26 $^{+9}_{-8}$ & 157 $\pm$30 & D\\
23 & - & - & 86 $^{+3}_{-2}$ & 70 $^{+3}_{-3}$ & 97 $\pm$9 & D\\
24 & - & - & 16 $^{+4}_{-3}$ & 75 $^{+6}_{-6}$ & 19 $\pm$4 & D\\
25 & - & - & 76 $^{+4}_{-4}$ & 74 $^{+5}_{-4}$ & 87 $\pm$12 & D\\
26 & - & - & 317 $^{+51}_{-57}$ & 38 $^{+19}_{-15}$ &  340 $\pm$30 & U\\
27 & - & - & 142 $^{+4}_{-4}$ & 62 $^{+3}_{-4}$ & 97 $\pm$19 & M\\
28$^{*}$ & 93 $\pm$ 14 & 53 $\pm$ 6 & 133 $^{+13}_{-12}$ & 52 $^{+13}_{-18}$ & 140 $\pm$5 & D\\
\noalign{\smallskip}
\hline
\end{tabular}
\tablefoot{Columns 2 - 5: values of the position and inclination angles, as derived by \galfit\ using the HST data (when available) and \cannubi\ using the ALMA total-flux map. Column 6: kinematic position angles derived by \bba. Column 7: kinematic class (KC) of each ALPAKA source (D: disk, U: uncertain, M: merger). Galaxies for which the HST filter covers the rest-frame wavelength at 
$\lambda_{\mathrm{rest, eff}} \lesssim 4000 \AA$ (Table~\ref{tab:mstar}) 
are marked with an asterisk $^{*}$. In these cases, the geometric parameters ($PA_{\mathrm{HST}}$ and $i_{\mathrm{HST}}$) may be not representative of the bulk of stellar emission.}
\end{center}
\end{table}

\subsubsection{Assumptions for the kinematic fitting}
In this section, we describe the assumptions we made to run \bba\ and fit the kinematics of the ALPAKA sample:
\begin{itemize}
    \item \textit{Mask.} Before fitting the data, \bba\ uses the source finder derived from {D{\sc {uchamp}}} \citep{Whiting_2012} to build a mask around the regions that are identified as containing the emission from the target. Within \bba, different parameters can be used to define how to build the mask. For our sample, we selected the parameters \texttt{SNRCUT} and \texttt{GROWTHCUT} that define the primary and secondary SNR cuts applied to the data.
    Once the emission pixels with flux above a threshold defined by \texttt{SNRCUT} are identified, the algorithm increases the detection area by adding nearby pixels that are above some secondary threshold defined by \texttt{GROWTHCUT} and not already part of the detected object. For ALPAKA galaxies, we used \texttt{SNRCUT} values of 2.5 - 4 and \texttt{GROWTHCUT} of 2 - 3 depending on the quality of each data cube. We note that the best-fit parameters are robust against the shape of the mask. Performing the fitting with masks obtained with the \texttt{SMOOTH\&SEARCH} task results in values of $V$ and $\sigma$ consistent within the uncertainties with the values reported in this paper. 
    \item \textit{Radial separation.} To keep the number of modeled rings as close as possible to the number of resolution elements, we used a radial separation between rings close to $0.5$ - 1 times the angular resolution (see Sect.\ref{sec:disks}).  
    This assumption ensures that the rotation curves and the velocity dispersion profiles are sampled with almost independent points, with the number of fitted rings ranging between 2 and 5 per galaxy.
    \item \textit{Center.} We fixed the galactic centers to the values obtained with \cannubi. For 18 out of the 23 galaxies with HST data, the center from \cannubi\ are consistent within $\lesssim0.1\arcsec$ with the ones found by \galfit\ on the HST data. The only exceptions are ID3, 14, 17, and 22 (see Sects.~\ref{sec:geometry} and \ref{sec:details} for further details on these targets). In some cases, we changed the coordinates of the center by less than 2 pixels\footnote{The ALMA data are imaged using pixel size equal to at least 1/5 of the beam size.} after visually inspecting the position-velocity diagrams (PVDs). The latter are cuts of the cubes along the major and minor axis (see Sect.\ref{sec:kinclassification} for details). The coordinates of the final adopted centers are listed in Table \ref{tab:tab1}.
    \item \textit{Inclination angle.} As discussed in Sect.\ref{sec:geometry}, for 21 galaxies with HST data, we fixed the inclination angles to the ones found by \galfit\ fitting, $i_{\mathrm{HST}}$, while for the remaining 7 galaxies, we fixed them to the values $i_{\mathrm{ALMA}}$ found by \cannubi.
\end{itemize}
Using the assumptions described above, we run \bba\ and we fit the rotation velocity $V_{\mathrm{rot}}$, the velocity dispersion $\sigma$ and the position angles. For each galaxy, the number of free parameters is equal to $N \times V_{\mathrm{rot}} + N \times \sigma + PA_{\mathrm{kin}} = 2 \times N + 1$, where $N$ is the number of rings over which the galaxy disk is divided. With the angular resolution of the ALPAKA galaxies, $N$ range from 2 to 5. We note that, despite some galaxies show evidence of a warp, we assume a constant $PA_{\mathrm{kin}}$ across the galaxy disks as the quality (SNR and angular resolution) of the data is not sufficient for constraining the change of $PA$ as a function of radius. 

\subsubsection{Outputs}
\begin{figure*}[htbp!]
    \begin{center}        \includegraphics[width=0.95\textwidth]{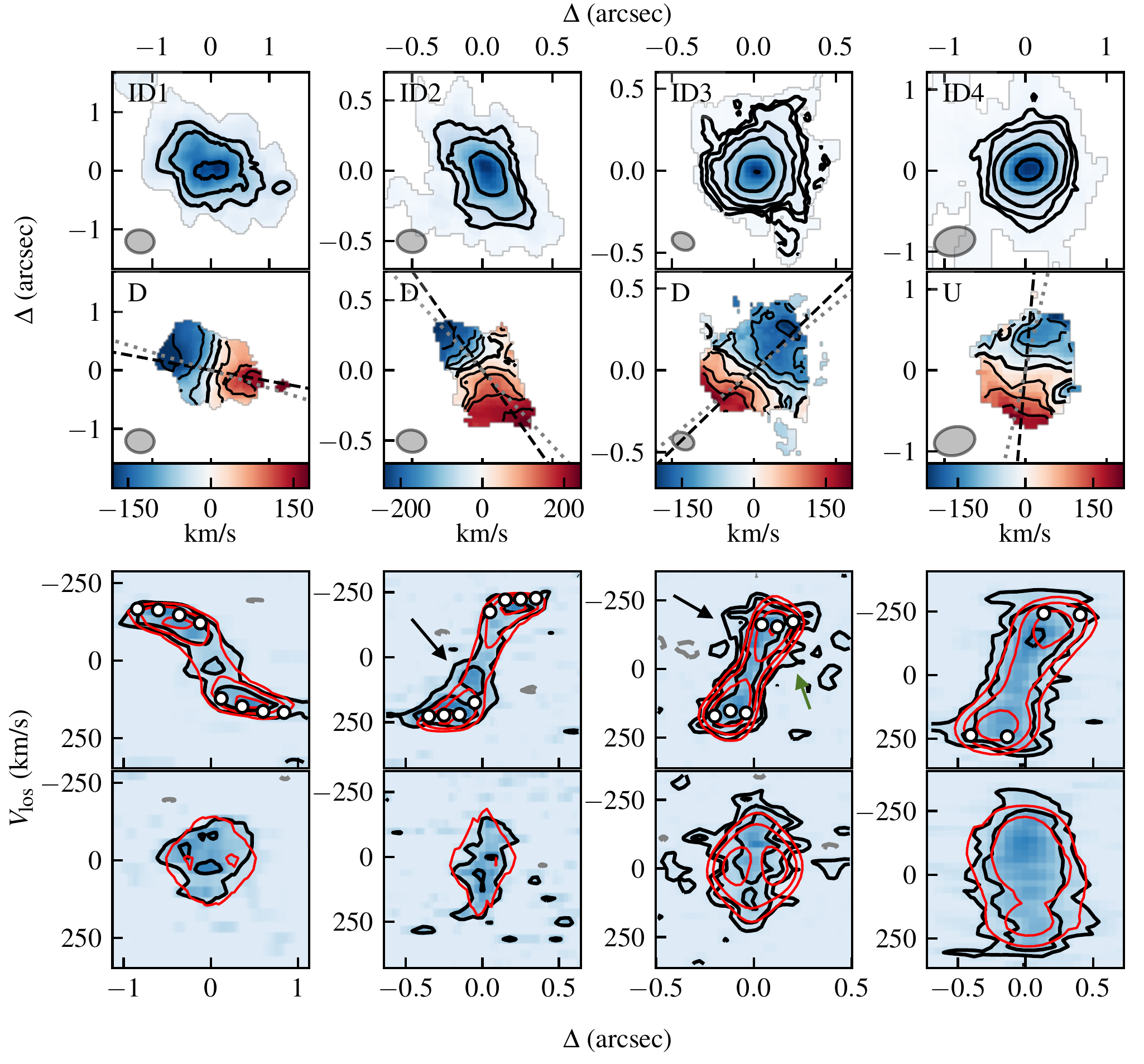}
        \caption{For each target with ID in the upper left, we show from top to bottom: the total-flux map and the velocity field and the major and minor-axis PVDs. In the total-flux maps, the first external contour is a "pseudo-contour" (see Appendix B in \citep{Roman_2023} for details) at 4 RMS. In the velocity field, the black lines show the iso-velocity contours, with the thickest one indicating the systemic velocity. The black dashed and gray dotted lines are the kinematic and morphological position angles, respectively. For the morphological position angle, we show the one obtained from fitting HST data when available and the total-flux map otherwise. The beam is shown in the bottom left. In the PVDs, the y-axis shows the line-of-sight velocities centred on the systemic velocity or redshift and the x-axis shows the distance with respect to the center of the kinematic model. The contours for the data (solid black) and the model (red) are at [1, 2, 4, 8, 16, 32] $\times$ 2.5 RMS. The gray dotted contours are at -2.5 RMS. The white circles are the best-fit line-of-sight rotation velocities. For each target, we indicate in the velocity field whether it is classified as a disk (D), merger (M) or uncertain (U). The arrows show kinematic anomalies that are described and discussed in detail in Sect.~\ref{sec:details}.} 
        \label{fig:pv1}
    \end{center}
\end{figure*}

For each galaxy, we show in Fig. \ref{fig:pv1} and Figs. \ref{fig:pv2} - \ref{fig:pv4}, the total-flux and velocity field maps and the PVDs along the major and minor axis for the data (dashed black contours) and the model (red solid contours). In Appendix \ref{sec:ch_maps}, we show 7 representative channel maps of the data, model, and residuals for each ALPAKA galaxy. The 2D kinematic maps (i.e., velocity field and velocity dispersion field) of the models are not shown here because, as extensively discussed in \citet{Rizzo_2022}, these are not as informative as the PVDs and channel maps. The velocity field and velocity dispersion fields are, in fact, strongly affected by the angular resolution and SNR of the data.
\\
The best-fit $PA_{\mathrm{kin}}$ as fit by \bba\ are listed in Table \ref{tab:PA}. In Fig. \ref{fig:ipa}, we compare these values with $PA_{\mathrm{HST}}$ and $PA_{\mathrm{ALMA}}$. As shown in the scatter plots in the central panel, the difference between the morphological PA and the kinematic ones are consistent, within 1.5-$\sigma$, with $\pm$ 30 $\deg$ for 20 of the 23 galaxies with HST data, indicating a general remarkable regularity of the sources. The only exception are ID13, 14, and 28, with $PA_{\mathrm{HST}} - PA_{\mathrm{kin}}$ in the range 40 -  53 $\deg$. Similarly, the absolute difference between $PA_{\mathrm{ALMA}}$ and $PA_{\mathrm{kin}}$ is consistent, within 1-$\sigma$, with $\pm 30 \deg$ for 27 out of the 28 ALPAKA galaxies, while ID27 has a difference of $\sim 45 \deg$. In Sect. \ref{sec:kinclassification}, we discuss potential explanations of the discrepancies between $PA_{\mathrm{kin}}$, $PA_{\mathrm{ALMA}}$ and $PA_{\mathrm{HST}}$ for this subsample of ALPAKA galaxies. A detailed description of the best-fit rotation velocity and velocity dispersion values is, instead, given in Sect.~\ref{sec:disks}.

\subsection{Kinematic classification} \label{sec:kinclassification}
The velocity fields of the ALPAKA galaxies are characterized by a smooth gradient. Therefore, we could, in principle, conclude that all ALPAKA galaxies are smooth rotating disks. However, the velocity map of a merging system can be similar to that of a smooth rotating disk as, due to the angular resolution of observations, the irregularity and asymmetries are smoothed out \citep[e.g.,][]{Simons_2019, Kohandel_2020, Rizzo_2022}. Further, even outflows could result in velocity gradients that could be erroneously interpreted as rotation \citep{Loiacono_2019}. In this Section, we will discuss how we identify any non-circular motions in the ALPAKA targets and build a subsample of galaxies where the presence of a rotating disk can be considered robust.
\subsubsection{Visual inspection} \label{sec:visclass} 
For rotating disks with non solid-body rotation curves, the PVDs have specific features: the major-axis PVD has an S-shape profile, and the minor-axis PVD has a diamond shape, symmetric with respect to the axes defining the center and the systemic velocity. Using mock ALMA data of simulated galaxies, \citet{Rizzo_2022} show that these features are imprinted in the data even for barely resolved observations. At the typical resolutions of high-$z$ observations and for galaxies with flat rotation curves, the flux distribution along the major-axis PVD has two symmetric brightest emission regions in the approaching and receding sides, along the horizontal parts of the S-shape. In addition, using geometrical arguments, it can be shown that for an axisymmetric rotating disk, the kinematic position angle should be aligned with its projected morphological major axis. Differences between these two angles of $\gtrsim 30 \deg$ can be ascribed to a 
 variety of reasons -- e.g., presence of outflows \citep[e.g.,][]{Lelli_2018, Hogarth_2021} or non-axisymmetric structures \citep[e.g., bar or interaction features;][]{Krajnovic_2011}. By visually inspecting the PVDs of the data and models, the channel maps and comparing the morphological and kinematic position angles, we identify three classes of galaxies:  
\begin{itemize}
    \item \textbf{Disk} These are galaxies with PVDs and channel maps typical of rotating disks, and alignment between the kinematic and morphological position angles, with $PA_{\mathrm{kin}} - PA_{\mathrm{ALMA}} \lesssim 30 \deg$. Some ALPAKA galaxies host disks rotating with remarkable regularity (e.g., ID1, 13, 18, 23, 24, see Sect.~\ref{sec:details} for details), while others have a few features indicating the presence of kinematic anomalies mostly identified at low SNR: asymmetries along the minor axis (e.g., ID3, 6, 7, 11), excess emission at high velocities in the inner regions (e.g., ID 28, see Sect.~\ref{id28} for details). The former can be ascribed to disturbances driven by environmental effects (e.g., ram pressure stripping) or gravitational perturbations; the latter are likely due to emission from outflows. 
    Overall, 19 out of the 28 ALPAKA galaxies are disks. 
    \item \textbf{Merger} These are systems where either there are two interacting galaxies that can be clearly identified in the PVDs, channel maps, and integrated-flux maps (ID17) or galaxies for which a rotating disk model does not reproduce the emission in the PVDs and channel maps, and the velocity fields are strongly irregular (ID27). The latter case likely indicates the presence of a late-stage interaction.
    \item \textbf{Uncertain} We choose to be conservative in this classification. Therefore, if a galaxy does not fit into the above classes for any reason, it is classified as uncertain. The 7 ALPAKA galaxies in this class have distorted iso-velocity contours and asymmetries in the PVDs. In most cases, the angular resolution of the observations is not sufficient for definitely identifying their nature (e.g., ID5, 16) and discriminating between irregularities due to the presence of two interacting/merging galaxies or kinematic anomalies driven by non-circular motions within a disk structure.  
\end{itemize}
The kinematic class for each ALPAKA galaxy is listed in Table~\ref{tab:PA} and shown in Fig.~\ref{fig:pv1} and Figs.~\ref{fig:pv2} - \ref{fig:pv4} with a letter: D for disk, M for merger or interacting system, U for uncertain. Details regarding the motivation of the kinematic class assigned to each target are provided in Sect. \ref{sec:details}.

\subsubsection{PVsplit analysis}
\begin{figure*}[htbp!] 
    \begin{center}
        \includegraphics[width=\textwidth]{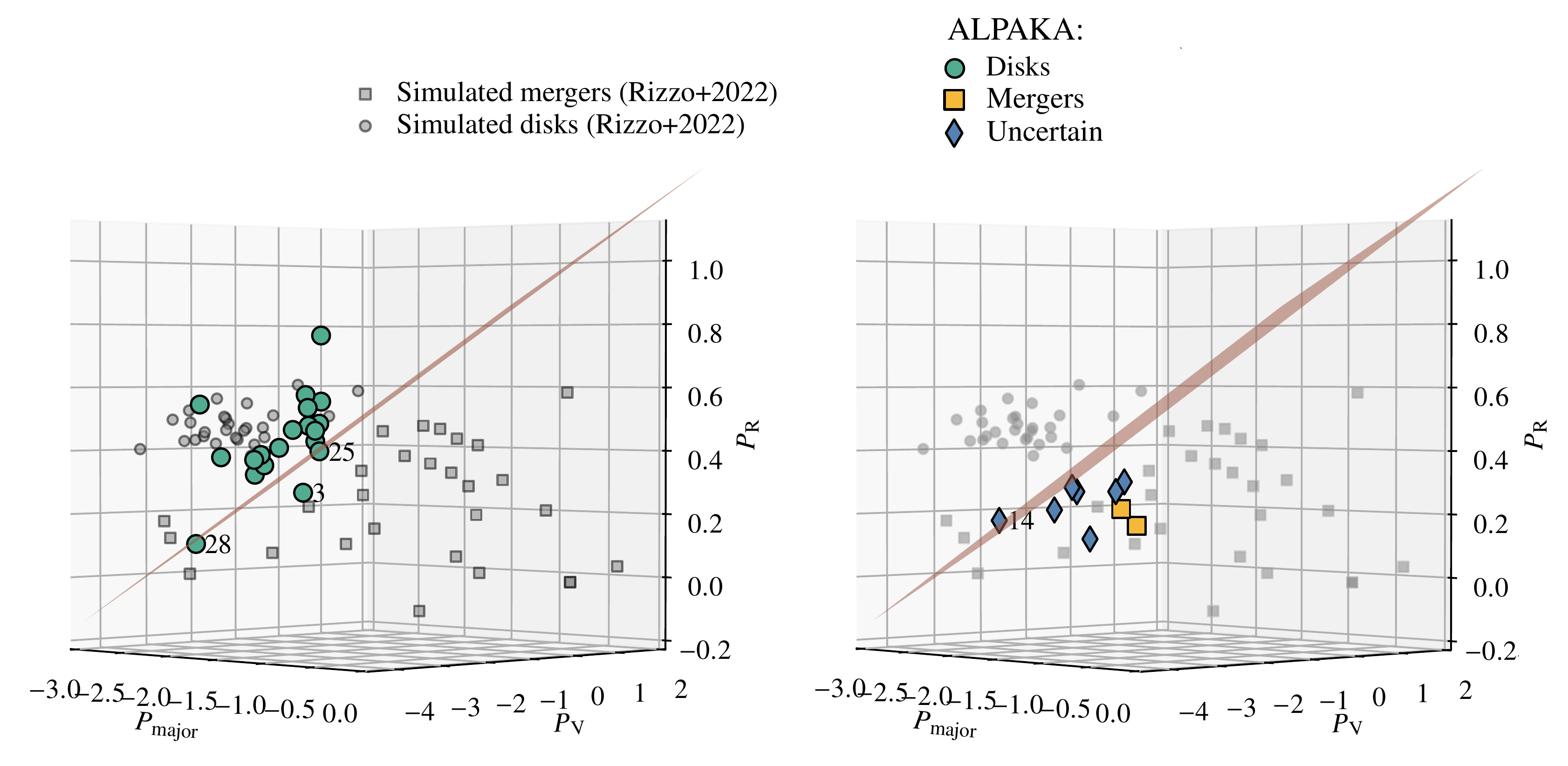}
        \caption{Distribution of the ALPAKA galaxies in the PVsplit parameter space. The gray circles and squares show simulated disks and mergers, respectively from \citet{Rizzo_2022}. The red plane divides the regions occupied by the two kinematic classes. The colored markers show ALPAKA galaxies classified as disks, merger and uncertain (right panel) based on the visual inspection of their PVDs and channel maps. Two slightly different projections are shown for better visualizing the position of all ALPAKA galaxies.} 
        \label{fig:pvsplit}
    \end{center}
\end{figure*}

Most of the kinematic classification methods rely on the analysis of the moment maps. However, recent studies show that at the typical resolution and SNR of current observations, the success rate of these methods can be as low as 10\% \citep{Rizzo_2022}. On the other hand, \citet{Rizzo_2022} present a new classification method -- "PVsplit''-- that relies on the morphological and symmetric properties of the major-axis PVDs, quantitatively defined by three parameters: $P_{\mathrm{major}}$, $P_{\mathrm{V}}$, and $P_{\mathrm{R}}$. The parameter $P_{\mathrm{major}}$ quantifies the symmetry of the PVD with respect to the systemic velocity: a rotating disk with an S-shape profile should have a completely symmetric PVD and a values of $P_{\mathrm{major}} = 0$. The parameters $P_{\mathrm{V}}$ and $P_{\mathrm{R}}$ define the position of the centroid of the brightest regions in the major-axis PVDs with respect to the line-of-sight velocity and center position, respectively. The PVsplit method has been tested using both low-$z$ systems and ALMA mock data of simulated galaxies \citep{Kohandel_2020, Pallottini_2022}, known to be disks, disturbed disks, and interacting systems. \citet{Rizzo_2022} find that disks and mergers occupy different locations in the 3D space defined by the PVsplit parameters (gray circles and squares in Fig.~\ref{fig:pvsplit}) and \citet{Roman_2023} define a plane that divides these two kinematic classes (red plane in Fig.~\ref{fig:pvsplit}). Simulated disks with disturbances driven by interactions are located both in the disk and merger PVsplit sections. \\
Despite the high quality of the ALPAKA data allowing for the accurate derivation of kinematic properties and identification of merger features, we applied the PVsplit method to get a further quantitative confirmation that the three classes described in Sect.\ref{sec:visclass} are reliable. In Fig.~\ref{fig:pvsplit}, we show the location of the galaxies that we classified as disks (green circles), mergers (yellow squares), and uncertain (blue diamond) according to the visual inspection of the spectral channels, PVDs, and residuals. Overall, ALPAKA disks and non disks (mergers and uncertain) lie in the two distinct regions of the PVsplit diagram. Some ALPAKA disks with kinematic anomalies (e.g., ID 3, 28) lie in the merger section or close to the dividing plane, similarly to the simulated perturbed disks analyzed in \citet{Rizzo_2022}. 

\section{ALPAKA in detail} \label{sec:details}
In this Section, we summarize the main physical properties of each ALPAKA target based on previous results from the literature, along with a description of the kinematic fitting and properties. The 2D maps (total flux and velocity fields), the PVDs of the data and models are shown in Figs.~\ref{fig:pv1}, \ref{fig:pv2}-\ref{fig:pv7}. 

\subsection{ID1}
ID1 is part of the IMAGES survey that investigated the [OII]3726, 3729 $\AA$ kinematics using the FLAMES-GIRAFFE multi-object IFU \citep{Yang_2008}. Based on their kinematic analysis, \citep{Yang_2008} classified ID1 as a regular disk. \\
The HST data for this galaxy clearly show that ID1 is a spiral galaxy (Fig.~\ref{fig:galfit1}). The residuals in the HST fitting are due, indeed, to the presence of spiral arms (Fig.~\ref{fig:galfit1}). The channel maps and PVDs from the CO(2-1) ALMA data have the typical features of a rotating disk (e.g., see the S-shape profile in Fig.~\ref{fig:pv1}). With our analysis, we find a rotation velocity for ID1 of 200 km\,s$^{-1}$, consistent within the $1\sigma$ uncertainties with the value of $230\pm 33$ km\,s$^{-1}$ found by the IMAGES collaboration \citep{Puech_2008}. 

\subsection{ID2}
This galaxy is part of the PHIBBS2 survey with ID: XV53. Based on the morphology derived by visually inspecting the HST \textit{I}-band (HST/ACS F814W) images, ID2 is classified as a disk with asymmetric features \citep{Freundlich_2019}. The HST image shows, in fact, a smooth disk and a clump in the south east direction (Fig.~\ref{fig:galfit1}). \citet{Lackner_2014} interpreted the presence of these two clumps as due to two interacting systems and classify this galaxy as a late-stage minor merger. Here, we model the HST image using two off-center Sérsic components. The CO(3-2) emission is compact and does not extend to the south-east peak. Overall, the channel maps and the PVD are well reproduced by a rotating disk model. However, in the velocity field, the iso-velocity contours on the receding side of the velocity fields are more distorted than the approaching side. This is also visible in the PVDs: along the major-axis, there are two bright peaks at positive velocities, one symmetric with respect to the negative side at $\sim 250$ km\,s$^{-1}$ and the other located at $\sim$ 50 km\,s$^{-1}$, possibly corresponding to the interacting companion (black arrow in Fig.~\ref{fig:pv1}); the minor axis PVD is asymmetric. Considering the disturbances on one side of ID2, we fitted $V_{\mathrm{rot}}$ and $\sigma$ only using the approaching side of the galaxy.  
To perform such a fitting, we fixed the $PA_{\mathrm{kin}}$ and the systemic velocity to the values obtained by using both sides of the galaxy.

\subsection{ID3}
This galaxy, also know as PACS819 is a \textit{Hershel} detected galaxy \citep{Rodighiero_2011} whose global properties are extensively studied in the far-infrared wavelength range \citep{Silverman_2015, Silverman_2018, Chang_2020}. In addition, by examining the BPT diagram \citep{Baldwin_1981}, \citet{Silverman_2015} show that PACS-819 is close to the line separating star-forming and AGN galaxies due to the strong [NII] emission line \citep{Kewley_2013}. In the rest-frame UV HST image, ID3 shows two bright clumps, while the center of the CO(5-4) emission is located between them. Due to the different morphology between the CO and UV data, for ID3 we fixed the inclination angle to the one obtained with \cannubi. 

The iso-velocity contours and the PVDs show that ID3 is kinematically lopsided, meaning that the velocity gradient in its approaching and receding sides are different from each other \citep{Swaters_1999, Bacchini_2023}. We therefore fit the approaching and the receding sides separately (note that in Fig.~\ref{fig:pv1} we show the receding-side model). The rotating disk models reproduce the bulk of the emission from ID3. However, the disturbances of the two external contours at 2.5 and 5$\sigma$ along the minor and major axis PVDs (see arrows in Fig.~\ref{fig:pv1}), as well as the residuals in the channel maps (Fig.~\ref{fig:channel1}), indicate the presence of non-circular motions, in particular at negative velocities. In Fig.~\ref{fig:pv1}, we show with a green arrow the gas that is moving at lower rotation than those predicted by the model. The black arrow shows, instead, gas that is rotating faster than predicted by the model, in particular in the inner regions. Such kinematic anomalies are usually attributed to extra-planar gas that is inflowing or outflowing from the disk \citep[e.g.,][]{Fraternali_2001, Heald_2011, Marasco_2019}. The minor-axis PVD is not well reproduced by the model: the contours of the data and the model do not overlap and the shape of the model appears rounder than the data. This is because \bba\ tries to fit the features at anomalous velocities, visible in the major-axis PVD, by increasing the velocity dispersion values. Therefore, for ID3, we consider the $\sigma$ values obtained by \bba\ as upper limits.

\subsection{ID4 - 9} \label{sec:id4_8}
These galaxies are part of the XMMXCS J2215.9-1738 cluster detected in the XMM Cluster Survey \citep{Romer_2001} at $z = 1.46$. Global and spatially resolved properties are studied using multi-wavelength observations \citep{Hayashi_2017, Hayashi_2018, Ikeda_2022}. Based on our kinematic analysis, we classify ID4 and 5 as uncertain, and ID6 - 9 as disks. 

The velocity gradient in the velocity field of ID4 is likely due to the presence of rotation, but the distorted iso-velocity contours may be ascribed to the presence of a bar \citep[e.g.,][]{Hogarth_2021}. However, due to the angular resolution of these data discriminating between radial motions driven by a bar or other mechanisms is not feasible. The best-fit Sérsic index obtained by modeling the HST data is consistent with a disk ($n \sim 1$) rather than a bar structure ($n \lesssim 0.5$).
In addition, the HST data of ID4 show a clump in the north-west direction that is not detected in CO and it is an [OII] emitter at $z \sim 1.46$, detected using a narrow-band filter \citep{Hayashi_2014, Ikeda_2022}.

\begin{figure*}
    \begin{center}
        \includegraphics[width=0.88\textwidth]{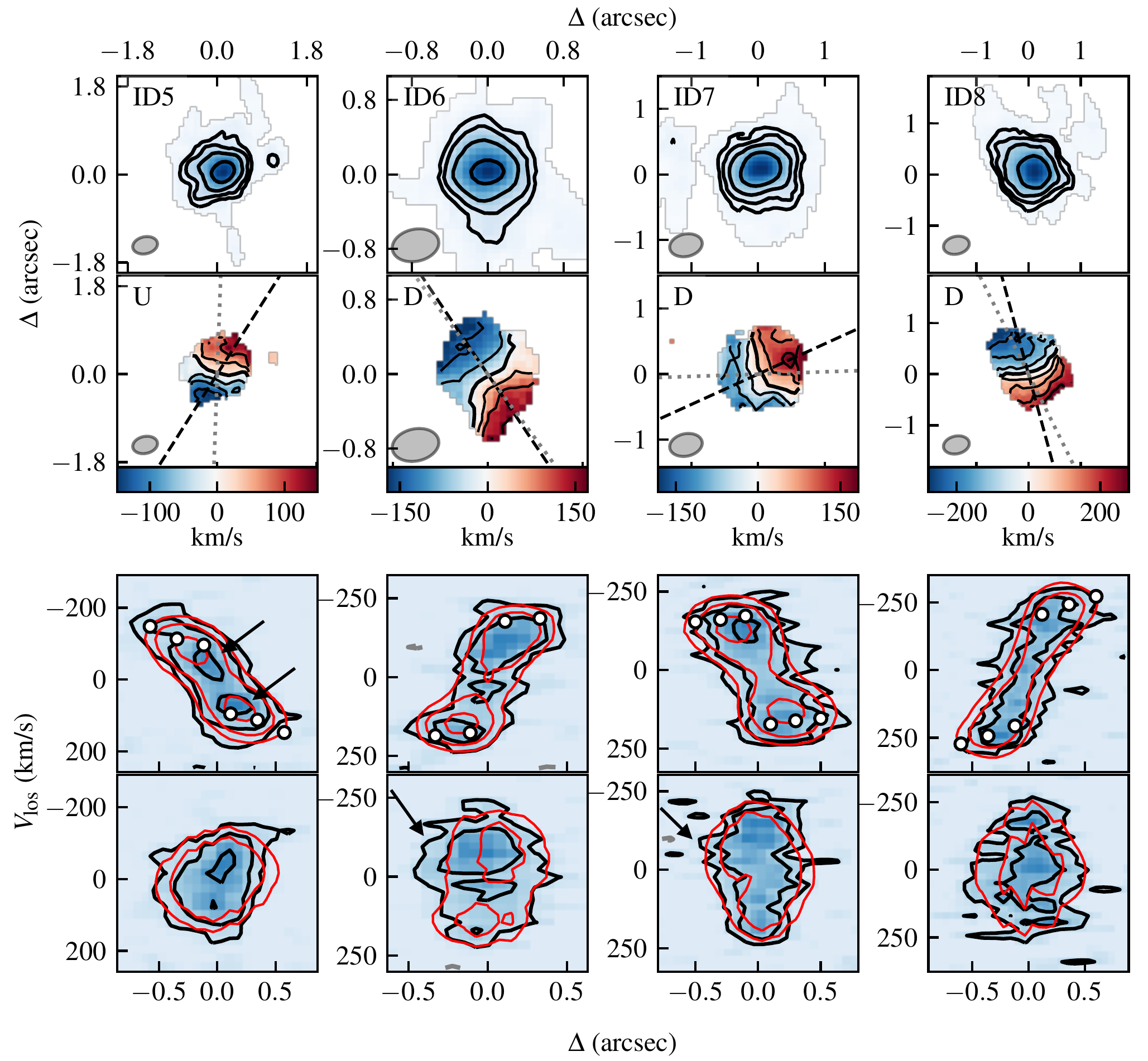}
        \caption{Same as Fig.~\ref{fig:pv1}, but for ID5 - 8.}  
        \label{fig:pv2}
    \end{center}
\end{figure*}

ID5 has strong asymmetric features in the PVDs.
The brightest regions of the major axis PVD are located at 0 and 100 km\,s$^{-1}$, respectively; they may be the nuclei of two unresolved interacting galaxies (see black arrows in Fig.~\ref{fig:pv2}). However, no clumpy structures are identified in the HST data. The minor axis PVD of ID5 is asymmetric and not reproduced by the rotating disk model. Another potential explanation of the irregularity of the ID5 PVDs is that this galaxy is a lopsided disk \citep{Noordermeer_2001}. Follow-up observations at higher angular resolution are needed to robustly identify its nature.

Galaxies 6 - 9 have, instead, the typical features of rotating disks. However, they show some asymmetries, especially along the minor-axis PVD (see arrows in Fig.~\ref{fig:pv2}), and distorted iso-velocity contours in the velocity fields that may be driven by environmental effects \citep[e.g., ram pressure stripping][]{Lee_2017, Zabel_2019, Cramer_2020, Bacchini_2023}, gravitational interactions \citep{Boselli_2022} or gas accretion \citep{deBlok_2014}. Due to the quality of the current data, discriminating between these mechanisms is challenging. For instance, at low-$z$, a clear indication of ram-pressure stripping is a distortion of the gas distribution with respect to the stellar component \citep{Lee_2017, Boselli_2022}. On the contrary, it is expected that gravitational interactions have an impact on both the gas and stellar components. For ID6 - 9, the total-flux maps appear smooth in all cases, likely because of the resolution of the observations, and aligned with the stellar distribution (Fig.~\ref{fig:hstalma1}). However, the HST data of ID7 show a close-by source (Fig.~\ref{fig:hstalma1}) that is not visible in the ALMA data. For ID7, the asymmetric features visible in the velocity field and in the minor-axis PVD (see black arrow in Fig.~\ref{fig:pv2}) are in the same direction of the nearby source, indicating that the asymmetries of the ID7 disk could be due to gravitational interactions. \bba\ tries to reproduce these kinematic perturbations by inflating the velocity dispersion (see the comparison between the models and the data in the minor-axis PVDs). For this reason, in Sect.~\ref{sec:discussion}, we consider the $\sigma$ values derived by \bba\ for ID7 as upper limits. 
ID8 has, instead, a close-by source that is visible both in HST and in CO(2-1) at low significance \citep[yellow controur in Fig.~\ref{fig:hstalma1}; see also][]{Ikeda_2022}. To recover this source using the 3D source finder described in Sect.~\ref{sec:barolo}, we assumed \texttt{SNRCUT} of 2 and \texttt{GROWTHCUT} of 1.8. This object is located at an angular distance of about $\sim 1.0 \arcsec$ (8 kpc) in projection from the main galaxy while its spectral distance from the systemic redshift of ID8 is of $\sim 800$ km s$^{-1}$. We note that the SNR of the ID8's companion is not sufficient to study its kinematics.

\subsection{ID10}
This galaxy is part of the SHiZELS survey \citep{Swinbank_2012, Molina_2017, Gillman} that map the H$\mathrm{\alpha}$ emission line with VLT/SINFONI. The ALMA observations of the CO(2-1) line used here were previously analyzed by \citet{Molina_2019}, who classify ID10 as a disturbed dispersion-dominated disk. Here, we confirm the presence of a smooth gradient in the velocity field, but the strong disturbances and asymmetries in both PVDs, the systematic residuals in the channel maps, and the presence of bright emission only on one side of the major-axis (see arrow in Fig.~\ref{fig:pv3}) lead us to classify ID10 as uncertain. 

\subsection{ID11 - 12}
\begin{figure*}[htbp!]
    \begin{center}
        \includegraphics[width=0.88\textwidth]{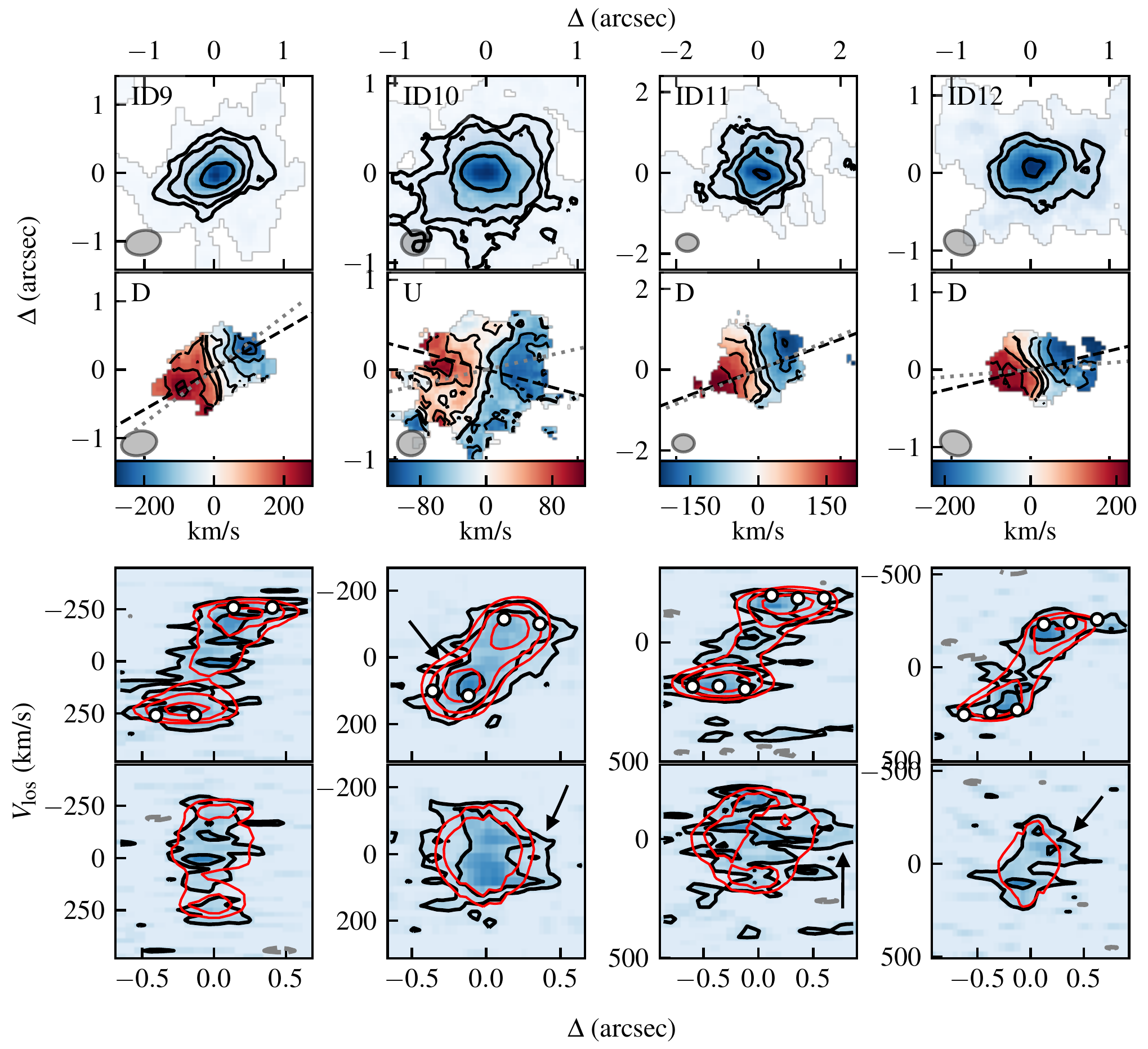}
        \caption{Same as Fig.~\ref{fig:pv1}, but for ID9 - 12.}  
        \label{fig:pv3}
    \end{center}
\end{figure*}

ID11 and 12 are members of two galaxy clusters that are recently studied in \citet{Cramer_2022} (see also \citet{Noble_2019} for ID11). Both galaxies have the typical features of a rotating disk (e.g., S-shape major-axis PVD) but they show some asymmetries in the minor-axis PVDs -- despite at low significance, see arrows in Fig.~\ref{fig:pv3} -- and residuals in the channel maps, likely due to tidal disturbances or ram-pressure stripping. 

\subsection{ID13}
This galaxy is a protocluster member. Its global properties (stellar masses, SFR, dust content) are studied in numerous papers \citep[e.g.,][]{Hung_2016, Zavala_2019}. In the HST cutout, two additional sources are visible, one located in the north east and the other in the south east directions. However, the redshifts of these sources are unknown and no dust or CO emission are detected with the ALMA data employed in this paper. The PVDs of ID13 show the typical features of a rotating disk. Despite this, ID13 has a relative difference between $PA_{\mathrm{kin}}$ and $PA_{\mathrm{HST}}$ of 53 $\deg$. Two reasons may be responsible for such a large difference: 1. being almost face-on, a robust estimation of the morphological position angle is not straightforward; 2. the light contamination from the two sources in the HST cutout may bias the estimation of the \galfit\ parameters.

 \subsection{ID14}
ID14, also known as BX610, has been extensively studied both in optical/near-infrared \citep{Forster_2009, Forster_2014, Tacchella_2018} and far-infrared wavelength \citep{Aravena_2014, Bolatto_2015, Brisbin_2019}. Its kinematics was previously analyzed using H$\alpha$ data from VLT/SINFONI survey \citep{Forster_2018}. Based on H$\alpha$, ID14 is described as a rotating disk containing massive star-forming clumps \citep{Forster_2011, Forster_2018}. It also exhibits signatures of gas outflows driven by a weak or obscured AGN in the center and by star formation at the location of the bright southern clump visible both in H$\alpha$ and UV \citep{Forster_2014}. Recently, \citet{Brisbin_2019} suggest that the high CO(7-6)/L$_{\mathrm{IR}}$ ratio measured in BX610 can be ascribed to shock excitation caused by a recent merger event. 

The data cube for ID14 contains both CO(4-3) and \CIi. However, we show here only the model obtained by fitting CO(4-3) which has higher SNR than \CIi. The CO(4-3) data show the presence of strong kinematic anomalies: the major axis PVD has the brightest emission only in the central regions; the minor-axis PVD is strongly axisymmetric, the iso-velocity contours are distorted in the south-east direction where a peak in the CO(4-3) distribution is visible (see arrows  in Fig.~\ref{fig:pv4}). The latter may be due to an interacting companion. In addition, ID14 has a large difference of 65 $\deg$ between $PA_{\mathrm{HST}}$ and $PA_{\mathrm{kin}}$. This misalignment is a further indication that ID14 is not a regularly rotating disk. However, the CO(4-3) kinematics is not sufficient to discriminate between the presence of a merging system and a rotating disk where the CO emission is contaminated by non-circular motions, likely driven by outflows. Therefore, in this paper, we put ID14 into the uncertain class. The combination of H$\alpha$ and CO kinematics and the multi-wavelength morphological analysis will allow us to constrain the nature of ID14 (Deveraux et al. in prep.).  

\subsection{ID15}
\begin{figure*}[htbp!]
    \begin{center}
        \includegraphics[width=0.88\textwidth]{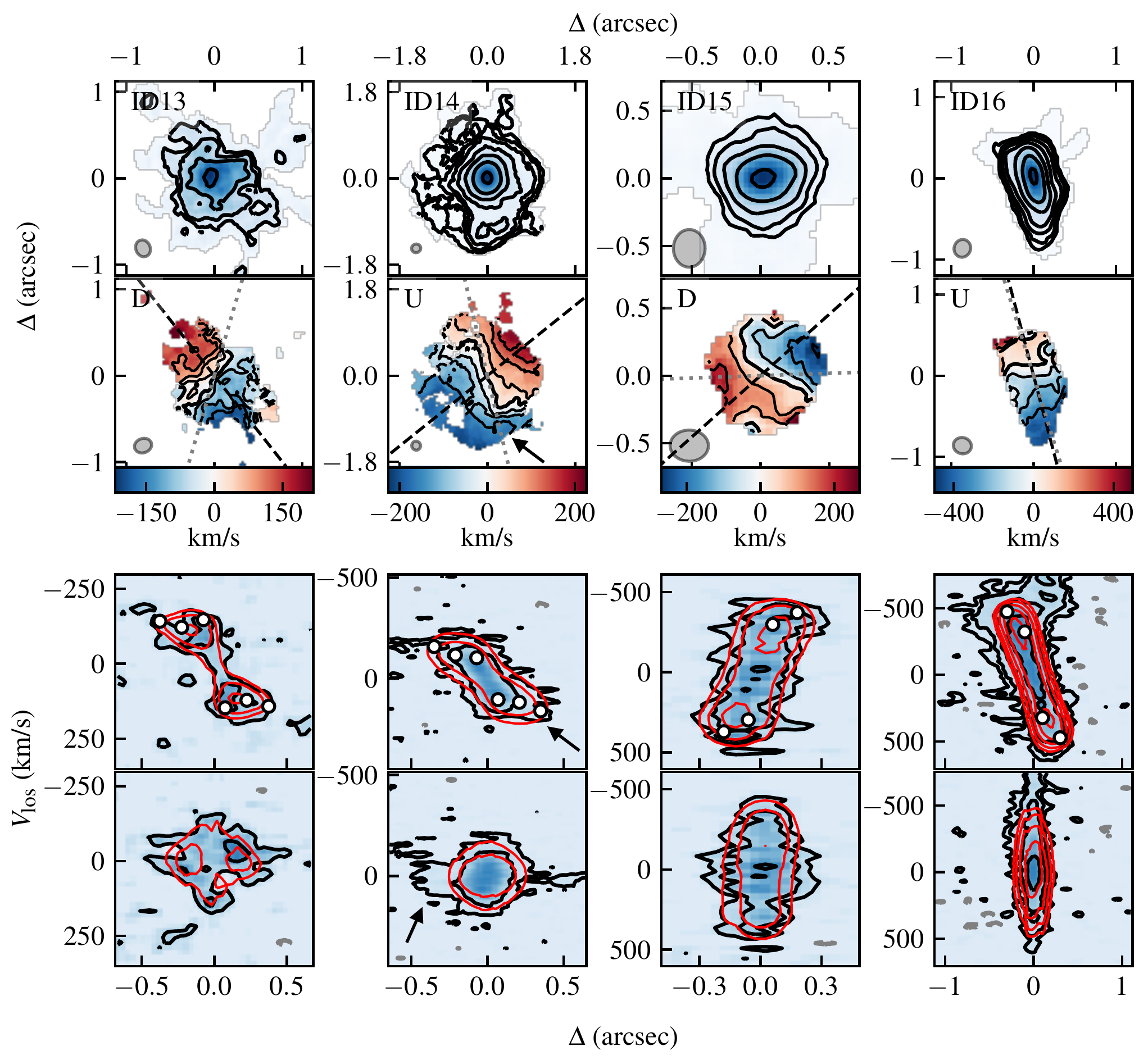}
        \caption{Same as Fig.~\ref{fig:pv1}, but for ID13 - 16.}  
        \label{fig:pv4}
    \end{center}
\end{figure*}

ID15 is known as the prototypical example of a compact star-forming galaxy that is rapidly consuming its gas reservoir and is expected to evolve into a quiescent galaxy \citep{Popping_2017}. The kinematics of the CO(6-5) emission line is studied in \citet{Talia_2018} that identify the presence of a rotating disk with a velocity $V = 320^{+92}_{-53}$ km\,s$^{-1}$. Further, \citet{Loiacono_2019} analyzed the H$\alpha$ and [OIII] kinematics, finding evidence of rotation motions, with $V \sim 200 ^{+17}_{-20} $ km\,s$^{-1}$ for an inclination of 75 $\deg$. Here, we assume an inclination of 60 $\deg$ derived from modeling the HST data. The CO(3-2) data clearly show the presence of a compact rotating disk. 
However, similarly to previous studies \citep{Talia_2018, Loiacono_2019}, the velocity dispersion values of ID15 are poorly constrained, having uncertainties of 50-60\% at all radii because of the low angular resolution of ALMA data. We note that ID15 has distorted iso-velocity contours (Fig.~\ref{fig:pv4}) and residuals in the channel maps that may indicate the presence of a warp or non-circular motions. However, higher angular resolution data are needed to robustly analyze them and constrain the driving mechanisms. 

\subsection{ID16 - 17}
\begin{figure*}[htbp!]
    \begin{center}
        \includegraphics[width=0.88\textwidth]{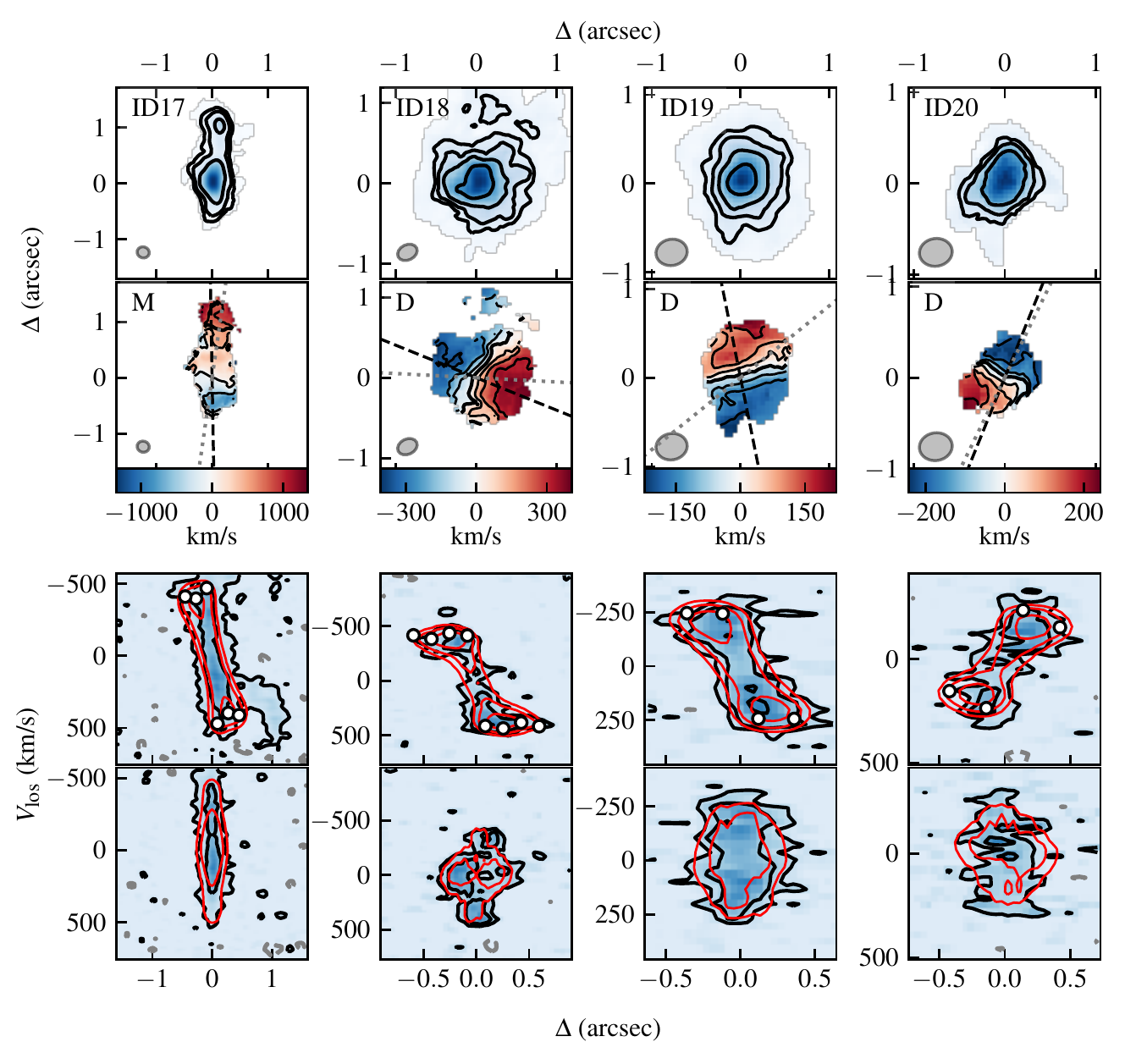}
        \caption{Same as Fig.~\ref{fig:pv1}, but for ID17 - 20.}  
        \label{fig:pv5}
    \end{center}
\end{figure*}

ID16 and 17 are two galaxies, separated by 3$\arcsec$ and connected by a bridge of gas and dust \citep{Fu_2013}. These sources are mildly magnified by two foreground galaxies, with magnification factors of 1.5. Due to their close proximity, ID16 and 17 were first identified as a unique bright SMG, HXMM01, in the \textit{Hershel} Multi-tiered Extragalctic survey \citep{Oliver_2012}. Subsequently, higher-angular resolution observations resolve HXMM01 into three sources, ID16, 17 and its companion \citep{Fu_2013}, also visible in the total-flux maps in Fig.~\ref{fig:hstalma1}. The bright clump visible in the HST image at the southern end of ID17 has an SED that is consistent with either a less obscured galaxy at $z = 2.3$ or a physically unrelated contaminating source \citep{Fu_2013}. The ALMA cube employed here contains two emission lines, CO(7-6) and \CIii\ that slightly overlap along the frequency axis. In Figs. \ref{fig:pv4} and \ref{fig:pv5}, we show only the CO(7-6), characterized by higher SNR than \CIii. A kinematic analysis of this complex system was already presented in \citet{Xue_2018} who interpreted ID16 and 17 as two rotating disks. Despite ID16 showing some symmetric features along the minor-axis PVD (Fig.~\ref{fig:pv4}), the major-axis PVD is strongly asymmetric, being bright only at negative velocities. Similarly, ID17 is strongly asymmetric as it is interacting with a companion, clearly visible in the total-flux map (Fig.~\ref{fig:hstalma1}) and major-axis PVD (Fig.\ref{fig:pv5}) located at 1$\arcsec$ in the north direction with respect to the ID17 center. Due to the presence of strong disturbances, we classify ID16 as uncertain and 17 as an interacting system. 

\subsection{ID18} 
ID18 lies in a well-studied protocluster. This structure was originally found in the largest extragalactic \textit{Hershel} survey (H-ATLAS). \citet{Ivison_2013, Ivison_2019} find that HATLAS J084933.4+021443 hosts an exceptionally high SFR $\sim 7000$ M$_{\odot}$\,yr$^{-1}$, spread over 5 galaxies (W, T, C, M, E), confined within a scale of $\sim$100 kpc . The ALMA data set used here maps galaxy W and T at high resolution and SNR, while the data quality is not sufficient for performing a kinematic analysis on galaxies C, M, and E. Further, galaxy T is not included in the ALPAKA sample as it is a strongly gravitationally lensed source, while galaxy W, that is, ID18 is not lensed \citep{Ivison_2013}. \\
The CO(4-3) data analyzed in this paper indicates that ID18 is a regularly rotating disk. 

\subsection{ID19 - 21} \label{id21}
These galaxies are members of the forming cluster CLJ1001 at $z = 2.51$ \citep{Wang_2016, Gomez_2019, Champagne_2021}. The CO(3-2) kinematics of ID19 - 21, along with another galaxy member of the same cluster (130901), are recently studied by \citet{Xiao_2022} using the same data employed here. We exclude 130901 from our analysis due to the low SNR. \\
For the kinematic fitting, \citet{Xiao_2022} used \texttt{GALPAK$^{\mathrm{3D}}$} \citep{Bouche_2015} and they fit simultaneously the kinematic and geometric parameters, finding values of the inclinations of 24, 45 and 19 $\deg$ and position angles of 12, 148 and 10 $\deg$. For fitting the kinematics with \bba, we used, instead, inclination angles from HST data modeling of 56, 45 and 26 $\deg$. The kinematic position angles (32$\pm$15, 133$\pm$8, 13$\pm$5 $\deg$) are consistent at 1.5-$\sigma$ level with the values reported by \citet{Xiao_2022}. 

The comparison between the \bba\ model and the data indicates that the bulk of the motions in ID19 and 20 are reproduced by a rotating disk. The value of the velocity dispersion for ID19 is consistent within 1.5-$\sigma$ with the one found by \citet{Xiao_2022}, while we find a value of rotation velocity which is a factor of 1.5 smaller than the value found by \citet{Xiao_2022}. This discrepancy is due to the different inclination angles used to recover the intrinsic rotation velocity. On the contrary, for ID20 we find rotation velocities consistent with those found in \citet{Xiao_2022}, while the velocity dispersion values are not well constrained, having relative uncertainties $\gtrsim 60\%$, due to the low angular resolution and SNR of the data. 

\begin{figure*}[htbp!]
    \begin{center}
        \includegraphics[width=0.88\textwidth]{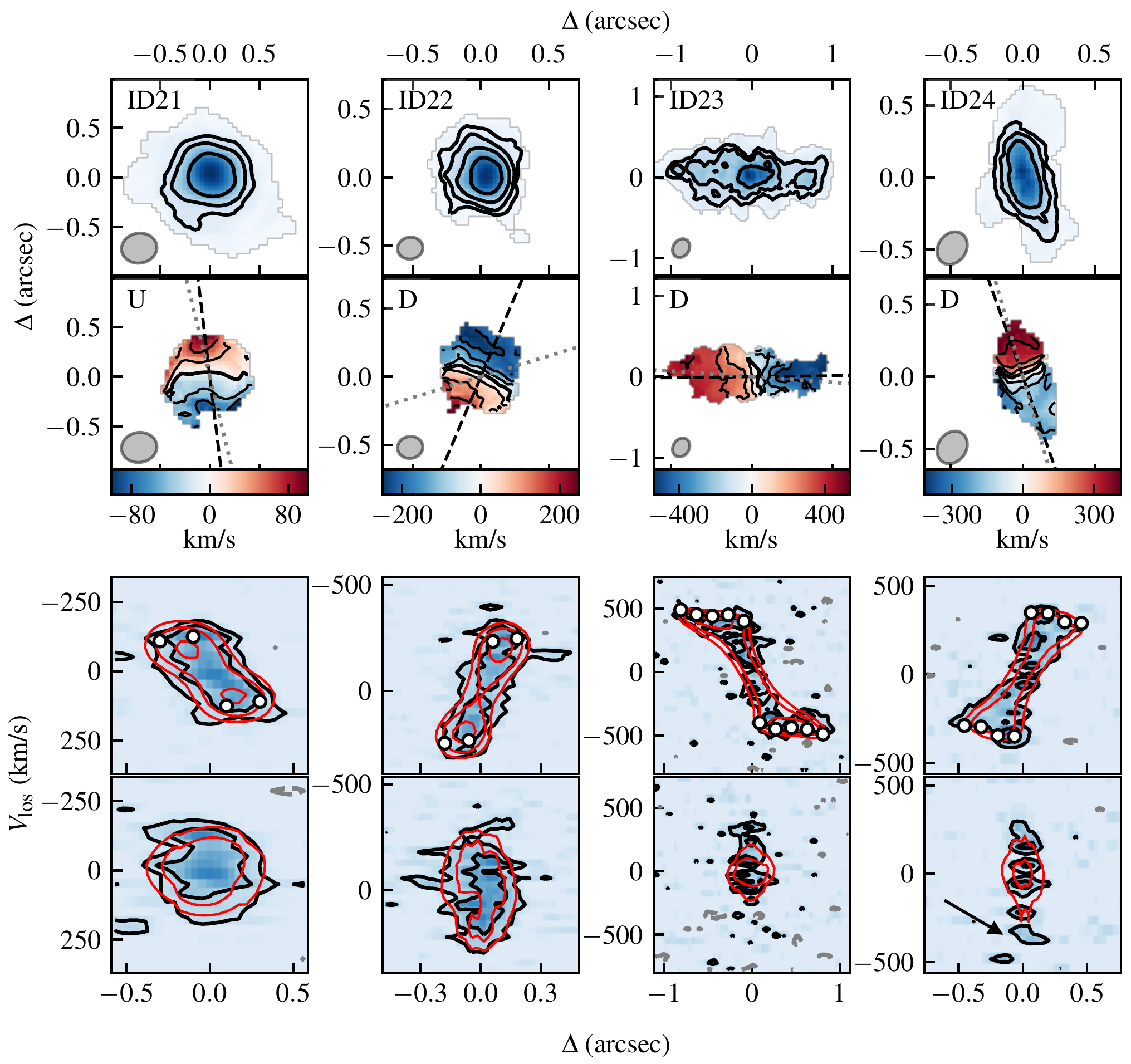}
        \caption{Same as Fig.~\ref{fig:pv1}, but for ID21 - 24.}  
        \label{fig:pv6}
    \end{center}
\end{figure*}

ID21 has, instead, asymmetries in the minor-axis PVD, while its major-axis PVD lacks the bright features typical of rotating disks. One potential explanation for this latter feature is that ID21 has a rotation curve that is slowly rising in the inner regions. However, since the angular resolution of the data employed in this paper does not allow us to robustly test this scenario, we classify this target as an uncertain galaxy. \citet{Xiao_2022} interpreted, instead, the gradient in the velocity map of ID21, also visible in Fig.~\ref{fig:pv6}, as due to the presence of a rotating disk. 

\subsection{ID22, 26 - 27} \label{id26}
These galaxies are recently studied by \citet{Cassata_2020} using ALMA observations with angular resolutions of 0.6$\arcsec$. The 0.17$\arcsec$ observations of the CO(5-4) used in our work allow us to resolve their kinematic structures. A rotating disk is a good description of the data only for ID22, despite the strong misalignment between the morphological position angle derived from HST data and $PA_{\mathrm{kin}}$. However, we note that, similarly to ID3, the HST data for ID22 probe the rest-frame UV and, therefore, the morphological parameters may be biased due to the contamination from the light of young stars and/or by dust attenuation.   

ID26 has irregularities in the minor-axis PVD (see arrow in Fig.~\ref{fig:pv7}), systematic residuals in the channel maps and distorted iso-velocity contours in the approaching side of the galaxy. These features may be due to the presence of a warp in a disk. However, the angular resolution of the observations do not allow us to verify this scenario. Therefore, to be conservative, we classify ID26 as uncertain. On the contrary, we classify ID27 as a merger. This target has a very irregular PVDs and velocity field and a difference between $PA_{\mathrm{ALMA}}$ and $PA_{\mathrm{kin}}$ of 45$\deg$, another indication that it is, likely, a non-resolved interacting system.

\subsection{ID23 - 25} 
These galaxies are members of the SSA 22 protocluster \citep{Steidel_1998} that was extensively studied with ALMA \citep{Umehata_2015, Umehata_2017, Umehata_2018}. \citet{Lehmer_2009} identified X-ray luminous AGNs in ID 23 and 25 using observations from the \textit{Chandra} Space telescope. Extended Lyman-$\alpha$ emission from multiple filaments between galaxies within SSA 22 are identified using MUSE observations \citep{Umehata_2019} and are thought to be responsible for the accretion of gas within the protocluster and the growth of galaxies and their supermassive black holes. 

The high-angular resolution observations employed here allow us to clearly see the typical features of edge-on rotating disks in ID 23, 24, and 25. The three galaxies have very extended S-shape major-axis PVDs (Fig.~\ref{fig:pv4}). ID24 has gas emission at $-400$ km s$^{-1}$ along the minor-axis PVD that is not reproduced by the model (see arrow in Fig.~\ref{fig:pv4}). However, due to the high inclination of the galaxy, these asymmetries are difficult to interpret. 

\subsection{ID28} \label{id28}
This is an hyper-luminous dust-obscured AGN that was identified by the Wide-field Infrared Survey Explorer \citep[WISE,][]{Wright_2010} and extensively studied using ALMA observations \citep[e.g.,][]{Fan_2018, Diaz_2021, Ginolfi_2022}. Recently, \citet{Ginolfi_2022} report the presence of 24 Lyman-$\alpha$ emitting galaxies on projected physical scales of 400 kpc around ID28. 

\begin{figure*}[htbp!]
    \begin{center}
        \includegraphics[width=0.88\textwidth]{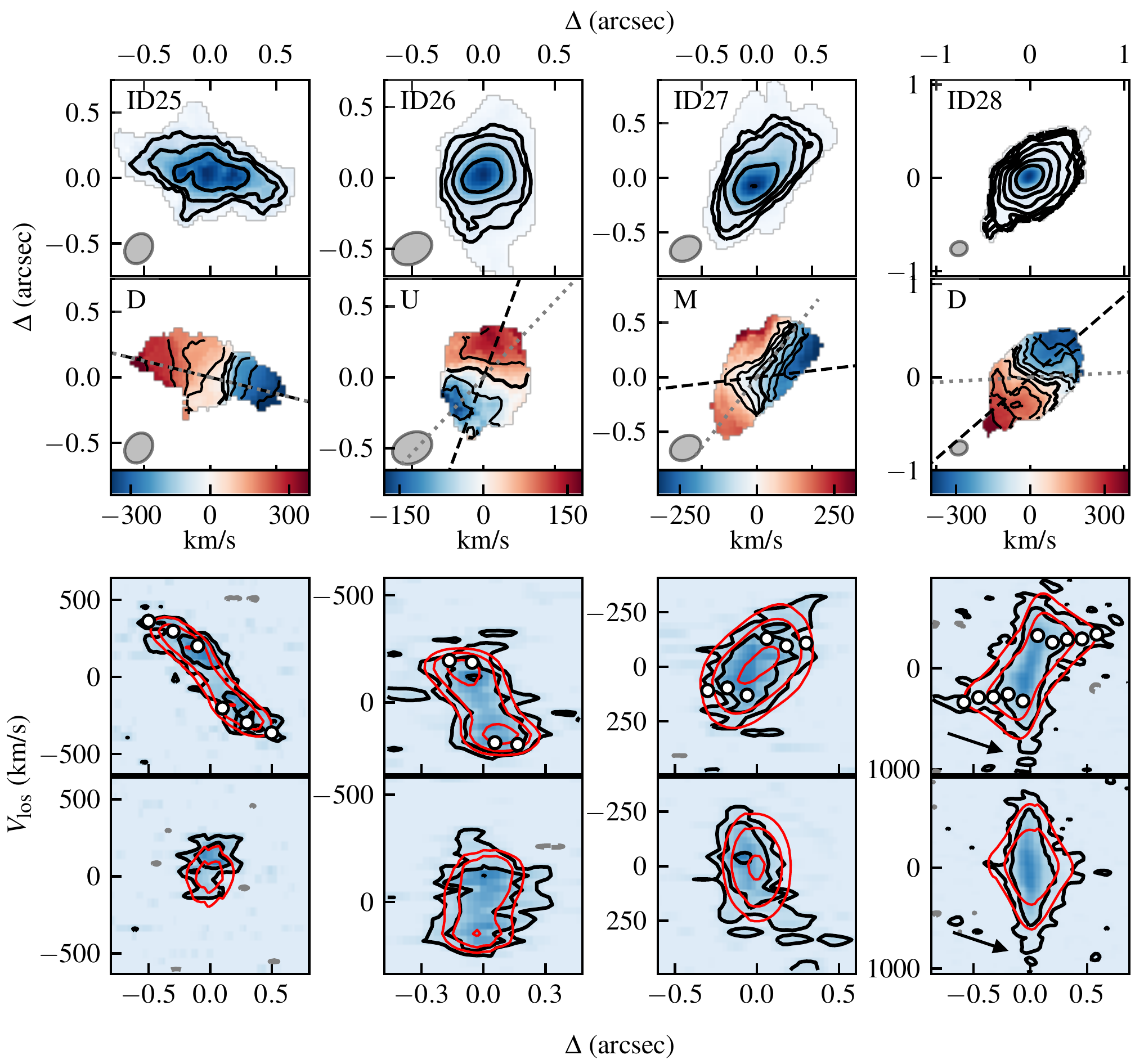}
        \caption{Same as Fig.~\ref{fig:pv1}, but for ID25 - 28.}  
        \label{fig:pv7}
    \end{center}
\end{figure*}

The kinematics of ID28 is peculiar as it shows the typical features of a rotating disk (e.g., S-shape along the major-axis PV, diamond shape along the minor axis PV) but also strong emission in the inner 1 kpc regions (see arrows in the PVDs, Fig.~\ref{fig:pv7}) that indicates the presence of gas moving at a line-of-sight velocity of 900 km s$^{-1}$ relative to the systemic velocity. This emission, not reproduced by the symmetric rotating disk model (see residuals in the channel maps, Fig.~\ref{fig:channel5}), can be explained in two ways. The first possibility is that the CO distribution is asymmetric between the approaching and the receding sides and there is an inner rise of the rotation curve caused by the presence of a compact bulge. The second possibility is that this emission is due to non-circular motions driven by outflows. We note that strong outflow motions were already identified in ID28 from the analysis of rest-frame UV spectrum \citep{Ginolfi_2022}.

The $V$ values in the inner regions for this galaxy should thus be taken with caution as the emission from the disk may be strongly contaminated by the one from the outflow. Furthermore, to reproduce the emission at high velocities in the inner regions, \bba\ inflates the velocity dispersion (e.g., the inner contours of the model are rounder than the data contour in the minor-axis PVD). For this reason, we consider the $\sigma$ values as upper limits. By
analyzing the same CO(6-5) data presented in this paper, \citet{Ginolfi_2022} find velocity dispersion and rotation velocity profiles consistent with those found here.

\section{Discussion} \label{sec:discussion}
\subsection{Kinematics of the disk subsample} \label{sec:disks}
\begin{figure*}[htbp!]
    \begin{center}
        \includegraphics[width=0.95\textwidth]{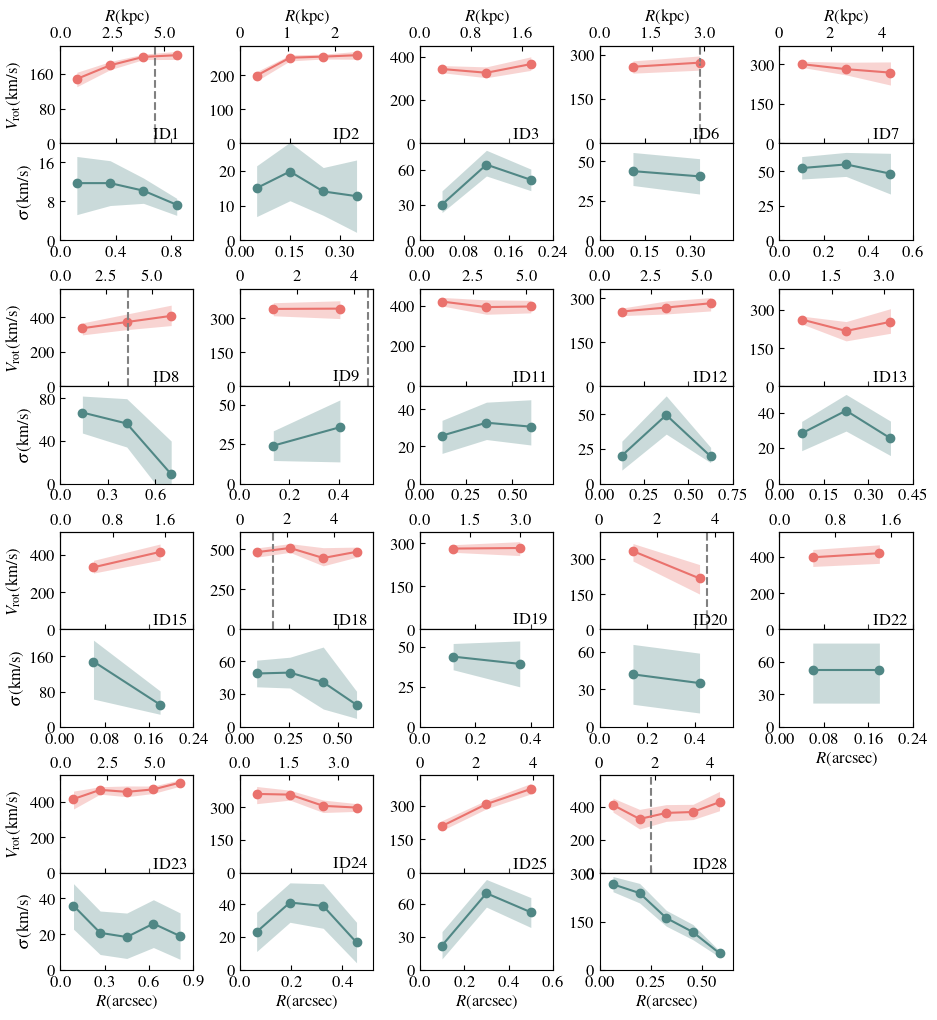}
        \caption{Rotation velocity (upper panels, pink) and velocity dispersion profiles (bottom panels, green) for the ALPAKA galaxies classified as disks. The circles show the location of the rings used for fitting the data with \bba\ and they can be considered independent from each other. The dashed vertical lines show the location of the optical/UV effective radius $R_{\mathrm{e}}$ for ALPAKA galaxies with HST data and with $R_{\mathrm{e}}$ comparable to or smaller than the extent of the CO/[CI] emission.} 
        \label{fig:vprofile}
    \end{center}
\end{figure*}

In this subsection, we discuss the kinematic properties of the 19 ALPAKA secure disks. The rotation velocities and velocity dispersions of mergers are, instead, unreliable as they are derived under the assumption that the observed kinematics is dominated by circular motions. As such, we decide not to show nor discuss them. For the same reason, the kinematic properties of the 7 galaxies in the uncertain class are not shown here. In Fig.\ref{fig:vprofile}, we show the profiles of the rotation velocity and velocity dispersion as derived by \bba\ (Sect.\ref{sec:barolo}). The angular resolution of the ALPAKA data allows us to sample the kinematic profiles with only 2 independent resolution elements for 6 ALPAKA galaxies and $\gtrsim 3$ for the remaining 13 targets. For the 16 ALPAKA disks with HST data, we quantified the relative extension of the kinematic profiles with respect to the rest-frame optical/UV effective radius $R_{\mathrm{e}}$ obtained with \galfit. The extension of the kinematic profile is estimated by adding half of the beam size to the outermost radius used for the kinematic fitting. The comparison between the CO/[CI] and rest-frame optical/UV extension is not straightforward as differences may depend not only on the heterogeneous sensitivities of ALPAKA data but also on the different emission lines that we are using to trace the kinematics. Low-J CO transitions may trace, in fact, more diffuse, extended molecular gas than high-J CO transitions \citep[e.g.,][]{Lagos_2012}. In addition, the values of $R_{\mathrm{e}}$ may vary with the rest-frame wavelength \citep[e.g.,][]{Vulcani_2014, Kennedy_2015}. Considering these caveats, we find that only 7 sources -- 4 with CO(2-1), 1 with CO(3-2), 1 with \CIii, 1 with CO(6-5) -- have kinematic profiles that extend up to radii $\gtrsim R_{\mathrm{e}}$ (see gray dotted lines in Fig.~\ref{fig:vprofile}), while for the others the ALMA data trace only the innermost regions. 

Both galaxies with compact and extended kinematic profiles show a variety of shapes: flat (e.g., ID12, 19), slowly increasing and then flattening (e.g., ID1, 2), declining in the inner regions and then flattening (e.g., ID13, 28). We caution the reader that the velocity profiles in Fig.\ref{fig:vprofile} show the rotation and not the circular speed. The latter is a direct proxy of the gravitational potential and can be derived from $V_{\mathrm{rot}}$ after applying the asymmetric drift correction which is a function of the velocity dispersion profile and gas distribution \citep[e.g.,][]{Iorio_2018}. Such estimates, as well as the derivation of the gravitational potential, will be part of a future paper in the ALPAKA series. 
 
For most of the ALPAKA disks, there is an indication of velocity dispersion profiles with a declining trend, where the $\sigma$ values are higher in the inner regions than at the outermost radius. To quantify this trend, we fitted the velocity dispersion profiles using a linear function and we find negative slopes for 14/19 disks. However, due to the large uncertainties of the velocity dispersion values, for 9/14 galaxies these negative slopes are consistent with 0 within the 1-$\sigma$ uncertainties.
The decrease of $\sigma$ with radius is also observed in the CO profiles of local disk galaxies \citep{Bacchini_2020}, and it is ascribed to the radial decline of the SFR surface density. The analysis of the ALPAKA galaxies shows, therefore, that caution must be taken when fitting the velocity dispersion profile assuming a constant value across the disks. This assumption, often used at high-$z$ \citep[e.g.,][]{Bouche_2015, Neeleman_2020}, may result in the estimation of velocity dispersion $\sigma$ that is mainly influenced by the emission from the bright regions at the center of a galaxy, in turn resulting in larger $\sigma$ measurements compared to the average values. On the other hand, another usually used assumption is to measure the velocity dispersion from the outer regions of the observed velocity dispersion field, where the contribution from the beam-smearing effect is minimized \citep[e.g.,][]{Wisnioski_2015, Harrison_2017}. In this latter case, it is not straightforward to evaluate whether the resulting $\sigma$ values overestimate or underestimate the average $\sigma$ across the galaxy disks, as it may strongly depend on the combination of galaxy size, angular resolution, shape of the rotation velocity and velocity dispersion profiles \citep[e.g.,][]{DiTeodoro_2015, Burkert_2016}.

\begin{figure}[htbp!]
    \begin{center}
        \includegraphics[width=0.95\columnwidth]{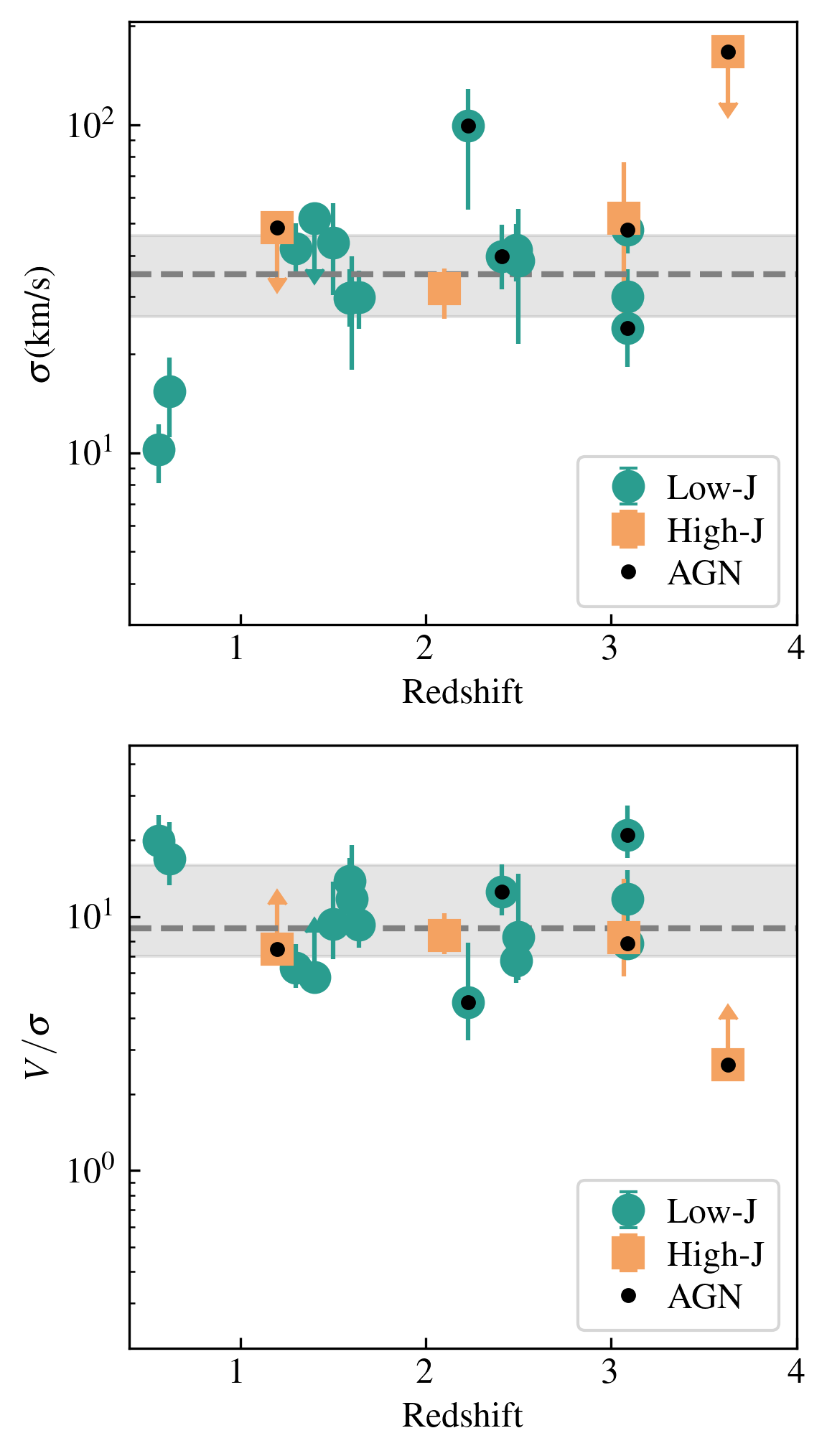}
        \caption{Distribution of the ALPAKA disks in the velocity dispersion - redshift plane (upper panel) and rotation-to-velocity dispersion ratio - redshift plane (bottom panel). The values of $\sigma_{\mathrm{m}}$ and $V_{\mathrm{max}}/\sigma_{\mathrm{m}}$ (Table~\ref{tab:vsigma}) are plotted here. The redshifts of ID3, ID6 - 9, and ID19 are  shifted by $|\Delta z| \lesssim 0.25$ for a better visualization of all the data points. Galaxies with kinematics from CO(2-1), CO(3-2) or \CIi\ are shown with green circles, while galaxies with CO(5-4), CO(6-5) or \CIii\ are shown with orange squares. The dashed gray line (and the gray area) show the median (and the 16th and 84th percentiles) $\sigma_{\mathrm{m}}$ and $V_{\mathrm{max}}/\sigma_{\mathrm{m}}$ values for the ALPAKA disks.}  
        \label{fig:vsigma}
    \end{center}    
\end{figure}

To facilitate the comparison with previous works, in Tab.~\ref{tab:vsigma}, we refactor our result in terms of the kinematic parameters mostly common in the literature: the maximum values of the rotation velocity $V_{\mathrm{max}}$; the radial average values of the velocity dispersion $\sigma_{\mathrm{m}}$; the values of velocities and dispersions computed by averaging the two outermost values in their profiles, $V_{\mathrm{ext}}$ and $\sigma_{\mathrm{ext}}$; the ratios $V_{\mathrm{max}}/\sigma_{\mathrm{m}}$ and $V_{\mathrm{ext}}/\sigma_{\mathrm{ext}}$. The latter are used to define the rotational support of galaxies: $V/\sigma \gtrsim 2$ is the threshold used to indicate rotation-dominated systems \citep{Forster_2018, Wisnioski_2019}. In Fig.\ref{fig:vsigma}, we show $\sigma_{\mathrm{m}}$ and $V_{\mathrm{max}}/\sigma_{\mathrm{m}}$ as a function of redshift for the ALPAKA disks. In this figure, galaxies hosting an AGN are indicated with black dots. As discussed in Appendix \ref{sec:kinclassification}, some ALPAKA disks have kinematic anomalies that \bba\ tries to reproduce by inflating the velocity dispersion. Therefore, for at least three galaxies (i.e., ID3, 7, 28), the velocity dispersion values can be considered upper limits (and thus lower limits for $V/\sigma$). 

As discussed in Sect.~\ref{sec:gloabl}, ALPAKA sample is biased towards galaxies hosting energetic mechanisms (e.g., stellar and AGN feedback, interactions due to the environment) that are expected to either boost the velocity dispersion values or completely destroy the disk structures \citep[e.g.,][]{Dubois_2016, Penoyre_2017, Gurvich_2022, Kretschmer_2022}. Despite this, we find that $\sigma_{\mathrm{m}}$ ranges between 10 and 52 km\,s$^{-1}$ for all galaxies, except for ID 15 and 28, two AGN hosts with values of $\sigma_{\mathrm{m}}$ of 99 km\,s$^{-1}$ and an upper limit of 167 km\,s$^{-1}$, respectively. The median value of $\sigma_{\mathrm{m}}$, obtained after excluding only the three upper limits is 35$^{+11}_{-9}$ km\,s$^{-1}$ (dashed gray line in Fig.~\ref{fig:vsigma}, upper panel). The $V_{\mathrm{max}}/\sigma_{\mathrm{m}}$ values range between 5 and 21, with a median value of 9$^{+7}_{-2}$ (dashed gray line in Fig.~\ref{fig:vsigma}, bottom panel). In the second paper of this series, we will compare the distribution of $\sigma$ and $V/\sigma$ of the ALPAKA sample with redshift-matched samples of galaxies with warm gas kinematics, and we will study the evolution of the cold gas kinematics across cosmic time.

\begin{table}[htbp!]
\begin{center}
\caption{Global kinematic parameters of the ALPAKA galaxies classified as disks. 
}
\label{tab:vsigma}
\begin{tabular}{ccccccc}
\\
\hline\hline \noalign{\smallskip}
ID   & $V_{\mathrm{max}}$ & $\sigma_{\mathrm{m}}$ & $V_{\mathrm{ext}}$ & $\sigma_{\mathrm{ext}}$ & $V_{\mathrm{max}}/\sigma_{\mathrm{m}}$ & $V_{\mathrm{ext}}/\sigma_{\mathrm{ext}}$\\
 & km\,s$^{-1}$ &  km\,s$^{-1}$ & km\,s$^{-1}$ & km\,s$^{-1}$\\
\noalign{\smallskip}
\hline
\noalign{\medskip}
1 & $ 204 \substack{+ 9 \\ -10 }$ &$ 10 \substack{+ 3 \\- 2 }$ &$ 202 \substack{+ 5 \\- 6 }$ &$ 9 \substack{+ 2 \\- 3 }$ &$ 19 \substack{+ 14 \\- 6 }$ &$ 21 \substack{+ 24 \\- 8 }$ \\
2 & $ 259 \substack{+ 11 \\ -12 }$ &$ 15 \substack{+ 4 \\- 4 }$ &$ 257 \substack{+ 7 \\- 7 }$ &$ 13 \substack{+ 6 \\- 6 }$ &$ 17 \substack{+ 6 \\- 4 }$ &$ 19 \substack{+ 15 \\- 6 }$ \\
3 & $ 366 \substack{+ 34 \\ -29 }$ &$ \lesssim 49 $ &$ 346 \substack{+ 22 \\- 18 }$ &$ \lesssim58 $ & $\gtrsim 7 $ & $\gtrsim 6 $ \\
6 & $ 273 \substack{+ 23 \\ -25 }$ &$ 42 \substack{+ 8 \\- 7 }$ &$ 266 \substack{+ 15 \\- 17 }$ &$ 42 \substack{+ 8 \\- 7 }$ &$ 6 \substack{+ 1 \\- 1 }$ &$ 6 \substack{+ 1 \\- 1 }$ \\
7 & $ 300 \substack{+ 10 \\ -12 }$ &$ \lesssim 52$ &$ 274 \substack{+ 24 \\- 26 }$ &$ \lesssim 52 \substack{+ 9 \\- 9 }$ &$ \gtrsim 6 $ &$ \gtrsim 5 $ \\
8 & $ 408 \substack{+ 62 \\ -56 }$ &$ 44 \substack{+ 14 \\- 14 }$ &$ 390 \substack{+ 39 \\- 36 }$ &$ 32 \substack{+ 19 \\- 18 }$ &$ 9 \substack{+ 4 \\- 3 }$ &$ 11 \substack{+ 11 \\- 4 }$ \\
9 & $ 343 \substack{+ 35 \\ -44 }$ &$ 30 \substack{+ 10 \\- 12 }$ &$ 342 \substack{+ 22 \\- 27 }$ &$ 30 \substack{+ 10 \\- 12 }$ &$ 12 \substack{+ 7 \\- 3 }$ &$ 12 \substack{+ 8 \\- 3 }$ \\
11 & $ 421 \substack{+ 21 \\ -22 }$ &$ 30 \substack{+ 7 \\- 5 }$ &$ 395 \substack{+ 24 \\- 24 }$ &$ 32 \substack{+ 9 \\- 7 }$ &$ 14 \substack{+ 3 \\- 2 }$ &$ 12 \substack{+ 4 \\- 2 }$ \\
12 & $ 283 \substack{+ 19 \\ -26 }$ &$ 30 \substack{+ 6 \\- 6 }$ &$ 275 \substack{+ 15 \\- 17 }$ &$ 35 \substack{+ 8 \\- 7 }$ &$ 9 \substack{+ 3 \\- 2 }$ &$ 8 \substack{+ 2 \\- 1 }$ \\
13 & $ 263 \substack{+ 14 \\ -17 }$ &$ 32 \substack{+ 5 \\- 6 }$ &$ 236 \substack{+ 32 \\- 30 }$ &$ 33 \substack{+ 7 \\- 8 }$ &$ 8 \substack{+ 2 \\- 1 }$ &$ 7 \substack{+ 2 \\- 2 }$ \\
15 & $ 417 \substack{+ 42 \\ -44 }$ &$ 99 \substack{+ 29 \\- 44 }$ &$ 375 \substack{+ 28 \\- 27 }$ &$ 99 \substack{+ 29 \\- 44 }$ &$ 5 \substack{+ 3 \\- 1 }$ &$ 4 \substack{+ 3 \\- 1 }$ \\
18 & $ 509 \substack{+ 27 \\ -27 }$ &$ 40 \substack{+ 10 \\- 8 }$ &$ 466 \substack{+ 35 \\- 30 }$ &$ 30 \substack{+ 17 \\- 14 }$ &$ 12 \substack{+ 3 \\- 2 }$ &$ 14 \substack{+ 11 \\- 5 }$ \\
19 & $ 283 \substack{+ 22 \\ -26 }$ &$ 42 \substack{+ 8 \\- 8 }$ &$ 282 \substack{+ 13 \\- 15 }$ &$ 42 \substack{+ 8 \\- 8 }$ &$ 7 \substack{+ 2 \\- 1 }$ &$ 7 \substack{+ 2 \\- 1 }$ \\
20 & $ 334 \substack{+ 32 \\ -43 }$ &$ 38 \substack{+ 17 \\- 17 }$ &$ 276 \substack{+ 34 \\- 40 }$ &$ 38 \substack{+ 17 \\- 17 }$ &$ 8 \substack{+ 6 \\- 3 }$ &$ 7 \substack{+ 5 \\- 2 }$ \\
22 & $ 420 \substack{+ 46 \\ -56 }$ &$ 52 \substack{+ 25 \\- 30 }$ &$ 409 \substack{+ 32 \\- 38 }$ &$ 52 \substack{+ 25 \\- 25 }$ &$ 8 \substack{+ 6 \\- 3 }$ &$ 8 \substack{+ 6 \\- 3 }$ \\
23 & $ 508 \substack{+ 16 \\ -20 }$ &$ 24 \substack{+ 6 \\- 6 }$ &$ 489 \substack{+ 12 \\- 15 }$ &$ 22 \substack{+ 9 \\- 9 }$ &$ 21 \substack{+ 7 \\- 4 }$ &$ 21 \substack{+ 15 \\- 6 }$ \\
24 & $ 362 \substack{+ 34 \\ -45 }$ &$ 30 \substack{+ 6 \\- 6 }$ &$ 303 \substack{+ 16 \\- 18 }$ &$ 28 \substack{+ 9 \\- 9 }$ &$ 12 \substack{+ 3 \\- 2 }$ &$ 11 \substack{+ 5 \\- 3 }$ \\
25 & $ 377 \substack{+ 25 \\ -22 }$ &$ 48 \substack{+ 7 \\- 7 }$ &$ 343 \substack{+ 15 \\- 16 }$ &$ 61 \substack{+ 9 \\- 9 }$ &$ 8 \substack{+ 2 \\- 1 }$ &$ 6 \substack{+ 1 \\- 1 }$ \\
28 & $ 428 \substack{+ 65 \\ -52 }$ &  $\lesssim$ 167  &$ 398 \substack{+ 40 \\- 35 }$ & $ \lesssim 85$ & $ \gtrsim 3$ & $ \gtrsim 5 $ \\
\noalign{\smallskip}
\hline
\end{tabular}
\tablefoot{$V_{\mathrm{max}}$ is the maximum rotation velocity, $\sigma_{\mathrm{m}}$ is the mean velocity dispersion, $V_{\mathrm{ext}}$ and $\sigma_{\mathrm{ext}}$ are the 
external rotation velocity and velocity dispersion, respectively. The latter are defined as the averages of the
last two radial points. }
\end{center}
\end{table}

\subsection{Selection effects and the impact of energetic mechanisms on the disk properties} \label{sec:selectioneffect}
The selection criteria used to build the ALPAKA sample are not based on the global physical properties of the sources, but on the quality of the available data in the ALMA archive. This, combined with the intrinsic faintness of CO transitions and the limited sensitivities of ALMA, results in a sample that is biased towards massive (i.e., $\gtrsim 10^{10}$M$_{\odot}$) main-sequence or starburst galaxies, mostly in overdense regions. Numerous studies show that the gas-to-stellar mass ratio are typically higher in cluster than in field galaxies at $z \gtrsim 1$ \citep{Noble_2017, Hayashi_2018, Tadaki_2019}. Further, at least 7 out of the 28 ALPAKA galaxies are known to host an AGN. According to current models of galaxy formation and evolution, the growth of high-$z$ massive star-forming galaxies is mainly driven by galaxy mergers and intense gas accretion which drive large amounts of gas towards their centers, boosting the SFR and the growth of supermassive black holes. In the following subsections, we discuss the impact of different astrophysical mechanisms on the dynamics of the ALPAKA disks. 

\subsubsection{Starbursts}
Among the 12 starbursts in ALPAKA, we find that 8 are rotating disks. 
The dynamical time scale of these disks is $\approx$ 10 Myr \footnote{The dynamical timescales are computed as $R_{\mathrm{ext}}/V_{\mathrm{ext}}$, where $R_{\mathrm{ext}}$ is the outermost radius for which we measured the rotation velocity.}, a factor of 10 smaller than the typical depletion timescales ($\sim$ 100s Myr) of starburst galaxies at these redshifts \citep{Scoville_2017, Liu_2019}. Despite the small number statistics, the presence of disks among the ALPAKA starbursts suggests that after a merger or an intense accretion event, galaxies quickly transition into a stable dynamical stage and form a disk with $V/\sigma \sim 10$. These high $V/\sigma$ values are consistent with recent findings of dynamically cold disks among dusty starburst galaxies at $z \gtrsim 4$ \citep{Rizzo_2020, Lelli_2021, Rizzo_2021, Roman_2023} and they are in contrast with previous studies reporting values of $V/\sigma \lesssim 3$ for this galaxy population \citep{Swinbank_2011, Alaghband-Zadeh_2013, Olivares_2016, Birkin_2023}. However, we note that the latter were mostly obtained with low-angular resolution observations and with no \citep{Alaghband-Zadeh_2013, Olivares_2016} or suboptimal beam-smearing correction \citep[][see Sect.~\ref{sec:intro}]{Swinbank_2011, Birkin_2023}. Another potential reason for the systematic difference with previous studies of $z \lesssim 4$ starbursts is the emission lines employed to trace the galaxy kinematics. With the exception of the target studied in \citep{Swinbank_2011} with CO transitions, the others works employ rest-frame optical emission lines \citep{Alaghband-Zadeh_2013, Olivares_2016, Birkin_2023} tracing warm, ionized gas. As discussed in Sect.~\ref{sec:intro}, the gas tracer may impact the derived rotation velocity and velocity dispersion. A detailed comparison with the results from the literature is beyond the scope of the paper and is part of a future study of the ALPAKA series.

\subsubsection{AGN feedback}
Among the subsample of 7 AGN-host galaxies, only one belongs to the uncertain class while the others are rotating disks. This result might indicate that AGN feedback does not prevent the formation of a rotating disk and it may have also a limited impact on the ISM properties of the host galaxies. With the exception of ID28 (see Sect.\ref{sec:disks}), whose kinematic properties are likely contaminated by outflows, all AGN hosts have velocity dispersions consistent the with rest of the disk subsample. Interestingly, ID28 is the only ALPAKA galaxy belonging to the rare, extreme population of hyper-luminous dust-obscured AGN \citep{Eisenhardt_2012, Wu_2012}. The latter are thought to represent a short-lived phase during which there is an evolutionary transition from dusty starburst galaxies to UV-bright quasars \citep[e.g.,][]{Wu_2018, Diaz_2021}.

\subsubsection{Overdense environments}
Previous results from the literature \citep[e.g.,][see Sect. \ref{sec:details} for details]{Ivison_2013, Umehata_2015} indicate that most ALPAKA galaxies are in overdense regions, clusters or groups (13/28) or protoclusters (6/28). In these dense environments, galaxy mergers and tidal interactions could be efficiently enhanced with respect to the field \citep[e.g.,][]{Merritt_1983, Moore_1996, Kronberger_2006, Cortese_2021}. Recent studies show that environmental effects, like ram-pressure stripping, are responsible for the kinematic asymmetries of the molecular gas distribution and kinematics in cluster galaxies at $z\sim0$ \citep[e.g.,][]{Lee_2017, Cramer_2020, Bacchini_2023}. Among the 13 ALPAKA galaxies in clusters and groups, we classify 1 as an interacting system, 4 as uncertain, and the remaining 8 as disks. With the exception of ID8 and 9, the remaining cluster disks have kinematic anomalies likely driven by environmental mechanisms. On the contrary, among the 6 protocluster galaxies, 5 are disks and 1 is a disk with outflow contamination (ID28). This finding suggests that environmental mechanisms may not have a relevant impact on the ISM of galaxies during the early stages of cluster formation. However, the statistics is too low to draw robust conclusions. 

\subsection{CO and [CI] transitions}
The ALPAKA sample comprises a large variety of CO and [CI] transitions. Overall, these lines trace the cold molecular gas (T $< 100$ K), a phase of the ISM different from the photoionized gas observed through forbidden (e.g., [OIII], [OII]) or recombination (e.g., H$\alpha$) lines. However, the different critical densities of the CO rotational transitions (e.g., $10^{4}$ cm$^{-3}$ for the CO(2–1) to $\approx 5 \times 10^{5}$ for the CO(7-6)) make them sensitive to distinct ISM conditions. Low-J CO transitions (J $\lesssim 3$) trace the diffuse molecular component at T $\approx$ 5 - 33 K, while higher J transitions trace increasingly dense gas at T $\approx$ 100 K \citep{Carilli_2013}. In addition, high rotational transition levels (e.g., CO(5-4) and higher) if excited, may trace the presence of gas heated by various mechanisms (e.g, X-rays, \cite{vanderwel_2010}; shocks induced by mergers, radio jets, and supernova- or AGN-driven outflows, \cite{Kamenetzky_2016, Vallini_2019}). Similarly, \CIi\ is more sensitive to the cold, diffuse gas, than \CIii\ \citep{Valentino_2020, Valentino_2021}. 

Among the ALPAKA targets, 20 have CO(2-1), CO(3-2), CO(4-3) or \CIi\ observations, with a large fraction (15/20) containing a rotating disk. Instead, among the 8 ALPAKA galaxies with CO(5-4), CO(6-5), CO(7-6) and \CIii\  observations, four are interacting or uncertain galaxies with strong asymmetries in their kinematics and 2/4 of the disks (i.e., ID3 and 28) have kinematic anomalies potentially driven by outflows. Despite the small statistics of the high-J subsample not allowing us to draw significant conclusions, the different fractions of regularly rotating disks in the low and high-J subsamples suggest that the high-J transitions tend to trace non-circular motions. On the other hand, the comparison between the distributions of $\sigma_{\mathrm{m}}$ and $V/\sigma$ for the low and high-J subsamples (green circles and orange squares in Fig.\ref{fig:vsigma}) suggest that gas components traced by these transitions are characterized by similar levels of turbulence. Follow-up observations of the two ALPAKA subsamples at high and low-J transitions, respectively, would be needed to gain insights into the impact of the specific transitions on the turbulence and the presence of non-circular motions.

\section{Summary and conclusion} \label{sec:conclusion}
Studying the kinematics of galaxies using cold gas tracers is crucial for gaining insights into the formation and evolution of structures across cosmic time.
Nevertheless, before the advent of the Next Generation Very Large Array \citep[ngVLA;][]{ngvla} in 2030s and the improvement of the ALMA sensitivity thanks to the correlator upgrade \citep{Carpenter_2020}, high-resolution observations of CO transitions at $z \gtrsim 0.5$ are going to be feasible only for a few bright targets, due to the long integration times required to achieve sensitivities sufficient to recover robust kinematic measurements. For this reason, systematic investigations of the cold gas kinematics at $z \sim 0.5 - 3.5$ are still lacking. In this context, the ALPAKA project aims at alleviating this limitation by providing a systematic derivation of the kinematic properties of a sample of 28 star-forming galaxies at $z \sim 0.5 - 3.6$ (median $z$ of 1.8). Data for the ALPAKA galaxies are obtained by collecting ALMA archive high-data quality observations (median angular resolution of 0.25$\arcsec$) of CO and [CI] transitions. 

In this paper, we present the ALPAKA galaxies and derive their global ($M_{\star}$, SFR), morphological and kinematic properties. We find that ALPAKA galaxies have high stellar masses ($M_{\star} \gtrsim 10^{10}$ M$_{\odot}$) and SFRs that range from 10 to 3000 M$_{\odot}$ yr$^{-1}$. A large fraction of ALPAKA are starbursts (12 out of the 25 galaxies with good estimates of $M_{\star}$ and SFR), while the remaining are on the Main sequence of star-forming galaxies. Further, after combining a heterogeneous set of works from the literature, we conclude that 19/28 ALPAKA galaxies lie in overdense regions (clusters, protoclusters, groups). We exploited the ALMA data cubes and modeled the kinematics of the ALPAKA targets using \bba, a forward-modeling tool which fits galaxy kinematics assuming a rotating disk. Our analysis and the main results can be summarized as follow.
\begin{itemize}
    \item We divide the 28 ALPAKA galaxies into three kinematic classes: rotating disks, mergers and uncertain. We base this classification on the visual inspection of the PVDs, channel maps, velocity fields and the comparison with the pure rotating disk model obtained with
    \bba. This kinematic classification is confirmed by PVsplit, a tool that discerns between disks and mergers based on the measurements of asymmetries of the major-axis PVD. We find that 19 targets are rotating disks, 2 are interacting systems, and 7 are uncertain. For the latter, the data quality is not sufficient to reliably determine their kinematic states. 
    \item We present the rotation curves of the disk subsample; their semi-major axes are sampled by at least 3 independent resolution elements for 13 out of the 19 disks. In these cases, we can trace the shape of the rotation curves, and we find that they flatten. 
    \item We present the velocity dispersion profiles of the 19 disks. For 14 of them, there is a marginal indication that the profiles are not constant and the $\sigma$ values are larger in the inner regions and decline up to a factor of 3 in the outskirts, indicating that there may be a radial change in the intensity of the mechanisms driving turbulence (e.g., decline in the SFR surface density).
    \item We present values for the global kinematics of the disk galaxies in ALPAKA: the maximum rotation velocity range from 204 to 509 km s$^{-1}$; the velocity dispersion averaged across the radii, $\sigma_{\mathrm{m}}$, ranges from 10 to 100 km s$^{-1}$; the $V/\sigma$ range from 5 to 21. The ALPAKA disk sample has a median $\sigma_{\mathrm{m}}$ of 35$^{+11}_{-9}$ km s$^{-1}$ and an overall large $V/\sigma$ with a median value of 9$_{-2}^{+7}$.
\end{itemize}
This work is the first of a series that will allow us to fully exploit the ALMA data presented here to characterize the ALPAKA galaxies. In particular, we plan to use the kinematic analysis described in this manuscript, in combination with the HST data presented in this paper and JWST observations that have been taken for some ALPAKA sources, to: systematically compare the dynamics from warm and cold gas tracers; investigate the evolution of cold-gas velocity dispersion with redshift and study the energy sources of turbulence; infer the dark matter halo properties and content within the ALPAKA sample; derive the dynamic scaling relations and study their evolution; constrain the impact of cold outflows and AGN feedback on the gas disks; compare the sizes and morphologies of the stellar, gas, and dust distributions. 

Considering the limitations of current telescopes in the mm and sub-mm wavelength range, the ALPAKA sample -- although biased towards massive, actively star-forming galaxies in overdense environment --  will constitute a legacy for high-angular resolution studies in the next decade. We note, indeed, that high-angular resolution observations of the cold gas kinematics of typical star-forming galaxies at $z \sim 2$ are not feasible with ALMA, even considering the current 50 hours limit for normal observing programs. For instance, a Milky-Way progenitor with stellar and gas masses of $1 \times 10^{10}~\mathrm{M_{{\odot}}}$ at $z \sim 2.2$ \citep{vandokkum_2013} is expected to have a CO(3-2) luminosity of 10$^9$ K\,km\,s $^{-1}$\,pc$^2$ and a flux of 0.08 Jy\,km\,s $^{-1}$, that is a factor of $\approx 10$ smaller than the CO(3-2) fluxes of ALPAKA galaxies at similar redshifts (e.g., ID19-21). To observe such a Milky-Way progenitor at the same level of detail we have for ID19 (i.e., similar SNR and angular resolution) an on-source integration time of 300 hours -- a factor of 100 times the on source time of ID19 -- would be required. On the other hand, ngVLA, with a factor of 6 increase in sensitivity over ALMA \citep{Carilli_2022}, will enable the study of the dynamics of Milky-like progenitors at $z \sim 2$ in 8 hours of integration time\footnote{ngVLA will cover the CO(2-1) line at $z \sim 2$. Therefore, we compute the comparison of the required integration times assuming $I_{\mathrm{CO(3-2)}} \approx I_{\mathrm{CO(2-1)}}$.}.

We make the data cubes of the ALPAKA targets and the kinematic profiles derived in this manuscript publicly available, trusting that the community will exploit this sample well beyond what we had originally envisioned.

\begin{acknowledgements}
We thank the anonymous referee for their careful feedback. FR is grateful to Melanie Kaasinen and Federico Lelli for useful comments and discussions.
FR acknowledges support from the European Union’s Horizon 2020 research and innovation program under the Marie Sklodowska-Curie grant agreement No. 847523 ‘INTERACTIONS’ and from the Nordic ALMA Regional Centre (ARC) node based
at Onsala Space Observatory. The Nordic ARC node is funded through Swedish Research Council grant
No 2017-00648. The Cosmic Dawn Center (DAWN) is funded by the Danish National Research Foundation under grant No. 140. The project leading to this publication has received support from
ORP, that is funded by the European Union’s Horizon 2020 research
and innovation programme under grant agreement No 101004719 [ORP]. FRO acknowledges support from the Dutch Research Council (NWO) through the Klein-1 Grant code OCEN2.KLEIN.088. MK acknowledge support from the ERC Advanced Grant INTERSTELLAR H2020/740120 (PI: Ferrara). Any dissemination of results must indicate that it reflects only the author’s view and that the Commission is not responsible for any use that may be made of the information it contains.
LDM is supported by the ERC-StG ``ClustersXCosmo'' grant agreement 716762 and acknowledges financial contribution from the agreement ASI-INAF n.2017-14-H.0.

This paper makes use of the following ALMA data: 
ADS/JAO.ALMA\#2017.1.01659.S; ADS/JAO.ALMA\#2017.1.00471.S; ADS/JAO.ALMA\#2017.1.01674.S; ADS/JAO.ALMA\#2015.1.00862.S; ADS/JAO.ALMA\#2017.1.01228.S; ADS/JAO.ALMA\#2018.1.00974.S; ADS/JAO.ALMA\#2017.1.00413.S; ADS/JAO.ALMA\#2019.1.01362.S; ADS/JAO.ALMA\#2017.1.01045.S; ADS/JAO.ALMA\#2013.1.00059.S; ADS/JAO.ALMA\#2018.1.00543.S; ADS/JAO.ALMA\#2015.1.00723.S; ADS/JAO.ALMA\#2018.1.01146.S; ADS/JAO.ALMA\#2016.1.01155.S; ADS/JAO.ALMA\#2017.1.01677.S; ADS/JAO.ALMA\#2018.1.01306.S; ADS/JAO.ALMA\#2017.1.00908.S.
ALMA is a partnership of ESO (representing its member states), NSF (USA) and NINS (Japan), together with NRC (Canada), NSC and ASIAA (Taiwan), and KASI (Republic of Korea), in cooperation with the Republic of Chile. The Joint ALMA Observatory is operated by ESO, AUI/NRAO and NAOJ.\\
We acknowledge usage of the Python programming language \citep{python3}, Astropy \citep{astropy}, Matplotlib \citep{matplotlib}, NumPy \citep{numpy}, and SciPy \citep{scipy}.


\end{acknowledgements}

\bibliographystyle{aa}
\bibliography{ms.bib}

\appendix

\section{GALFIT fitting} \label{sec:galfit}
To determine the structural parameters of the stellar emission for the 23 ALPAKA galaxies with HST data, we perform the fitting of the 2D light profile with \galfit. The latter convolves a parametric 2D model with an input PSF and finds the best-fit parameters with a $\chi^2$-minimization. For fitting the ALPAKA targets, we assume a single Sérsic profile and we add additional components only when the isophotes of the target overlap with close sources (e.g., ID4, 8, 13). The 2D Sérsic profile is defined by 7 parameters in total (position of the center, effective radius $R_{\mathrm{e}}$, Sérsic index, axis ratio, position angle). As \galfit\ input, we give cutouts with a sidelength of 8$\arcsec$, an empirically determined PSF, the noise image and a mask. To build the PSF, we stack high SNR point-like sources, while the mask was built based on the {SE{\sc {xtractor}}} source detection \citep{Bertin_1996}.

\begin{figure*}[htbp!]
    \begin{center}
        \includegraphics[width=0.9\textwidth]{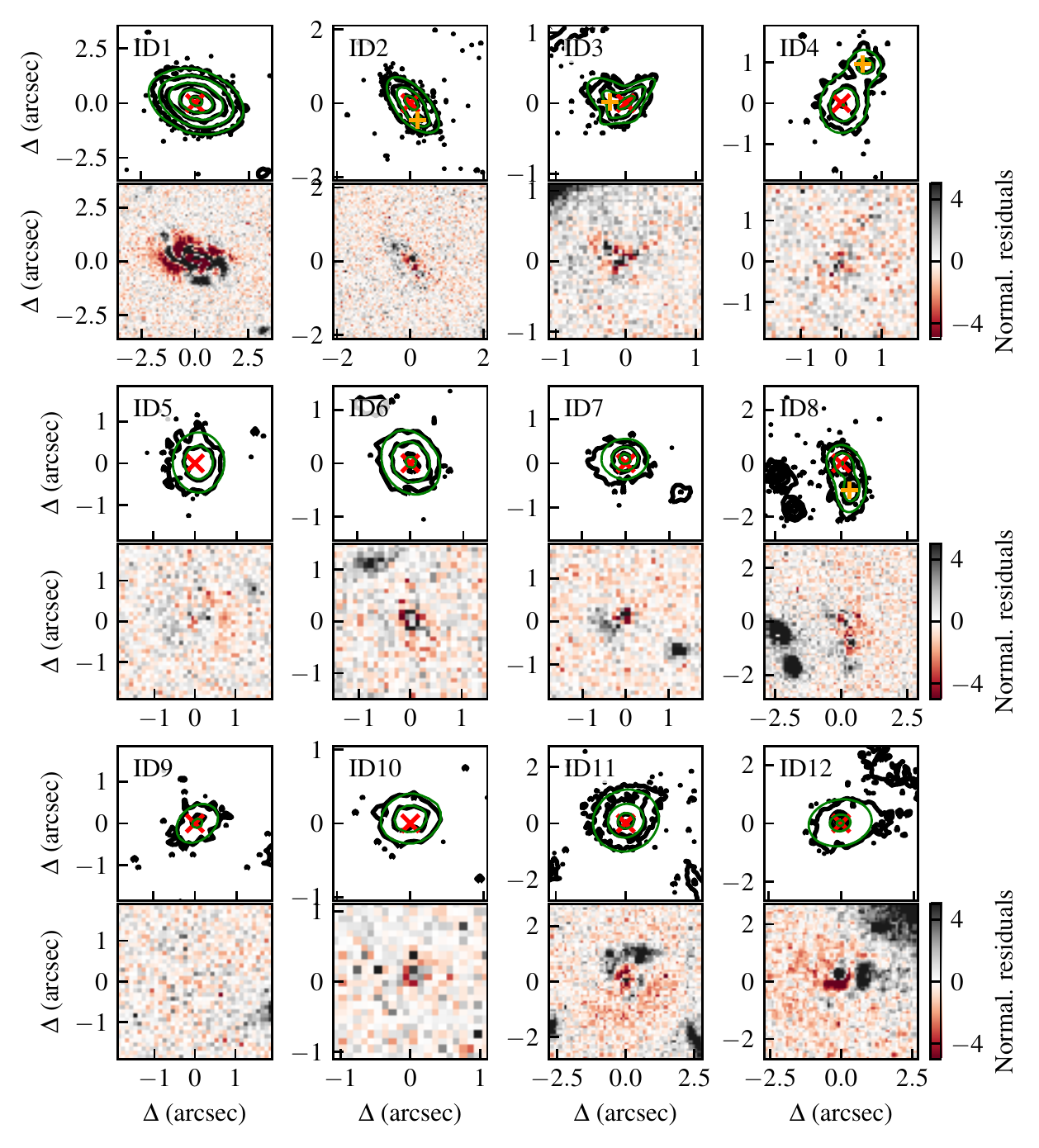}
        \caption{\galfit\ model and residuals. For each target, we show in the upper panel the contours of the HST data (black) and \galfit\ model (green). The levels of the contours are at [3, 9, 27, 81] $\times$ RMS.The bottom panel shows the residual normalized to the noise map. The red cross shows the center of the main galaxy and the orange crosses show the center of additional components that were added to the fitting, when necessary. }    
        \label{fig:galfit1}
    \end{center}
\end{figure*}

\begin{figure*} [htbp!]
    \begin{center}
        \includegraphics[width=0.9\textwidth]{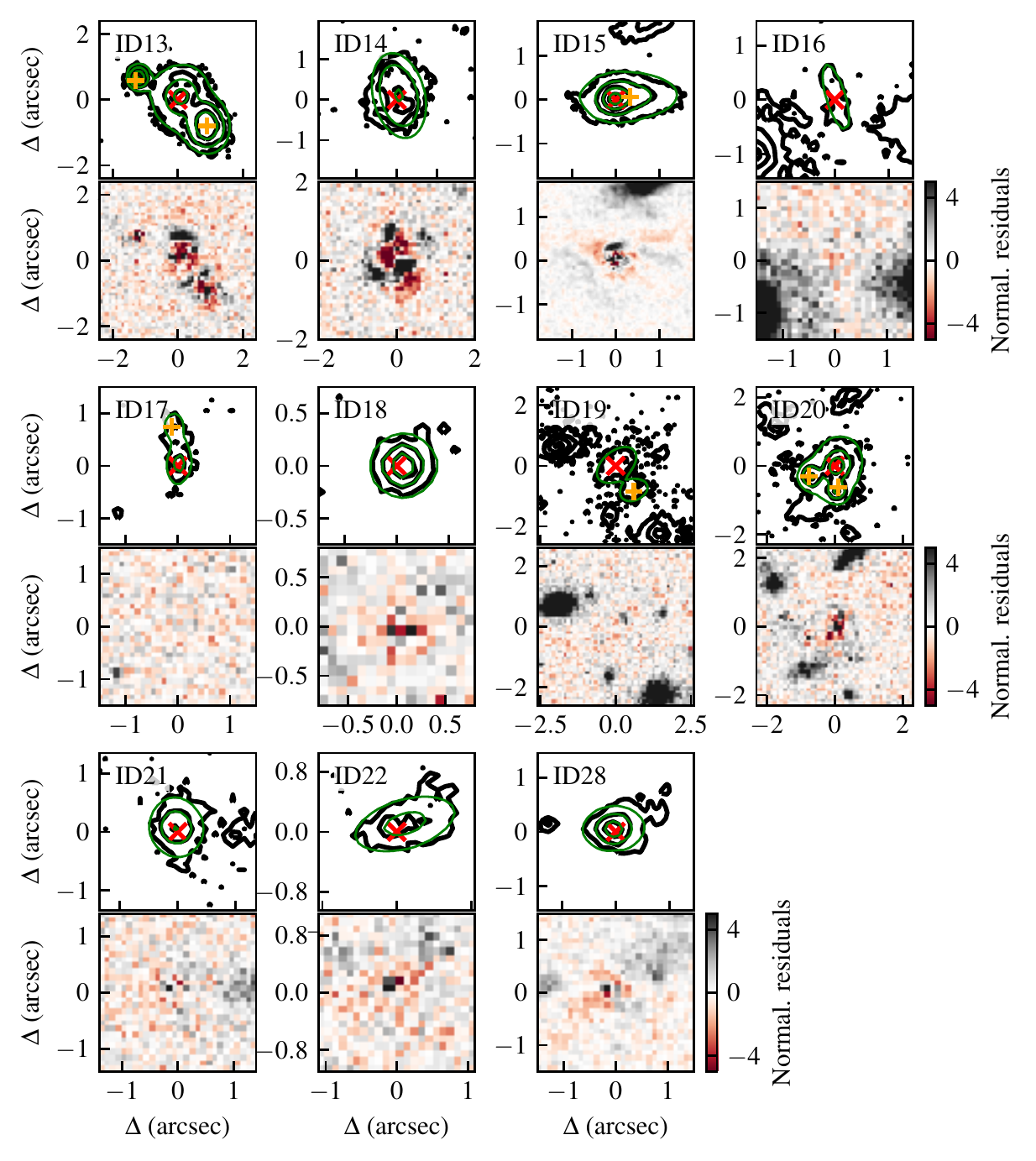}
        \caption{Same as Fig.~\ref{fig:galfit1}} 
        \label{fig:galfit2}
    \end{center}
\end{figure*}

In Figs.~\ref{fig:galfit1} and \ref{fig:galfit2}, we show the contours for the data and model (upper panels) and the residuals normalized to the RMS (bottom panels). The best-fit geometrical parameters (i.e., $i_{\mathrm{HST}}$, $PA_{\mathrm{HST}}$) are listed in Table \ref{tab:PA}. We note that, as discussed in Sect.~\ref{sec:geometry}), $i_{\mathrm{HST}}$ is derived by the fitted axis ratio $b/a$. To compute the parameter uncertainties, \galfit\ uses the covariance matrix produced during the least-squares minimization. These uncertainties would be correct if the residuals were only due to Poisson noise. However, this ideal situation does not occur often because the residuals are mainly due to systematics, such as the assumptions of the number and profile functions used in the fitting, the uncertainties in the PSF \citep[e.g.,][]{Peng_2010, Haussler_2007}. For some ALPAKA galaxies (e.g., ID1, 14), for example, the residuals have a symmetric pattern which indicate the presence of extra components. In ID1, spiral arms are clearly visible in the HST image, while ID14 has two bright clumps. However, since the main aim of this analysis is to fit the structural parameters that describe the bulk of the light emission, rather than to describe in detail the morphology, we decide to consider as fiducial the model obtained assuming a single Sérsic component. Considering that the PSF uncertainties are typically of the order of 5\% \citep[e.g.,][]{Tacchella_2015, Lustig_2021}, the major source of uncertainties is due, therefore, to the prior assumptions on the number of components and their profiles. To take into account these uncertainties, which we cannot quantify, we choose to make a conservative choice and consider a 15\% uncertainties on the parameters fitted with \galfit\ \citep[e.g.,][]{vanderwel_2012}. 

\section{Inclination angles}
\label{app:inclination}
The median value of the inclination distribution used for the kinematic fitting is 49$^{+20}_{-14} \deg$. This value is smaller, although within 1-$\sigma$, than the average inclination of 60 $\deg$ expected from the observation of randomly oriented galaxies \citep{Romanowsky_2012}. To assess this quantitatively, we apply the Kolmogorov-Smirnov test and find that the probability that these two inclination distributions (i.e., inclination of ALPAKA targets and randomly oriented galaxies) are consistent between each other is of 90\%. Although this falls below the standard threshold ($\sim95\%$) for confirming the consistency of a given sample with a reference probability distribution \citep{Hodges_1958}, such a result show that any potential effects that would affect the derivation of the inclination angle do not have a significant impact. Nevertheless, the slight discrepancy could be indicative of systematic biases.

For instance, a potential reason for this discrepancy can be due to an observational bias towards low-inclination galaxies. The line detection of galaxies is, indeed, more challenging for edge-on rather than face-on galaxies \citep[e.g.,][]{Kohandel_2019}. At fixed line luminosity, the Full Width at Half Maximum is larger in the edge-on case, pushing the peak flux below the detection limit. 

Another potential reason for the low inclination of the ALPAKA sample may be due to our assumption of the thickness of the disks. By assuming that the disks are thin, we derived the inclination angles from the measurements of the axis ratio $b/a$. However, if galaxies have a thick disk, these inclination angles are underestimated. A common assumption for deriving the inclination of thick disks is to estimate the inclination using the following equation, 
\begin{equation}
    \cos i = \sqrt{\frac{(b/a)^2 - Q^2}{1- Q^2}},
    \label{eq:cos}
\end{equation}
where $Q$ is the intrinsic axis ratio that is often assumed equal to 0.2 \citep[e.g.,][]{Forster_2018}. If we use Eq.~(\ref{eq:cos}) to compute the inclination angles, the values change by 2 to 7\%, with a median value of the distribution of 54 $\deg$, consistent with the one obtained with the thin disk assumption. Therefore, an inclination angle obtained using Eq.~(\ref{eq:cos}) would make no significant difference with respect to the thin-disk assumption adopted to infer the kinematic parameters presented in Sect.~\ref{sec:disks}.

\section{Channel maps}\label{sec:ch_maps}
\begin{figure*}
    \begin{center}
        \includegraphics[width=0.9\textwidth]{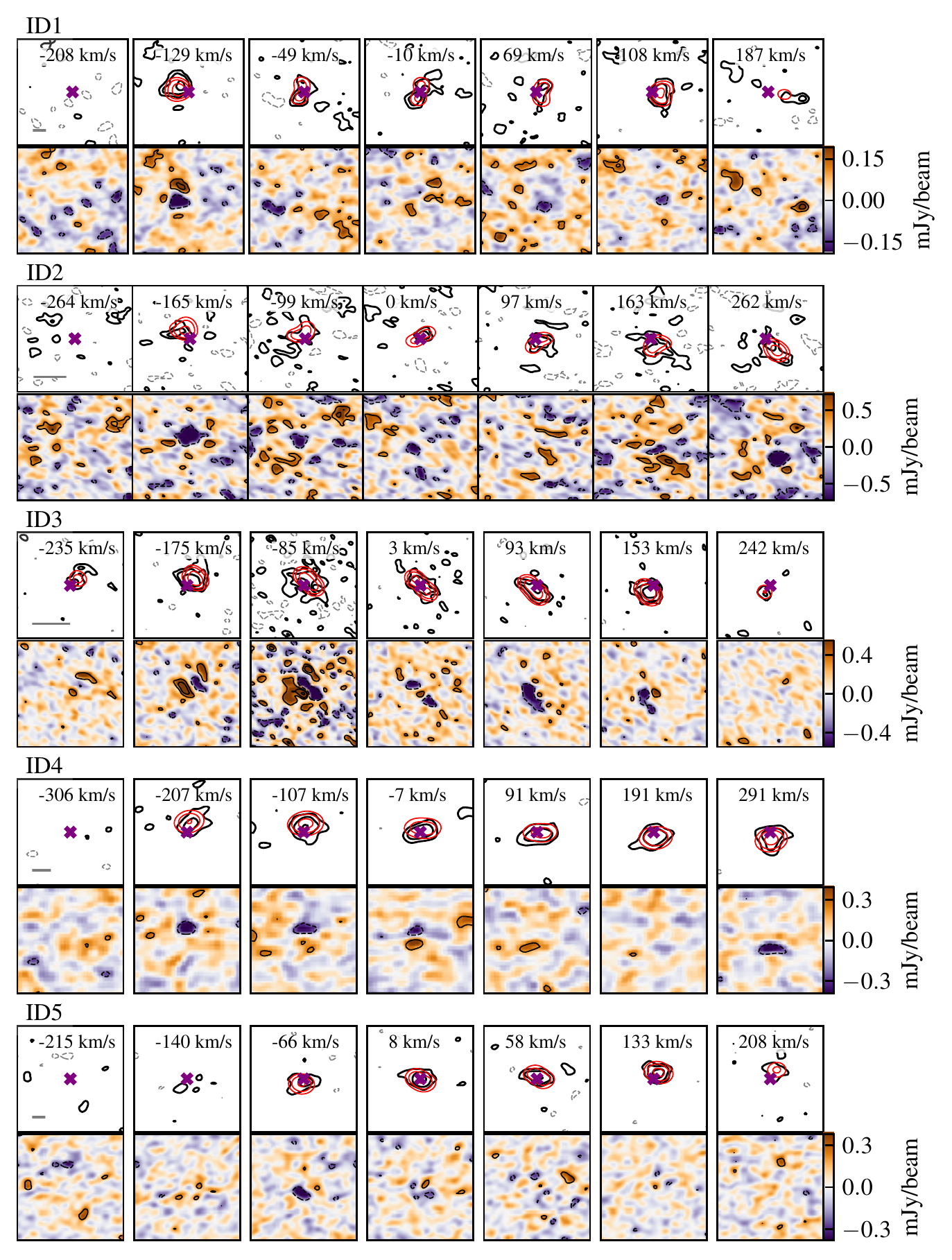}
        \caption{For each ALPAKA target, we show 7 representative channel maps for the data (black contours) and the rotating disk model (red contours) in the upper panels and the residuals in the bottom panels. The solid contours of the data and the models are at [1, 2, 4, 8, 16, 32] $\times$ 2.5 RMS, while the dashed gray contours are at -2.5 RMS. The purple cross shows the location of the center of the disk model. For the residuals, we show the emission over a scale of $\pm 5$ RMS. The gray bar in the first channel of each target (bottom left) shows a scale of 0.5$\arcsec$. }  
        \label{fig:channel1} 
    \end{center}
\end{figure*}

\begin{figure*}
    \begin{center}
        \includegraphics[width=0.9\textwidth]{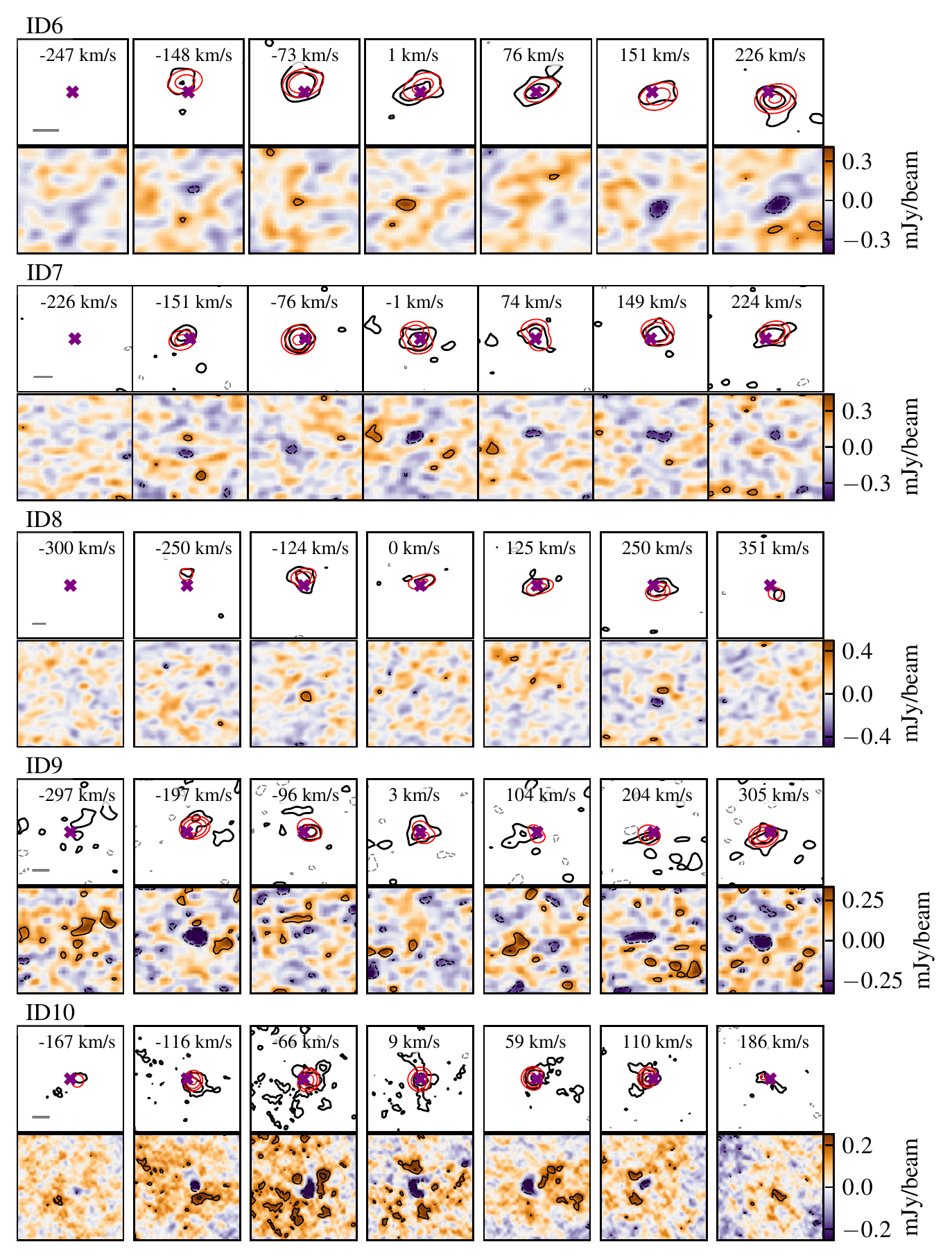}
        \caption{Same as Fig.~\ref{fig:channel1}, but for ID6 - 10.}  
        \label{fig:channel2} 
    \end{center}
\end{figure*}

\begin{figure*}
    \begin{center}
        \includegraphics[width=0.9\textwidth]{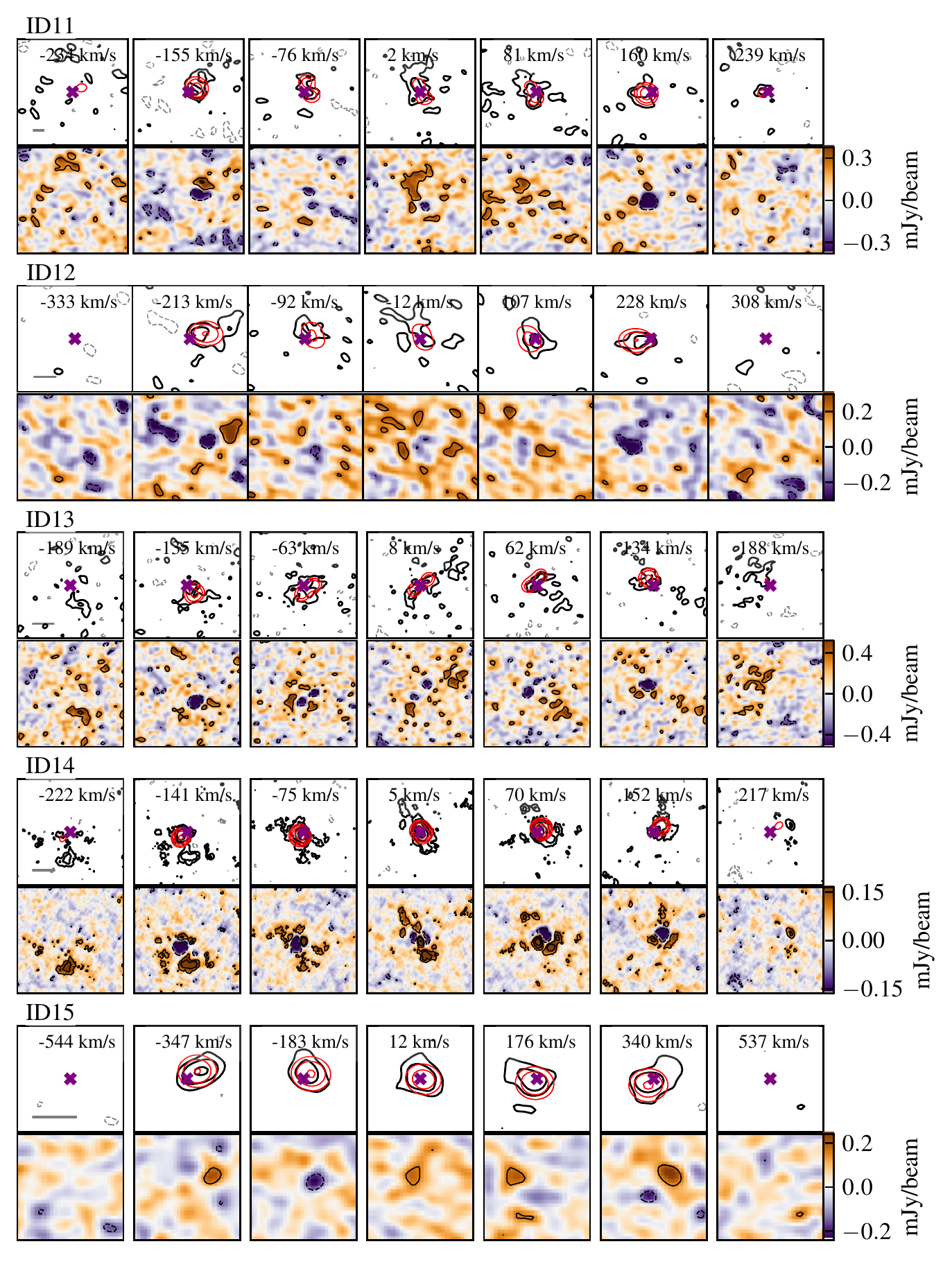}
        \caption{Same as Fig.~\ref{fig:channel1}, but for ID11 - 15.}  
        \label{fig:channel3} 
    \end{center}
\end{figure*}

\begin{figure*}
    \begin{center}
        \includegraphics[width=0.9\textwidth]{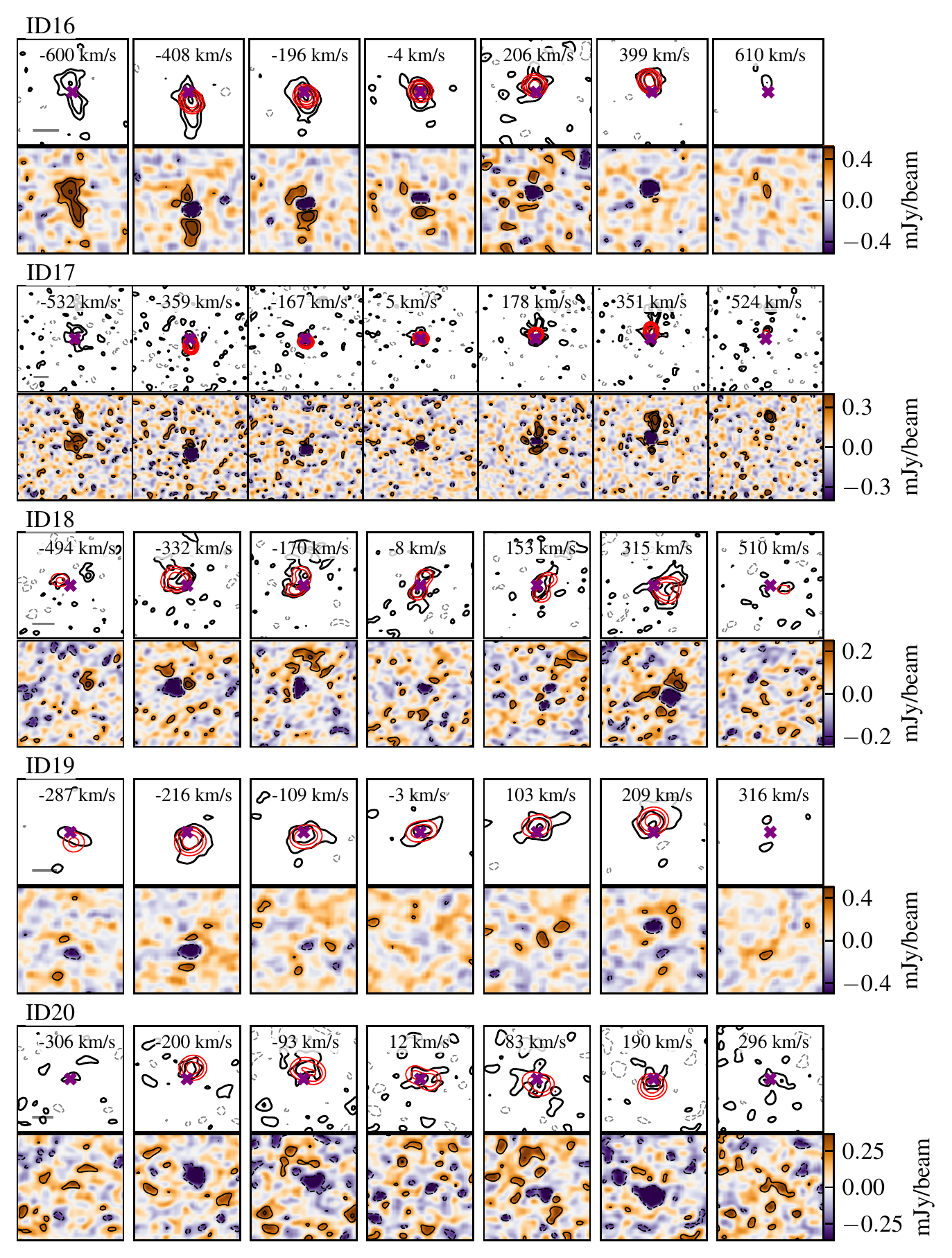}
        \caption{Same as Fig.~\ref{fig:channel1}, but for ID16 - 20.}  
        \label{fig:channel4} 
    \end{center}
\end{figure*}

\begin{figure*}
    \begin{center}
        \includegraphics[width=0.9\textwidth]{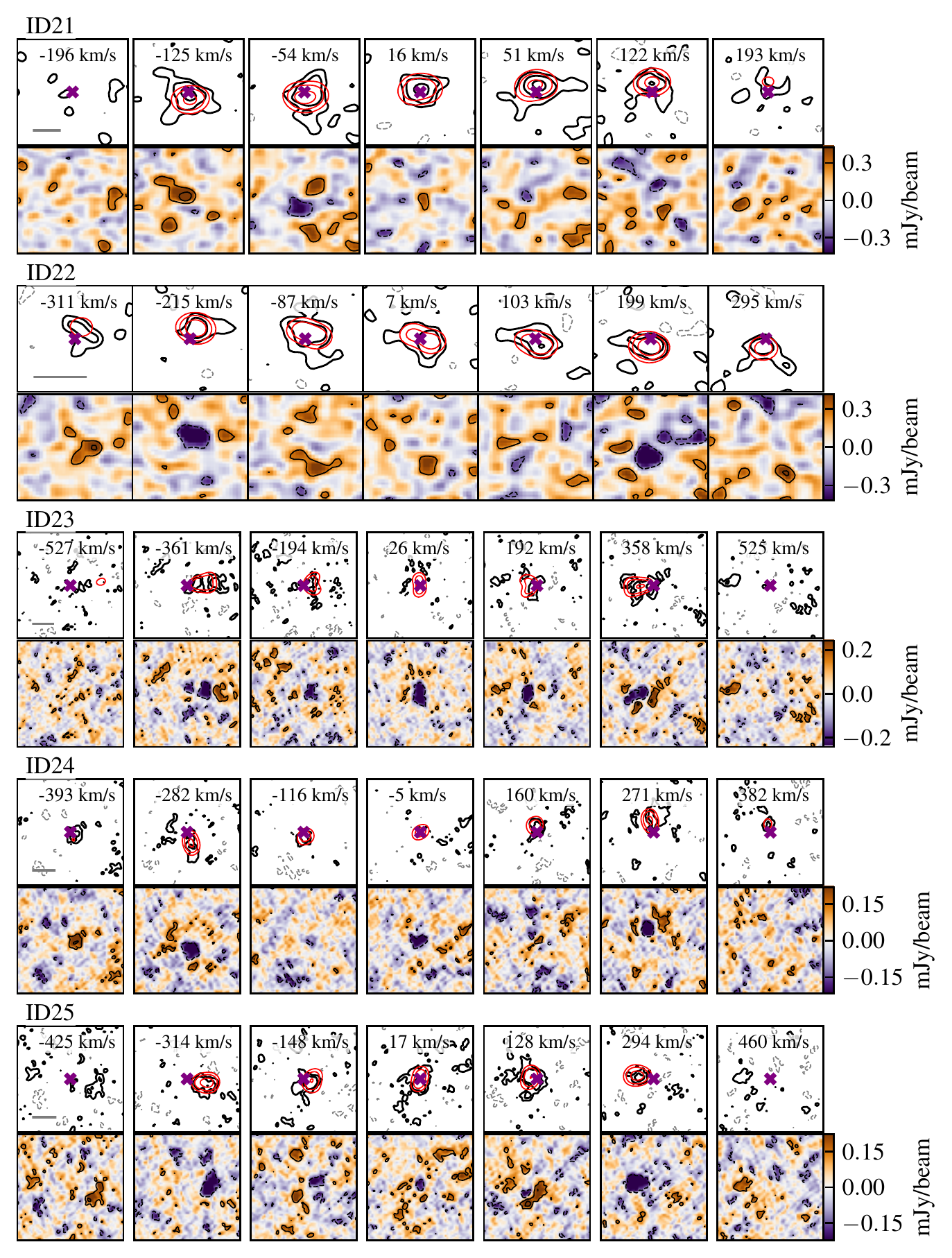}
        \caption{Same as Fig.~\ref{fig:channel1}, but for ID21 - 25.}  
        \label{fig:channel5} 
    \end{center}
\end{figure*}

In Figs.~\ref{fig:channel1}-\ref{fig:channel5}, we show representative channel maps for the data, the model and the corresponding residuals for the ALPAKA galaxies and the corresponding model created by \bba.

\begin{figure*}
    \begin{center}
        \includegraphics[width=0.9\textwidth]{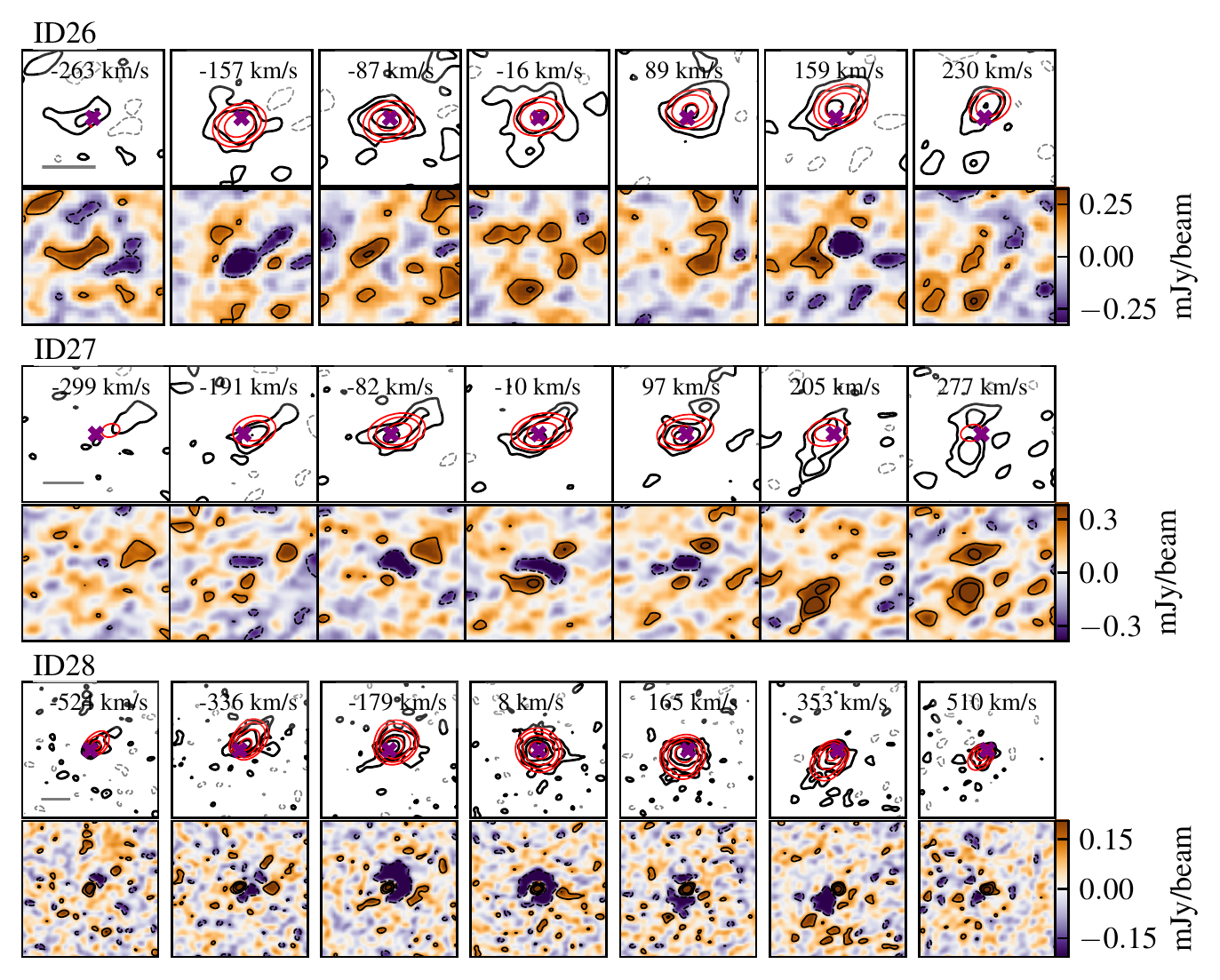}
        \caption{Same as Fig.~\ref{fig:channel1}, but for ID26 - 28.}  
        \label{fig:channel6} 
    \end{center}
\end{figure*}

\end{document}